\documentclass[a4paper,11pt]{article}
\pdfoutput=1 

\usepackage{jheppub} 
\usepackage[T1]{fontenc} 

\usepackage{textcomp}
\usepackage{lineno}

\usepackage{natbib} 

\title{A deep learning method for the trajectory reconstruction of cosmic rays with the DAMPE mission}

\author[a,1]{Andrii Tykhonov\note{Corresponding author.},} 
\author[a]{Andrii Kotenko,}
\author[a]{Paul Coppin,}
\author[a]{Maksym Deliyergiyev,}
\author[a]{David Droz,}
\author[b]{Jennifer Maria Frieden,}
\author[b]{Chiara Perrina,}
\author[a]{Enzo Putti-Garcia,}
\author[a]{Arshia Ruina,}
\author[a]{Mikhail Stolpovskiy,}
\author[a]{Xin Wu.}

\affiliation[a]{Department of Nuclear and Particle Physics, University of Geneva, CH-1211, Switzerland}
\affiliation[b]{Institute of Physics, Ecole Polytechnique F\'{e}d\'{e}rale de Lausanne (EPFL), CH-1015 Lausanne, Switzerland}

\emailAdd{andrii.tykhonov@unige.ch}

\newcommand{\micron}{$\text{\textmu m}$}

\newcommand{\mm}{$\text{mm}$}
\newcommand{\degree}{$^{\circ}$}

\abstract{
A deep learning method for the particle trajectory reconstruction with the DAMPE experiment is presented. The developed algorithms constitute the first fully machine-learned track reconstruction pipeline for space astroparticle missions. Significant performance improvements over the standard hand-engineered algorithms are demonstrated. Thanks to the better accuracy, the developed algorithms facilitate the identification of the particle absolute charge with the tracker in the entire energy range, opening a door to the measurements of  cosmic-ray proton and helium spectra at extreme energies, towards the PeV scale, hardly achievable with the standard track reconstruction methods. In addition, the developed approach  demonstrates an unprecedented accuracy in the particle direction reconstruction with the calorimeter at high deposited energies, above several hundred GeV for hadronic showers and above a few tens of GeV for electromagnetic showers. }

\begin{document}
\maketitle
\flushbottom

\section{Introduction}

The DArk Matter Particle Explorer (DAMPE) mission was launched on December 17, 2015, from the Gobi desert in China. It operates on a 500 km sun-synchronous orbit in sky survey mode, accumulating about two billion cosmic-ray events per year~\cite{dampe_mission}. The broad scientific program of DAMPE includes  the measurement of the cosmic electron and positron ($e^-+e^+$) spectrum in the energy range between a few GeV and about 10 TeV,  the measurement of the cosmic-ray proton and ion spectra in the particle kinetic energy range between 10 GeV and hundreds of TeV, and gamma-ray physics. The DAMPE instrument is composed of four main subdetectors: a bismuth germanium oxide (BGO) imaging calorimeter,  segmented in 14 layers of 22 bars in a hodoscope arrangement,  with a total thickness of about 32 radiation lengths for precise energy measurements and the electron/hadron separation~\citep{Wei:2019rep,Zhang:2016xkz,Zhang_2012}; a Silicon Tungsten tracKer-converter (STK) with 12 layers of single-side silicon trip detectors, 6  in the $x$  direction and 6 in the $y$ direction, for a precise cosmic-ray trajectory reconstruction, the absolute charge (Z) measurement, and the identification of the gamma-ray direction through the photon conversion into an electron--positron pair~\citep{Tykhonov:2018stq,Tykhonov:2017uno,Azzarello:2016trx}; 
 a Plastic Scintillator Detector (PSD) consisting of two double-layers of scintillator bars for the cosmic-ray absolute charge measurement and also serving as a veto for the gamma-ray detection~\citep{Ding_2019,Yu:2017dpa}; a NeUtron Detector (NUD) consisting of four boron-doped scintillator tiles enhancing the electron/hadron discrimination power~\citep{Huang:2020skz}. 
   Thanks to the fine-segmented thick calorimeter and relatively large acceptance, DAMPE is capable of detecting cosmic rays with very good energy resolution, $\sim$1.5\% for electrons and gamma rays~\citep{dampe_nature}, and $\sim$30\% for protons and ions~\citep{dampe_science}. 
  
Recent DAMPE results~\citep{dampe_prl,dampe_science} provide the most precise measurements of cosmic-ray proton and helium spectra at the highest particle kinetic energies reached by direct-detection experiments, with unprecedented statistical accuracy and energy resolution. They confirm the previously established hardening of both proton~\citep{CALET:2019bmh,Atkin:2018wsp,Yoon:2017qjx,ams_2015,PAMELA:2017bna,pamela_2014,ADRIANI2013219,pamela_2011,atic_2009} and helium~\citep{ams_2017,ams_2015,Yoon:2017qjx,Atkin_2017,ADRIANI2013219,pamela_2011,Ahn_2010,atic_2009} spectra at about a few hundred GeV per nucleon, and reveal a new spectral feature, a softening at about 14~TeV and 34~TeV particle kinetic energy for proton and helium, respectively. The measurements are compatible with the hypothesis of a particle charge dependence  of the softening, indicating the presence of  a nearby cosmic-ray source, like a supernova remnant (SNR), with the acceleration cutoff  corresponding to the softening energy. The complex spectral structures  challenge the long-standing paradigm of the SNR origin of galactic cosmic rays~\citep{Ohira:2015ega,Mollerach:2017idb}. In particular, it is not clear whether SNRs can accelerate cosmic rays to PeV energy or whether other sources are needed~\citep{Cristofari:2021jkl,TibetASg:2021kgt,HAWC:2020nvc,Ohira:2017bxa}. Future  cosmic-ray measurements towards the PeV scale with the existing particle detectors in space  (DAMPE, CALET~\citep{CALET:2019gex}, ISS-CREAM~\citep{iss_cream_mission_2014}) and the next generation calorimetric space observatory (HERD~\citep{herd_2021_icrc_gargano}) are therefore of paramount importance in order to shed light on the century-long puzzle of cosmic-ray origin. However, while calorimetric experiments provide unique opportunities to explore the TeV--PeV domain, the systematic uncertainties related to the  conventional data reconstruction techniques hinder such measurements.   
 
One of the key challenges in direct cosmic-ray detection comes from the  limited precision of the absolute charge identification. It is  directly linked with the accuracy of the particle trajectory reconstruction. In particular, considering  the cosmic-ray proton spectrum measured by DAMPE~\citep{dampe_science}, the systematic uncertainties of the  analysis grow  rapidly with energy  rendering the adopted cosmic-ray reconstruction and identification techniques insufficient for future measurement at a few hundred TeV and higher energies.  This problem is critical for any other cosmic-ray analysis, including the recently published helium spectrum~\citep{dampe_prl} and future measurements of heavier nuclei~\citep{dampe_carbon_2021,dampe_boroncarbon_2021}. 
 
The goal of this work is to develop a new method for the reconstruction of cosmic-ray tracks, in order to enable a reliable and accurate  absolute charge identification in the entire energy range, in particular at the TeV--PeV scale. Our approach deviates from the conventional combinatorial pattern recognition  adopted by the DAMPE Collaboration
~\citep{Tykhonov:2017uno,Tykhonov:2017rdq} and, in similar ways, in other major calorimetric and spectrometer experiments in space~\citep{bruel2018fermilat,fermi_mission_2009,calet_tracking_2017,ALPAT2010207,cream_tracking_2019}, toward the rapidly developing deep learning domain~\citep{RevModPhys.91.045002,lecun_2015_nature}. The paper is structured as follows. In Section~\ref{sec:data} we describe the data sample and the Monte-Carlo simulation used in this work. In Section~\ref{sec:reconstruction} we briefly introduce the existing DAMPE data reconstruction pipeline and identify the key challenges of the cosmic-ray track reconstruction and absolute charge identification. In the subsequent parts, we describe the constituent blocks of the deep learning machinery developed for the  particle trajectory reconstruction, which includes the shower axis direction finding in the calorimeter in Section~\ref{sec:calo},  and the particle direction reconstruction in the tracker in Section ~\ref{sec:tracker}. Next, in Section~\ref{sec:classifier} we introduce the neural network algorithm developed to reject particles inelastically interacting before the tracker, as a prerequisite for the reconstruction of the particle charge. Finally, in Section~\ref{sec:analysis}  we apply the developed methods on the proton and helium flight data samples and we compare the absolute charge identification performance based on the new approach with the one based on the standard DAMPE track reconstruction. The results and implications are discussed in Section~\ref{sec:conclusion}.

\section{Data and simulation}
\label{sec:data}
The data sample used for this analysis was collected by the DAMPE instrument during the period between December 2015 and December 2021. The raw data transferred from the satellite are processed offline with the standard software pipeline, including the energy and shower axis direction reconstruction with the calorimeter~\citep{Wu:2018gyd}, the track pattern recognition based on the Kalman filter approach~\citep{Tykhonov:2017rdq}, the reconstruction and correction of the signals in PSD~\citep{Dong:2018qof,Ma:2018brb}, and the internal alignment of STK~\citep{Tykhonov:2017uno}.

The full-scale Monte-Carlo (MC) simulation of the DAMPE detector is based on the GEANT4 toolkit  version 4.10.5~\citep{Allison:2016lfl}. It is used to perform the training of deep learning models, to optimize the event selection and to estimate the  performance of the developed algorithms. The proton and helium cosmic-ray spectra are generated in the particle kinetic energy range between 10 GeV and 1 PeV following a power-law distribution with the spectral index of -1. For the simulation up to 100~TeV, we use the FTFP\_BERT hadronic model~\citep{ANDERSSON1987289,NILSSONALMQVIST1987387,WRIGHT2015175}, while above 100 TeV we employ the EPOS-LHC~\citep{Pierog:2013ria} model from the CRMC package\footnote{Cosmic Ray Monte Carlo (CRMC): https://web.ikp.kit.edu/rulrich/crmc.html} linked to GEANT4 using a previously developed  interface~\citep{tykhonov_icrc_2019}.  The electron MC sample is generated in the energy range between 1 GeV and 50 TeV. To ensure a good match with the real data, the detector geometry is implemented in the simulation from Computer-Aided-Design (CAD) drawings, using a software conversion toolkit\footnote{https://github.com/tihonav/cad-to-geant4-converter}~\citep{Tykhonov:2017rdq}. For a fair comparison, the simulated data are processed with the same reconstruction and analysis chain as the flight data.  

In addition to GEANT4, an alternative DAMPE simulation based on the FLUKA version 2011.2$\mathrm{X}$.7~\citep{BOHLEN2014211} is also used at 100 TeV and higher energies. It incorporates the DPMJET3~\citep{Roesler:2000he} model for nucleus-nucleus interaction above 5 GeV/n. We profit from the FLUKA samples as they allow us to test the stability of the deep learning models and improve the model performance by extending the statistics of the training sample.

 \begin{table}
 \begin{tabular}{cc r@{--}l ccl}
\hline
Particle Type & Generator & \multicolumn{2}{c}{Energy Range} &  \multicolumn{3}{c}{Statistics (events) $\times10^6$} \\
\hline
 &  & 10 GeV~~&~1 TeV & & & 2402   \\
$p$ & GEANT4 & 1 TeV~~&~100 TeV & & & 100   \\ 
 &  & 100 TeV~~&~1 PeV & & & 13   \\ 
\hline
&  & 10 GeV~~&~1 TeV & & & 348   \\ 
$\mathrm{He}$  & GEANT4 & 1 TeV~~&~100 TeV & & & 388   \\ 
 &  & 100 TeV~~&~1 PeV & &  & 24   \\   
 \hline
$\mathrm{He}$  & FLUKA & 100 TeV~~&~500 TeV & &  & 16   \\   
\hline
 &  & 1 GeV~~&~100 GeV & & & 542   \\  
   $e^-$ & GEANT4 & 100 GeV~~&~1 TeV & & & 289   \\  
 &  & 1 TeV~~&~50 TeV & & & 122   \\  
\hline
\end{tabular}
\caption{Monte-Carlo (MC) samples used in this work and their total generated statistics.} 
\label{tab:mc_samples}
\end{table}

The list of all MC samples is given in Table~\ref{tab:mc_samples}. It is worth noting that less than 2\% of generated MC events correspond to particles that geometrically pass through all the DAMPE subdetectors. These events are analyzed and processed with the deep learning model training. This feature is explained by the fact that particles are generated on the surface of a sphere surrounding the DAMPE satellite in order to mimic the isotropic flux of cosmic rays~\citep{dampe_science,dampe_prl}.

\section{Cosmic-ray reconstruction and  identification}
\label{sec:reconstruction}

\begin{figure}
\begin{center}
\includegraphics[width=1.0\textwidth]{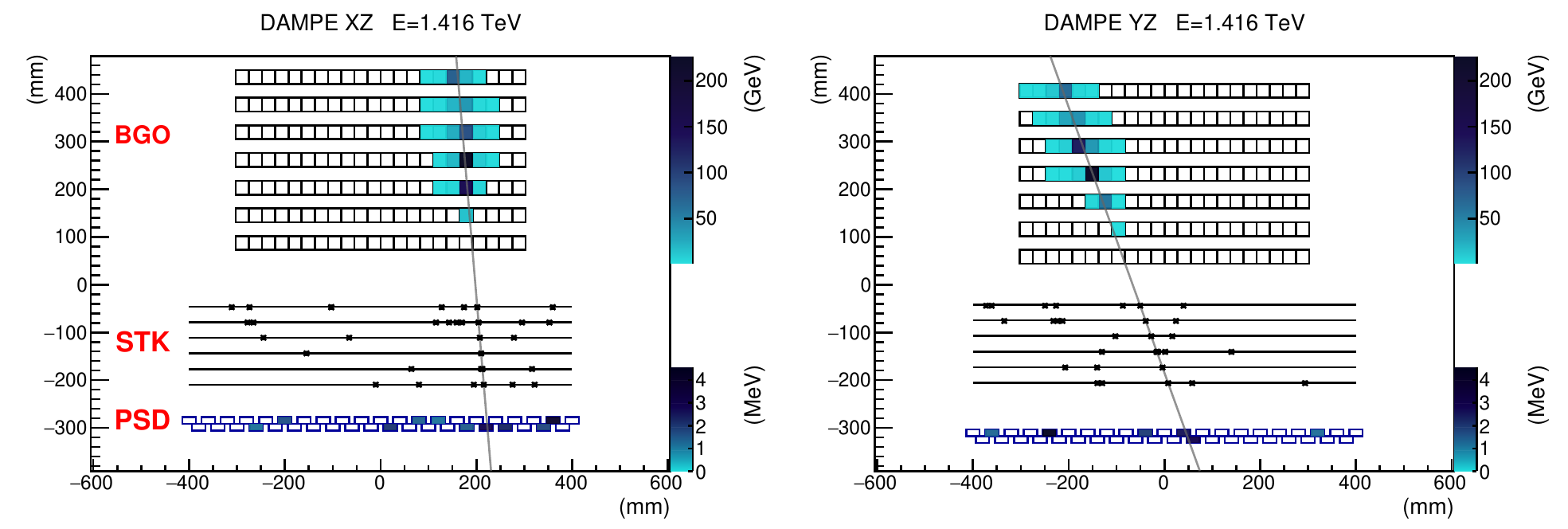}
\includegraphics[width=1.0\textwidth]{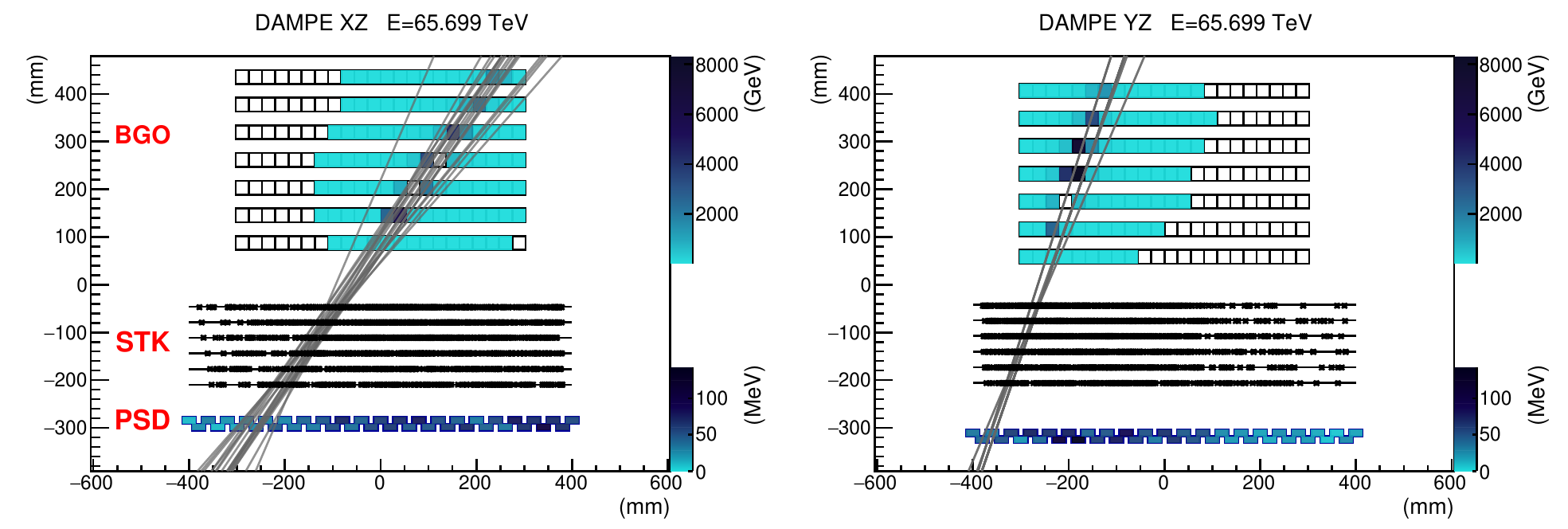}
\end{center}
\caption{Typical displays of simulated cosmic-ray protons in DAMPE. The proton kinetic energy is  3.8 TeV (top) and 179 TeV (bottom), respectively. Both events are shown in two orthogonal views of the detector (corresponding to the left and right subfigures). Three subdetectors can be seen, from top to bottom: these are the calorimeter (BGO), tracker (STK), and plastic scintillator detector (PSD).  Hits in the tracker are shown with black stars. Track candidates reconstructed with the standard algorithm~\citep{Tykhonov:2017uno,Tykhonov:2017rdq} are shown with gray lines. The total deposited (observed) energy in BGO is indicated on top of the figure.} 
\label{fig:event_displays}
\end{figure}

Passages of cosmic-ray particles through the DAMPE detector are illustrated in Figure~\ref{fig:event_displays}. The standard procedure for the trajectory reconstruction and particle identification can be grouped into the following steps:
\begin{itemize}
\item Reconstruction of the shower axis direction in the BGO calorimeter;
\item Track reconstruction in the STK using the  BGO shower axis direction as a seed; 
\item Selection of the best STK track from the ensemble of candidate tracks;
\item Projection of the STK track onto PSD, calculation of the path length; 
\item Measurement of the particle absolute charge with PSD using the STK track projection.
\end{itemize}

Normally, an additional selection criterion is applied requiring consistency between signals in different PSD bars along the path of the particle to ensure the correct absolute charge identification, which could otherwise be altered by the inelastic interactions of cosmic rays inside PSD~\citep{dampe_science,dampe_prl}. The particle track finding starts with the reconstruction of the shower direction in BGO, which is obtained from the fit of the energy-weighted ``cluster'' positions in  different calorimeter layers~\citep{dampe_mission}. A somewhat similar approach is adopted by other calorimetric experiments to date, including Fermi-LAT~\citep{fermi_mission_2009}, CALET~\citep{AKAIKE2010690}, and CREAM~\citep{cream_tracking_2019}. The direction of the shower axis reconstructed in the calorimeter provides the region of interest to look for candidate hits in STK, as well as the seed direction to be fed as a ``best guess'' to the Kalman filter algorithm. A further track finding  is done through the combinatorial search and simultaneous (Kalman) fitting of the potential track candidates in STK. Poor-quality and ghost tracks  are removed, keeping only those that pass certain quality criteria. It is the task of the further analysis  to identify which track best corresponds to the direction of the impinging cosmic ray. In other words, two conditions must be fulfilled in order to identify the real particle track: (a) it has to be reconstructed and present in the pool of the candidate tracks; (b) it has to be correctly selected on the analysis level. Once the track is identified, it is projected onto the PSD subdetector, which can provide the absolute charge measurement with high resolution in a broad dynamic range, up to nickel (Z=28)~\citep{Dong:2018qof,Ma:2018brb}. Moreover, it is worth mentioning that  the width of a PSD bar is 2.8 cm, which is much higher than the pointing resolution of the STK, 50--100~\micron~\citep{Tykhonov:2017uno}. Hence, the PSD measurement is not so vulnerable  to  potential errors in the STK track identification. In other words, given that the selected STK track candidate is relatively close to the real trajectory of the particle, even if it is wrongly identified, the PSD measurement is likely correct. With the argumentation above, the PSD rightfully serves as a major tool for absolute charge identification in DAMPE. However, the advantage of a relatively large PSD bar size turns into a weakness at high energies,  especially  in the context of proton and light ion identification, as described below.

Figure~\ref{fig:signal_psd_stk_truth} demonstrates the ultimate charge identification capacity of PSD and STK with respect to protons and helium nuclei, in different energy bins, up to 1 PeV. The distributions are obtained from simulation, using the true particle direction. In addition, a selection is applied on the MC truth level requiring no inelastic interaction inside PSD. The absolute charge measured with the PSD is defined as the minimum signal among PSD bars crossed by the particle\footnote{We also tested other algorithms, including mean and truncated mean, and found that the ``minimum'' algorithm shows optimal proton--helium separation in PSD.}. For the STK charge calculation, hits closest to the true direction in each layer are selected, with a maximum of 12 hits. If no hit within 0.3~\mm~is found in an STK layer, it is skipped\footnote{For STK we also tested a cut of 0.6~\mm~and found no significant difference in the results.}. The STK measurement is obtained as the average of the signals of associated hits, excluding layers with abnormally high signals coming from  particles that pre-shower inside the STK or due to the tail of the Landau distribution~\citep{1965417}. 
   
\newcommand{\phefiguresize}{0.48}
\begin{figure}
\begin{center}
\includegraphics[width=\phefiguresize\textwidth]{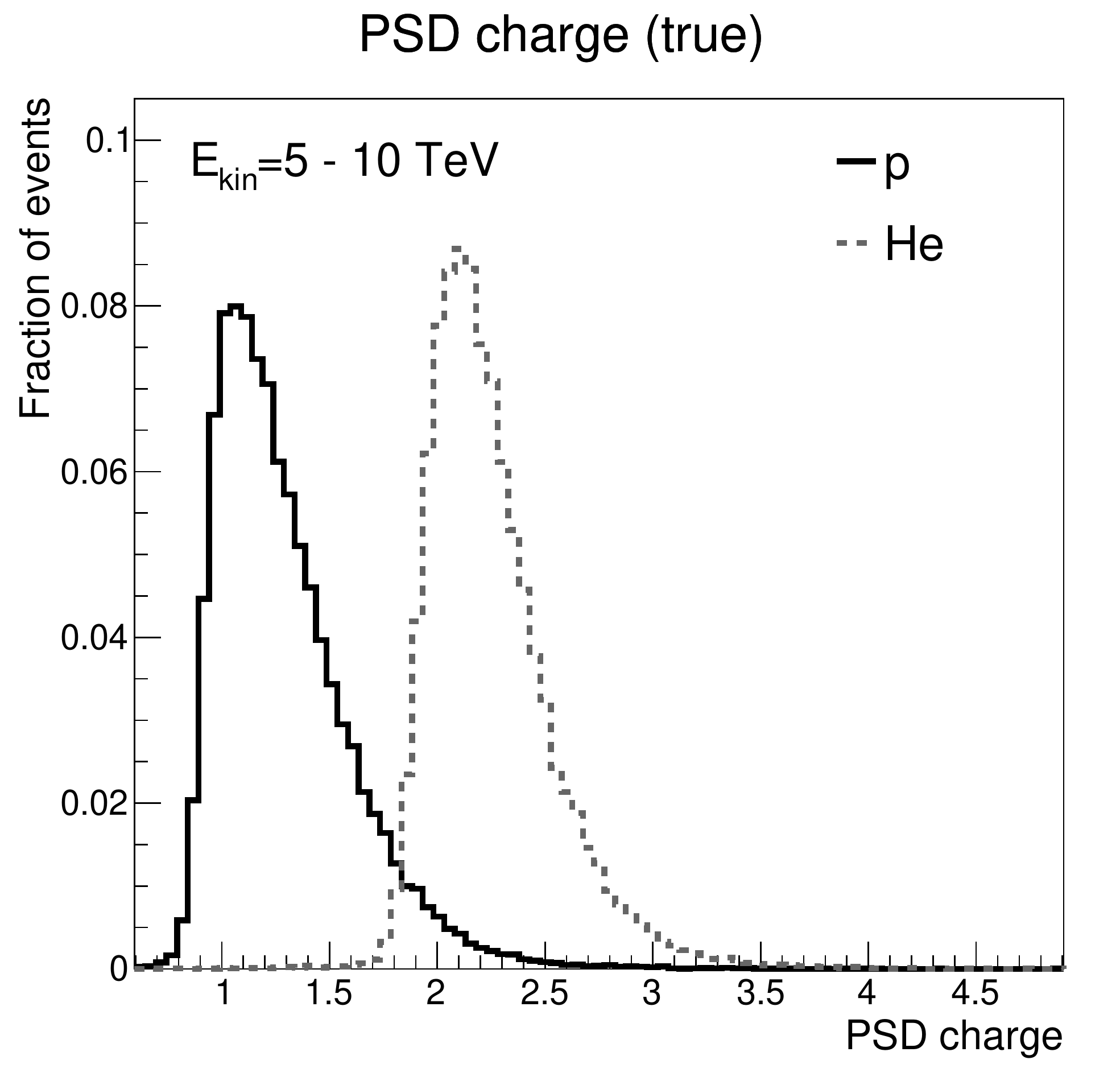}
\includegraphics[width=\phefiguresize\textwidth]{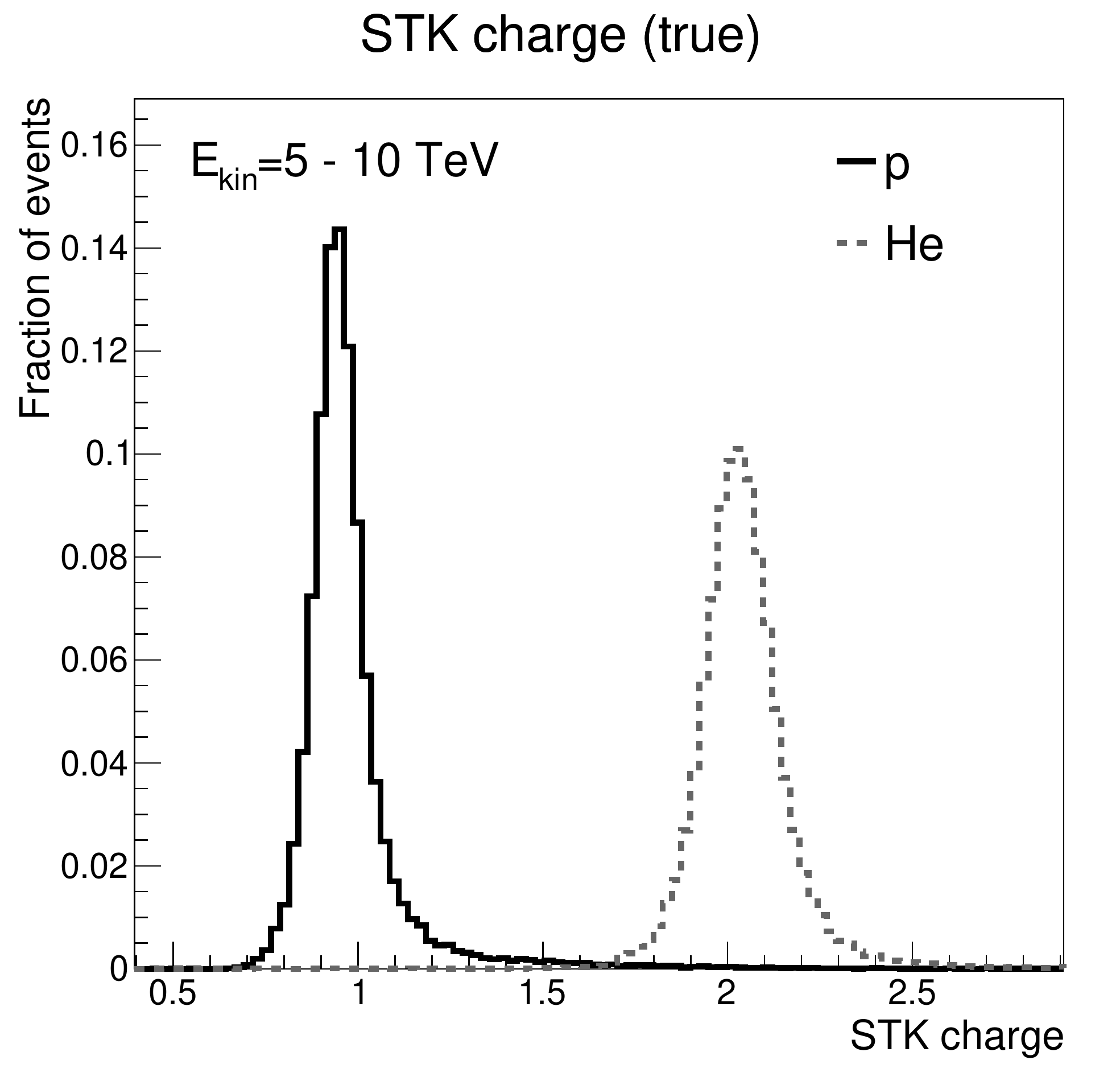}
\includegraphics[width=\phefiguresize\textwidth]{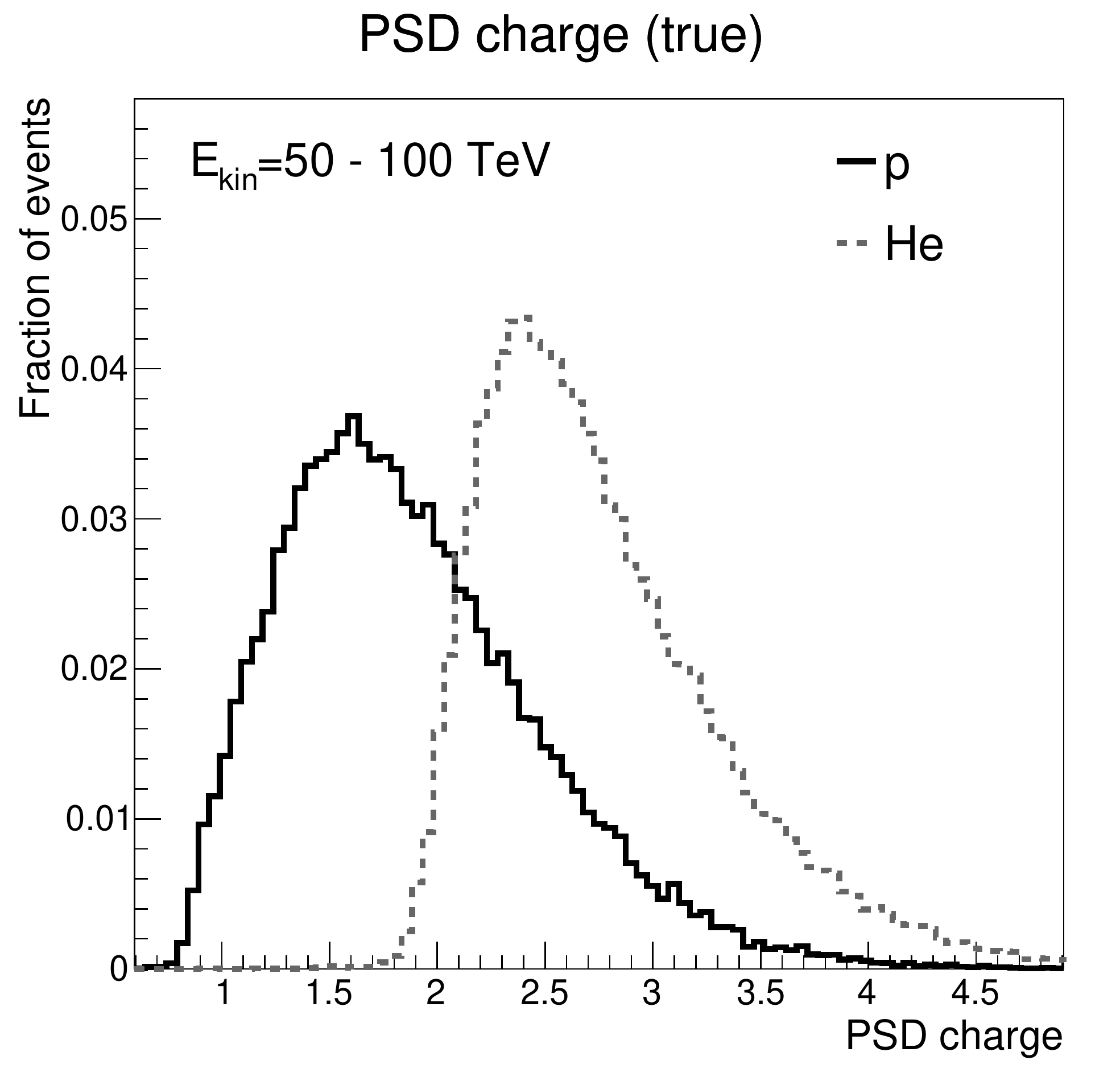}
\includegraphics[width=\phefiguresize\textwidth]{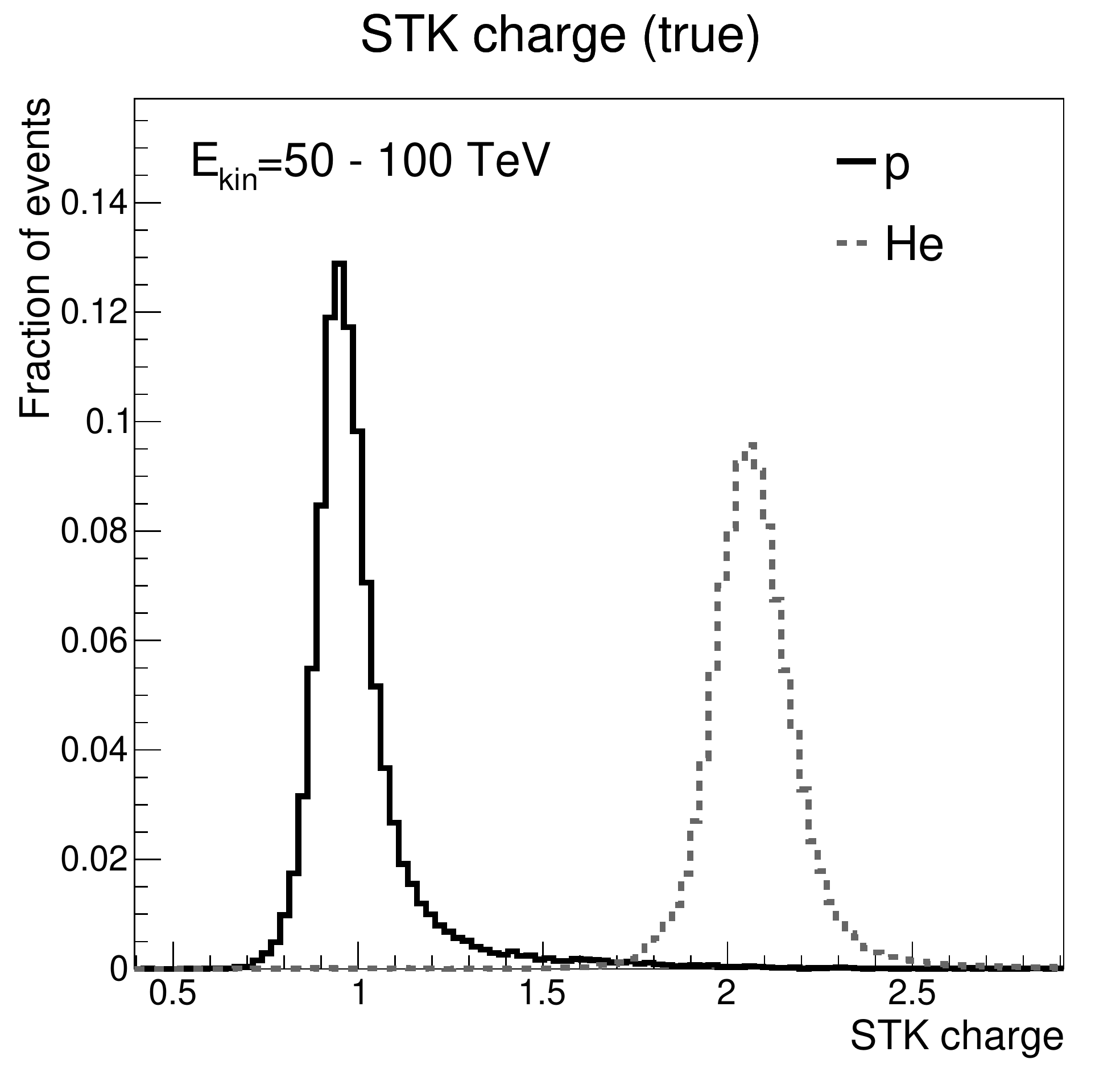}
\includegraphics[width=\phefiguresize\textwidth]{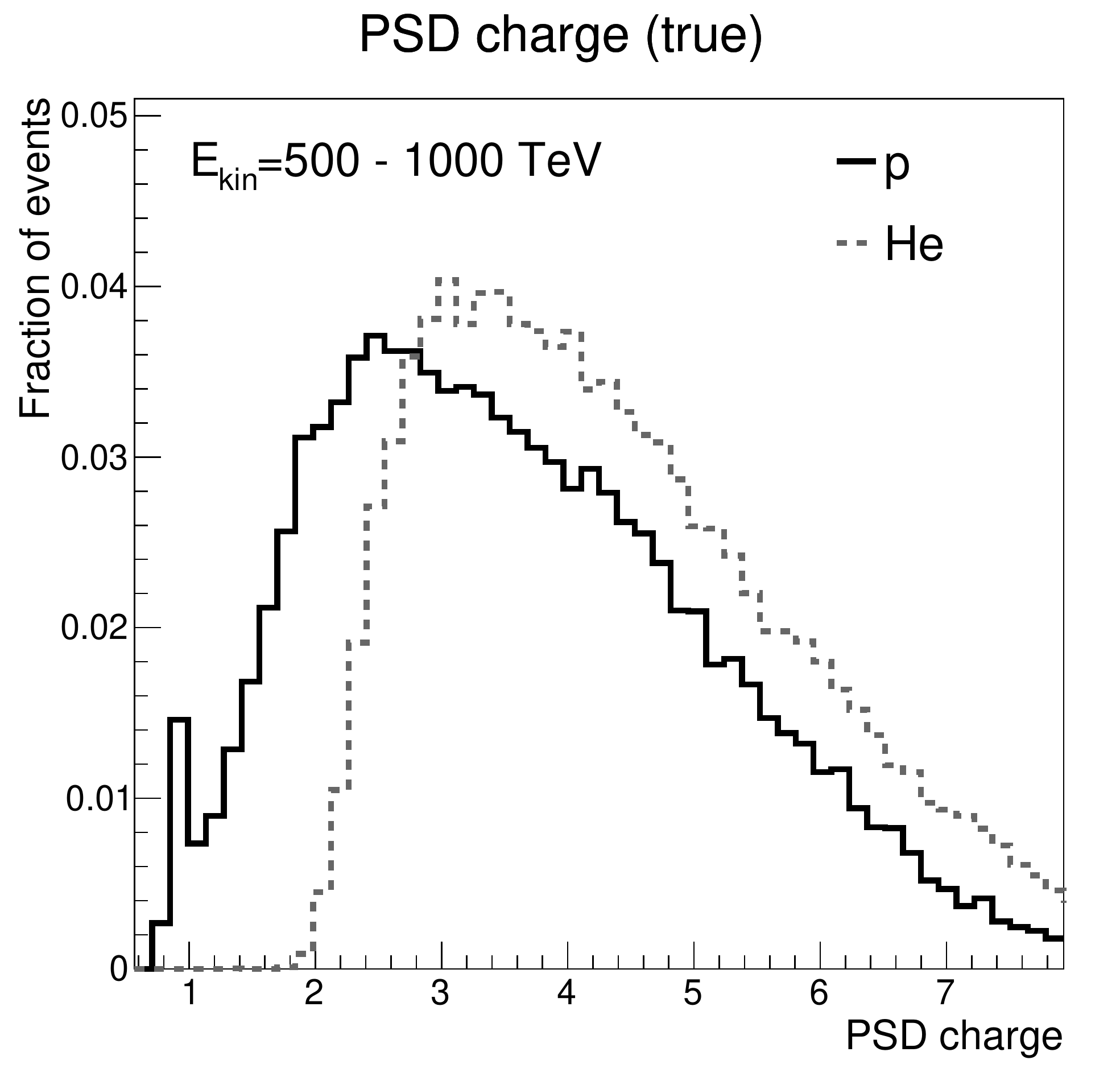}
\includegraphics[width=\phefiguresize\textwidth]{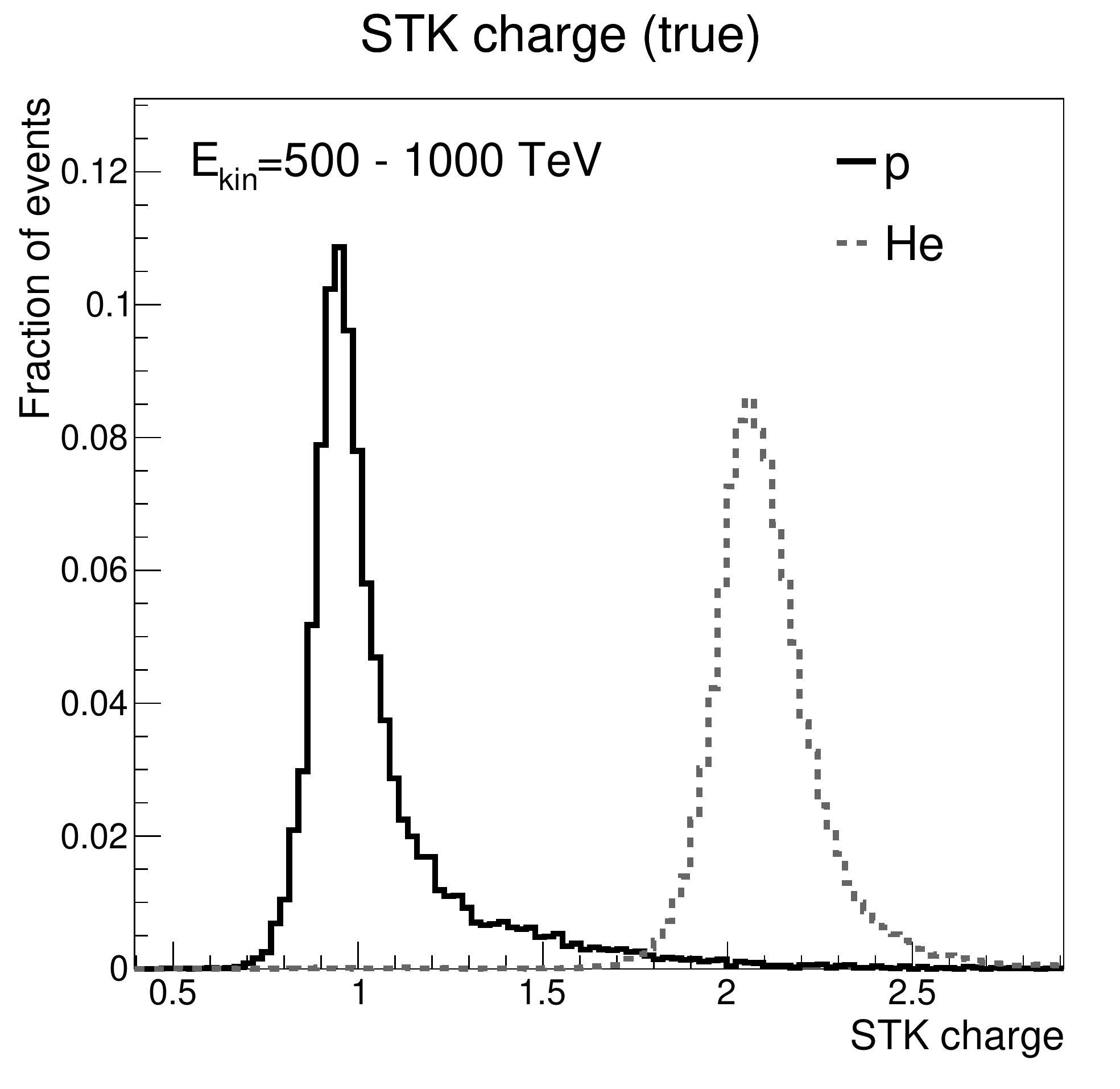}
\end{center}
\caption{Absolute charge of particles with different kinetic energies measured with  PSD (left) and STK (right) for MC events, obtained using the true particle direction. All distributions are normalized to unity.}
\label{fig:signal_psd_stk_truth}
\end{figure} 
  
At high energies, secondary particles originating in the calorimeter showers  severely contaminate the PSD, deteriorating the signal distributions and making the measurement towards PeV scales extremely hard (Figure~\ref{fig:signal_psd_stk_truth}--left). The tracker, on the other hand, can provide very precise measurements, nearly independent of particle energy (Figure~\ref{fig:signal_psd_stk_truth}--right), assuming, however,  that the particle track can be reliably identified. In particular,  given a 95\% containment of the STK proton distribution, the helium cross-contamination is less than 0.5\% at all energies; in turn, given a 95\% containment of the STK helium distribution, the proton cross-contamination is $\sim$1\% up to 100 TeV and  $\sim$2\% at PeV energies. Due to the relatively small silicon strip pitch and a large number of tracking layers, providing up to 12 signal points, the STK  measurement is much less affected by  the secondary particles that originate in the calorimeter shower and scatter back into the tracker. In particular, it can be seen that STK  provides very clean proton and helium peaks. These are the most abundant cosmic-ray species, naturally making them a target for the first direct cosmic-ray measurements at PeV energies. Notably, such measurements are also among the major physics goals of the High Energy Cosmic Radiation Detection Facility (HERD) -- the largest cosmic-ray instrument to be launched into space in the foreseeable future~\citep{herd_2021_icrc_gargano,herd_2021_icrc_Perrina,herd_2021_icrc_Pacini}. The HERD mission shares its design philosophy with DAMPE, consisting of a thick calorimeter surrounded by a  fine-segmented tracker.  Therefore, the task of precise tracking is crucial not only for DAMPE, but also  for the future success of cosmic-ray direct detection experiments.
   
\begin{figure}
\begin{center}
\includegraphics[width=\phefiguresize\textwidth]{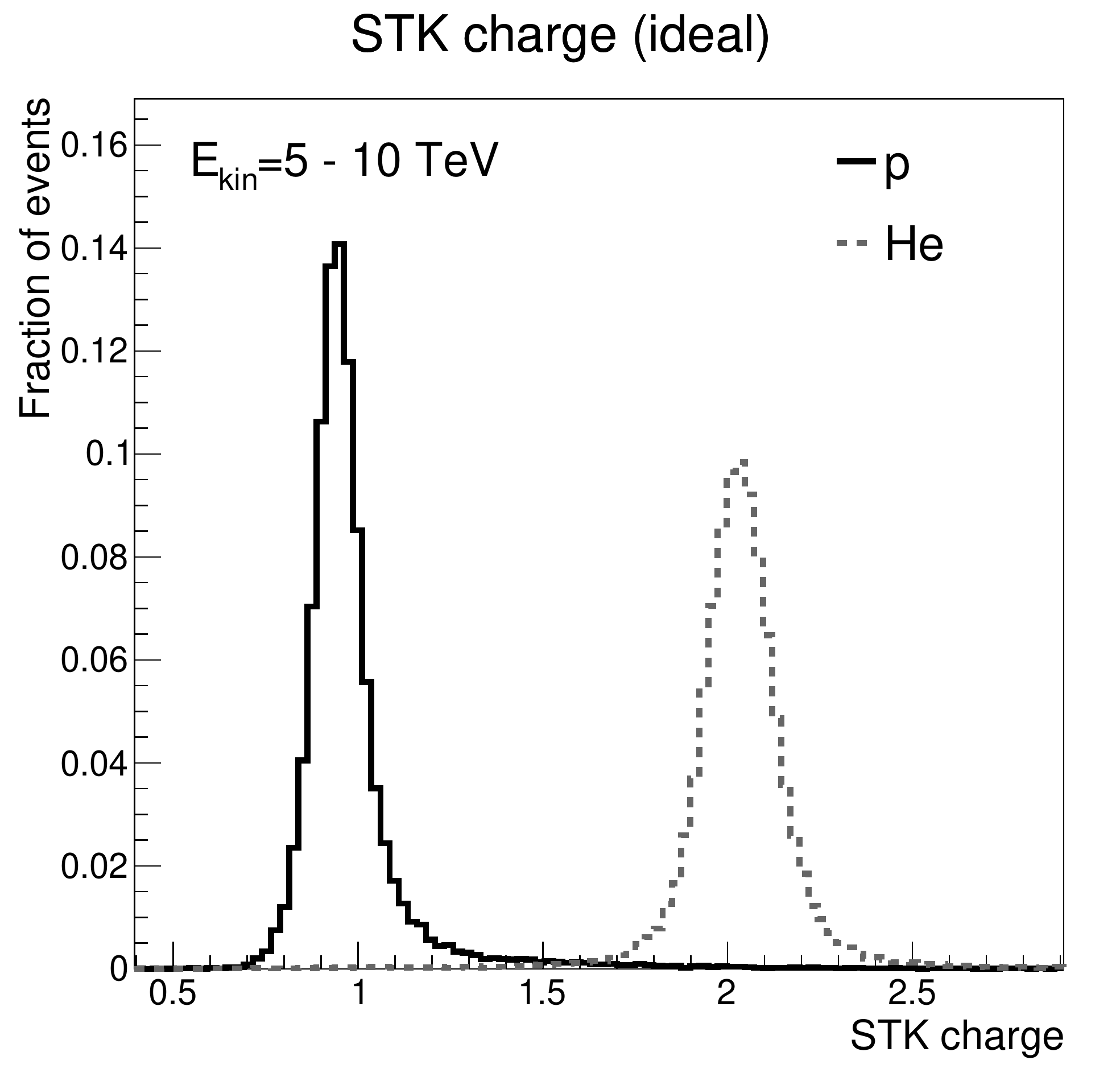}
\includegraphics[width=\phefiguresize\textwidth]{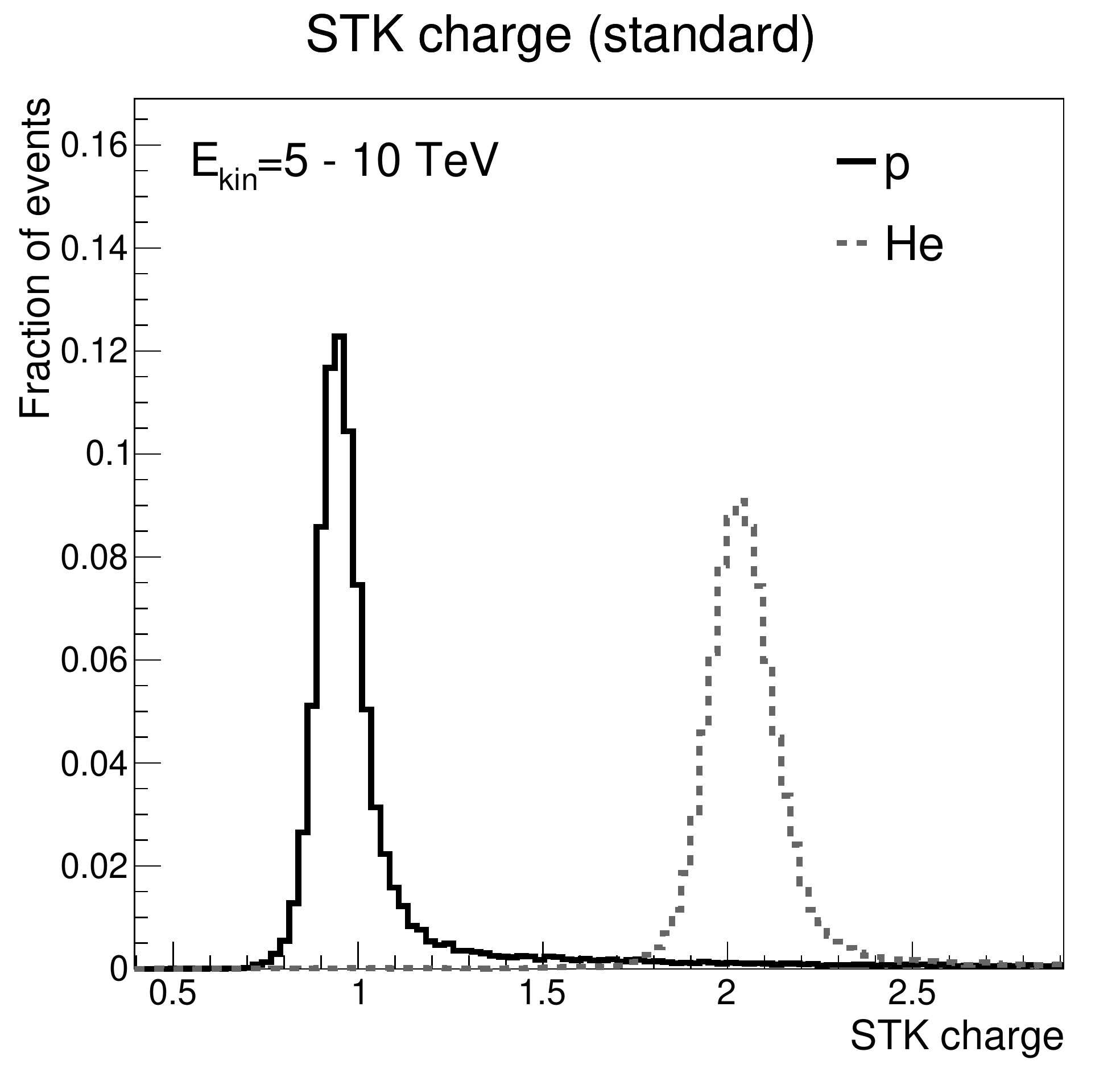}
\includegraphics[width=\phefiguresize\textwidth]{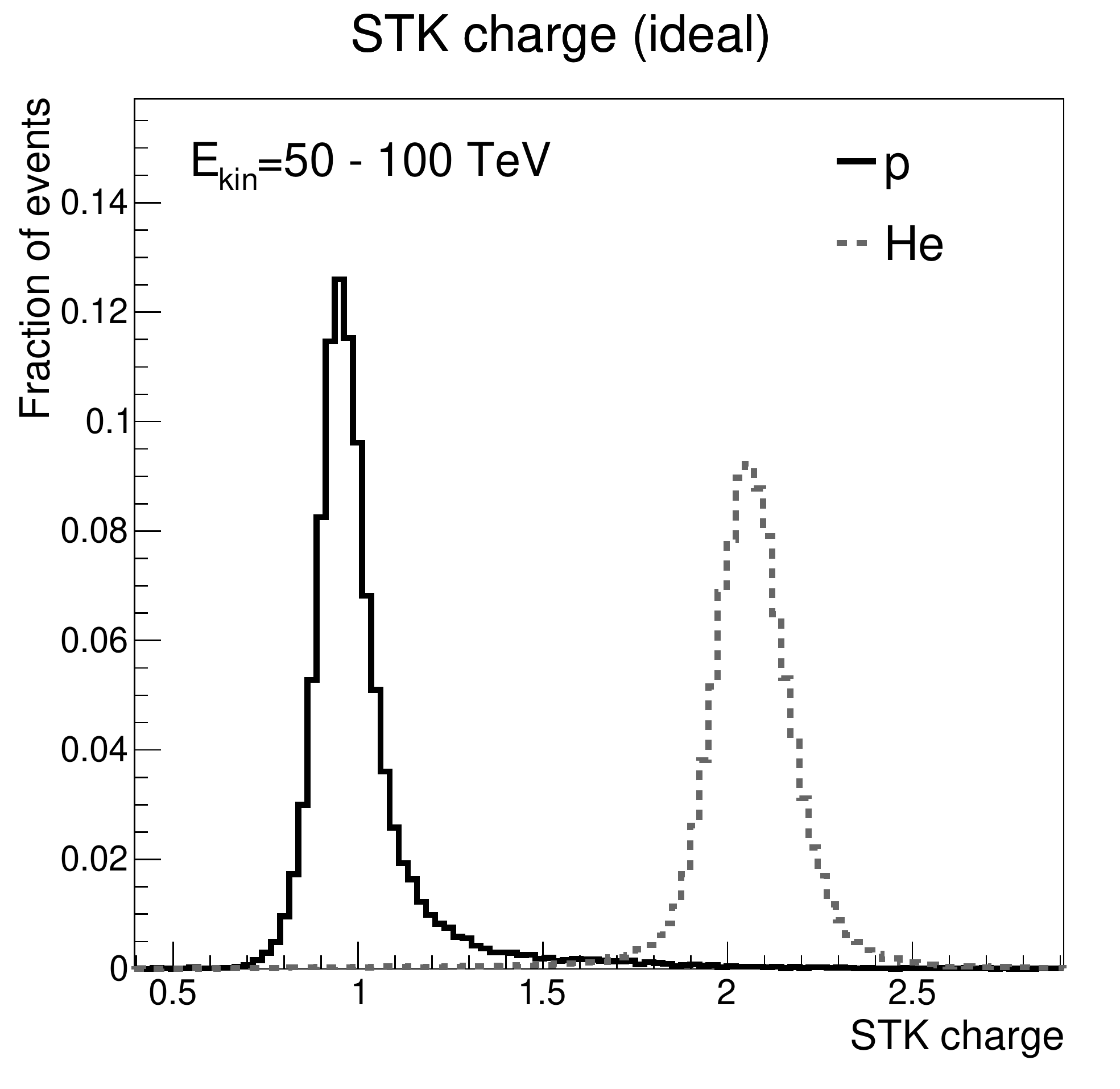}
\includegraphics[width=\phefiguresize\textwidth]{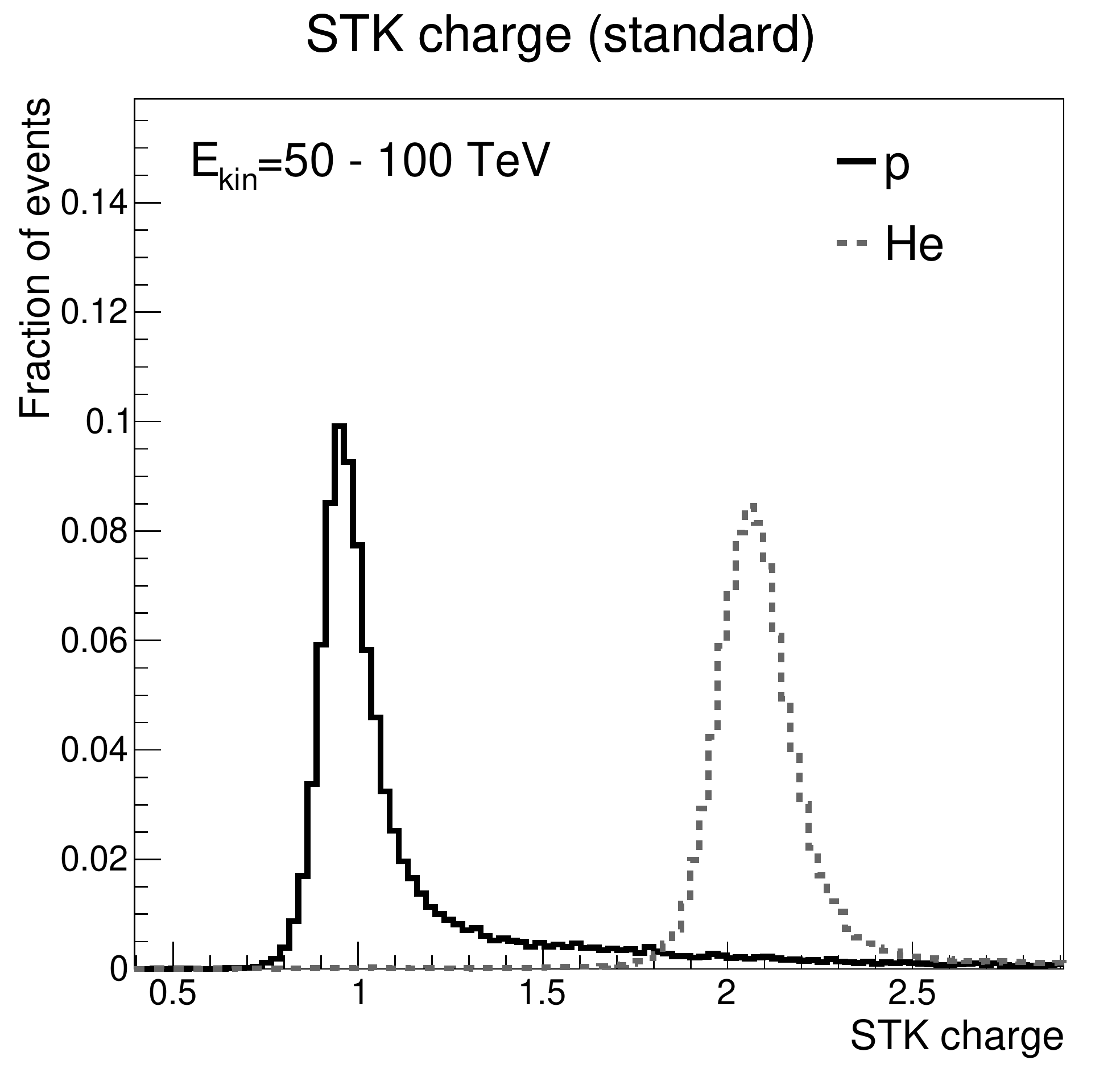}
\includegraphics[width=\phefiguresize\textwidth]{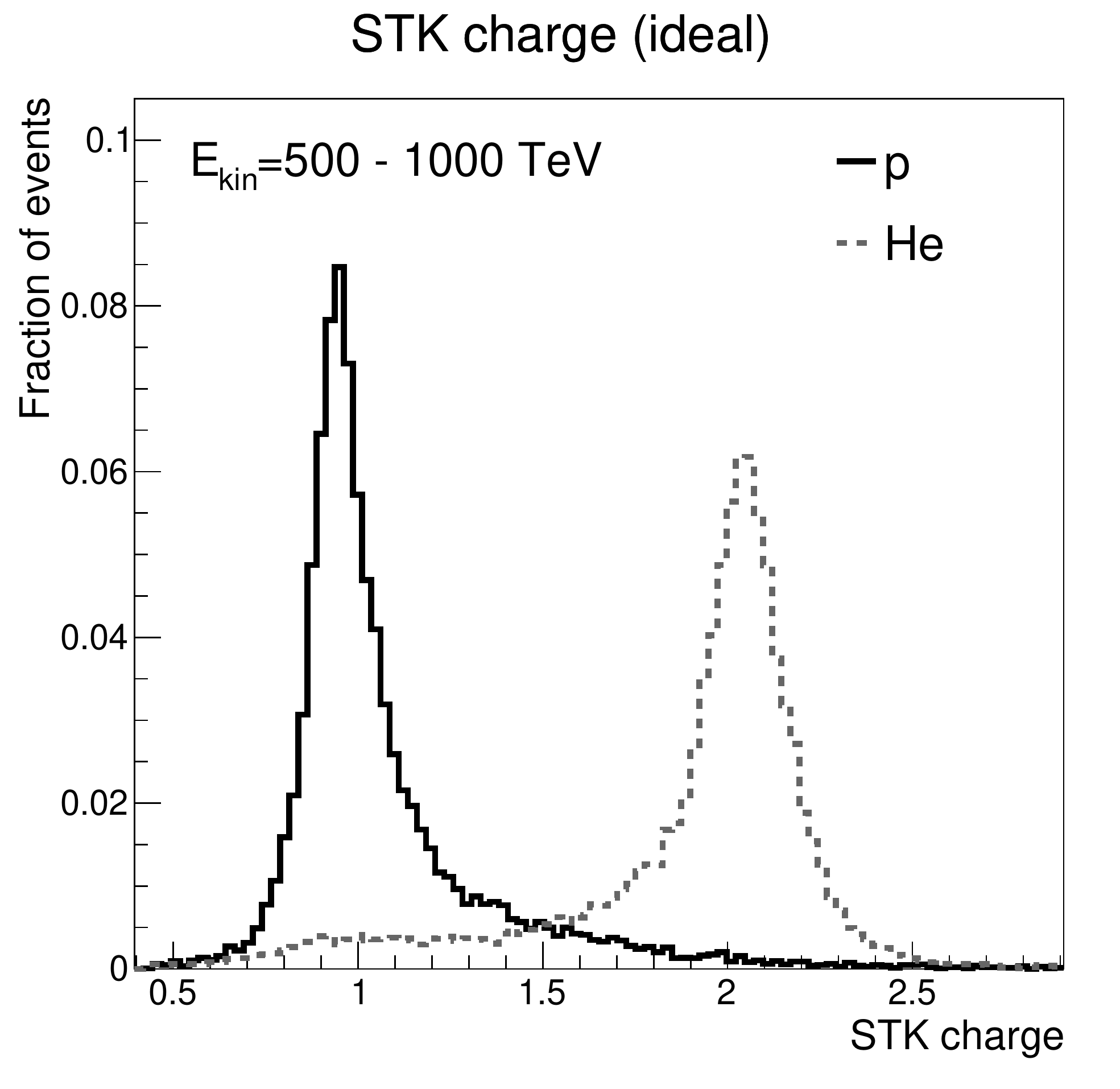}
\includegraphics[width=\phefiguresize\textwidth]{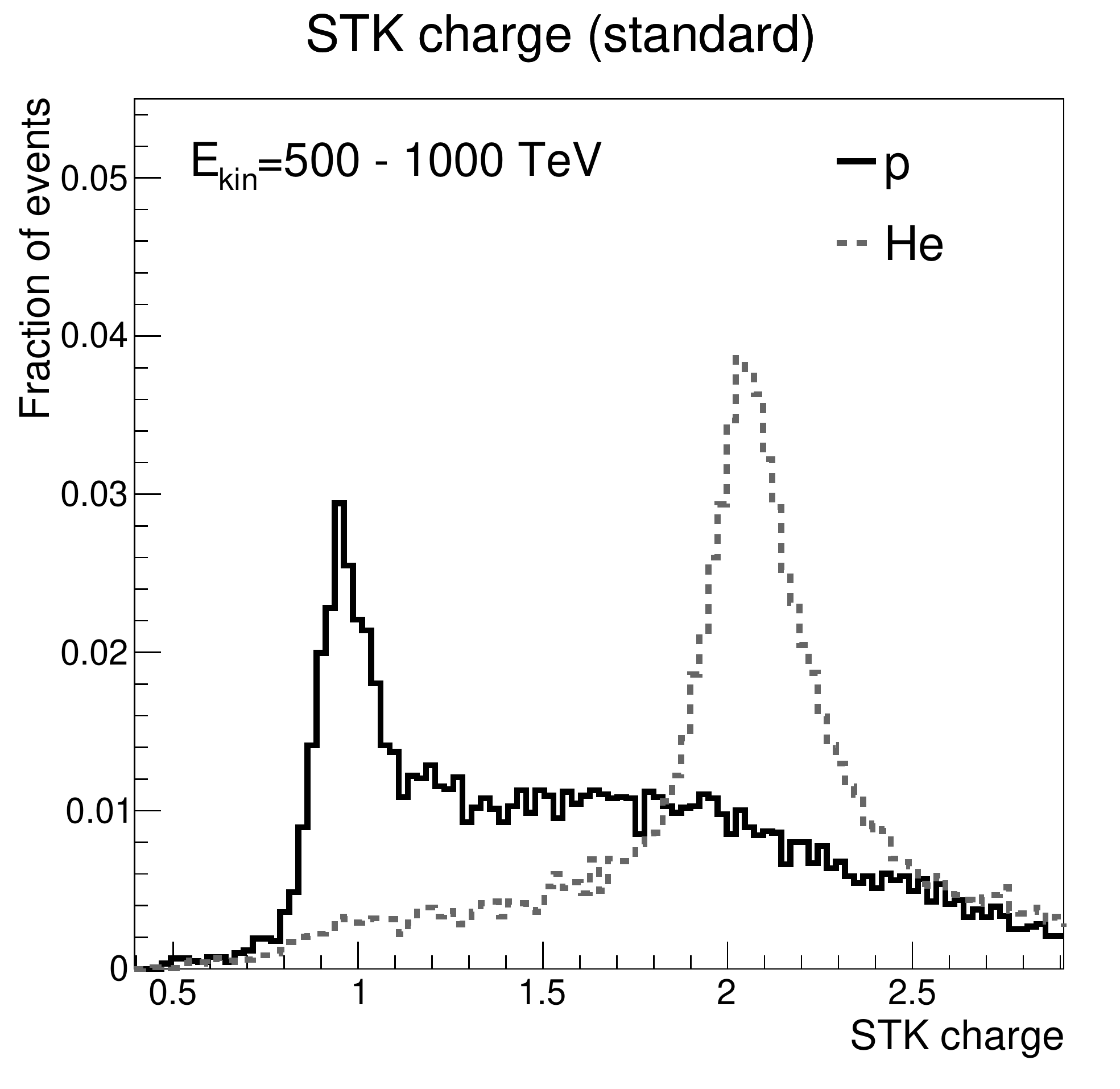}
\end{center}
\caption{Absolute charge of particles with different kinetic energies measured with STK for MC events, obtained using the STK track. Track reconstruction is done with the conventional DAMPE algorithm~\citep{Tykhonov:2017rdq}. The candidate track is selected using either the~\emph{ideal identification} (left) or the~\emph{standard identification} (right) method~\citep{dampe_prl}. All distributions are normalized to unity.}
\label{fig:signal_stk_standard}
\end{figure}

Figure~\ref{fig:signal_stk_standard} shows the charge identification capability of the STK using the standard track reconstruction algorithm with two different track identification methods. The first one is referred to as \emph{ideal identification}, the ``perfect'' algorithm that selects the reconstructed track best matching the true particle. The second one is the~\emph{standard identification} -- an algorithm optimized for the helium analysis which selects a track matching certain quality criteria and having the highest  signal compared to other candidate tracks~\citep{dampe_prl}. As can be seen from the distributions in Figure~\ref{fig:signal_stk_standard}, the reconstruction and identification of the track are particularly difficult at high energies. Due to the back-scattering and  pre-showers in the tracker, a  vast multiplicity of secondary hits arise, dramatically obscuring the signal of the primary impinging cosmic ray, as illustrated in Figure~\ref{fig:event_displays}--bottom. Moreover, as the  number of hits increases, the combinatorial pattern recognition turns computationally expensive -- the search for the primary track becomes a ``needle in a haystack'' problem. While the conventional  track reconstruction algorithm of DAMPE operates adequately up to about 100 TeV, the identification of the correct track remains an open task even in this case. At higher energies, however, not only the track identification is a challenge, but the standard algorithm also fails to reconstruct the primary track in a large fraction of events. The exact figures on the tracking efficiency with the standard approach and the one developed in this work will be provided further in the paper. 
  
Below, we develop a  novel deep learning regression approach for predicting the particle trajectory, using the calorimeter (BGO) and tracker (STK) data as an input. This approach enables to perform an accurate absolute charge measurement and allows to overcome one of the major difficulties for the first direct measurement of the cosmic proton and helium energy spectra at the PeV frontier.

\section{Neural network calorimeter direction prediction}
\label{sec:calo}
Up to now, conventional techniques based on the analytical fitting of the shower axis have been used for the calorimeter shower direction reconstruction in all major space experiments~\citep{dampe_mission,AKAIKE2010690,cream_tracking_2019,Gallucci:2295065,Atwood:2013rka,fermi_mission_2009}. These techniques  have a typical pointing resolution comparable to the granularity of the calorimeter, which is about $\sim$1~cm in the case of DAMPE. In this work, we propose a method based on Convolutional Neural Networks (CNNs), treating calorimeter data as  images. Profiting from low-level data without a bias of human  pre-processing, CNNs can learn  ``hidden'' features in the data which conventional ``analytic'' algorithms may not be sensitive to, potentially providing a better particle direction prediction~\citep{RevModPhys.91.045002}. Deep learning techniques including CNNs have already demonstrated their first successful applications in high-energy physics~\citep{ATLAS:2020iwa,Alimena:2020web,Metodiev:2017vrx,Polson_2022,PhysRevD.97.014021,Buhmann:2020pmy,Ghosh:2020kkt}, astroparticle and gamma-ray physics~\citep{Droz:2021wnh,wada_gaps,Finke:2020gbv,mikhail}, neutrino~\citep{Aurisano:2016jvx,Abbasi:2021ryj}, and extensive air shower detection experiments~\citep{PierreAuger:2021fkf,GUILLEN201912}. The bulk of applications belong to classification type problems~\citep{ATLAS:2020iwa,Alimena:2020web,Metodiev:2017vrx,Droz:2021wnh,wada_gaps,Finke:2020gbv,Aurisano:2016jvx}, yet the first examples of regression tasks have also emerged~\citep{Polson_2022,mikhail,Abbasi:2021ryj,PierreAuger:2021fkf,GUILLEN201912}. Interesting applications of neural networks and adversarial training in generative models are also attracting a growing attention  in the community~\citep{PhysRevD.97.014021,Buhmann:2020pmy,Ghosh:2020kkt}. Beyond this work, in DAMPE, the neural network paradigm is also actively being explored for electron/hadron particle discrimination~\citep{Droz:2021wnh} and calorimeter energy reconstruction~\citep{mikhail}. 

\begin{figure}
\begin{center}
\includegraphics[width=1.0\textwidth]{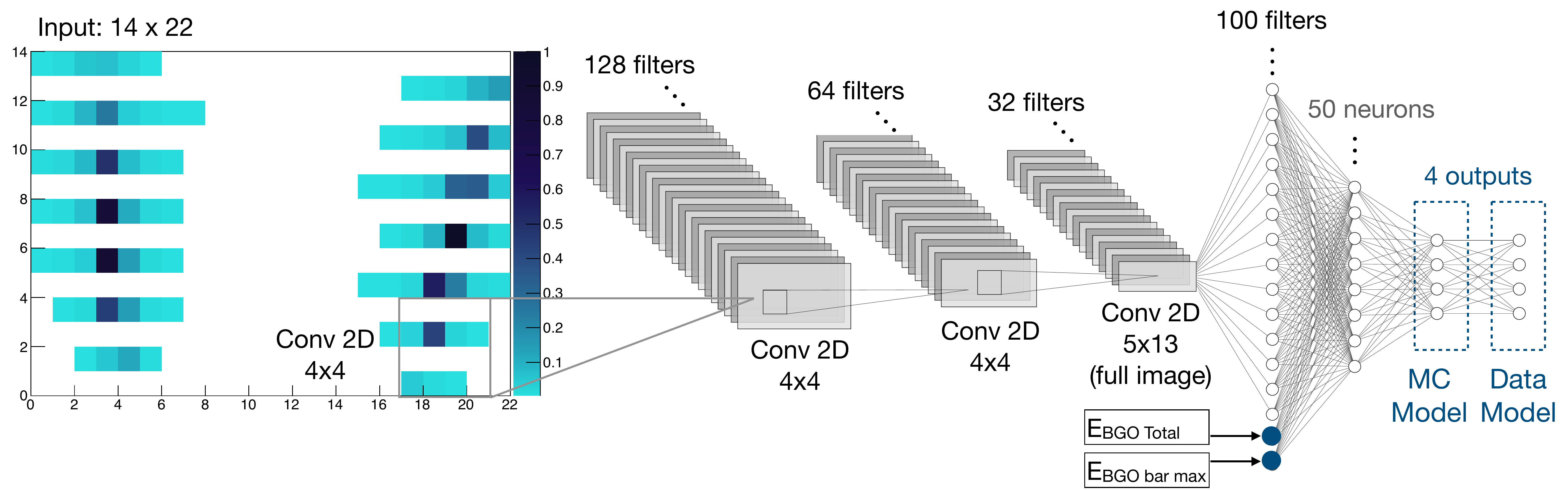}
\end{center}
\caption{Convolutional neural network  for the particle direction prediction in the BGO calorimeter. The output of the convolution layers is a set of 100 variables augmented with two additional variables: the total deposited energy and maximum-bar energy (both in units of TeV). It is followed by a fully-connected layer of 50 neurons, in turn, fully connected to the 4 output variables. Activation in all layers is done with the Rectified Linear Unit (ReLU) function, except for the last layer where the activation is linear.  An additional fully-connected layer of 4 outputs with linear activation is added to perform the data/MC correction (alignment).}
\label{fig:bgo_cnn}
\end{figure}

The chosen network architecture together with an example input calorimeter image is shown in Figure~\ref{fig:bgo_cnn}. The image is constructed as a~\emph{mixture} of two projections of the BGO calorimeter with a total dimension of 14$\times$22 pixels. Vertical-wise, even and odd layers of an image correspond to $xz$ and $yz$ projections, respectively. There are 7 layers per projection, according to their hodoscopic geometrical arrangement. We have also considered an alternative architecture, where the $xz$ and $yz$ projections are connected to separate CNNs, the outputs of which are then combined in a fully-connected neural network. We learned that the~\emph{mixed} image architecture has a better prediction accuracy compared to the one with the~\emph{disconnected} images. This result is expected since the two projections are not independent. The~\emph{mixed} image accounts for cross-correlations throughout the calorimeter in the vertical direction, as particles are traversing sequentially the 14 layers. The value in each pixel of an image is taken as the signal of the corresponding BGO bar divided by the signal of the maximum-energy bar in the event. In this way, the values are limited to the [0;1] range. We use 8-bit precision to decode signals in each pixel. We also tested 16-bit precision and found no significant difference in the network performance. The output of the network, $\hat{\mathbf{x}}$, is a vector of 4 variables, which correspond to particle coordinates ($x$ and $y$) in the first plane and the last plane of the STK. The choice of the output variables is motivated by the fact that the BGO direction prediction serves as a first approximation for the particle trajectory finding in the tracker.  As a target for training we use the mean squared error:
\begin{equation*}
L(\hat{\mathbf{x}},\hat{\mathbf{x}}_{tru}) = \frac{1}{4 N} \sum_{i=1}^{N}( \hat{\mathbf{x}} - \hat{\mathbf{x}}_{tru}  )^2,
\end{equation*}
where $\hat{\mathbf{x}}_{tru}$ is the corresponding vector of true particle coordinates in the first and the last planes of the STK and $N$ is the number of events in the batch.

The training of the network has been done with MC data consisting of simulated proton, helium, and electron particles passing through the DAMPE detector. The CNNs were trained separately for the low-, middle- and high-energy ranges, corresponding to particle kinetic energy between 10~GeV and 10~TeV, between 1~TeV and 1~PeV, and between 10~TeV and 1~PeV, respectively. This yields a better accuracy compared to the case when a single model is trained on the entire energy range. We intentionally choose highly overlapping energy intervals for the three models in order to facilitate smooth transitions between the models. The output accuracy of the low- and middle-energy range models overlap in the region in which the deposited energy is about a few hundreds of GeV, while the accuracy of middle- and high-energy models overlap in the region of few tens of TeV. Hence, the transition thresholds  were chosen  at 300~GeV  and 20~TeV, respectively\footnote{Depending on the value of deposited energy, one of the 3 models is used for the  inference.}. The fact that the multiple energy range training shows a better performance compared to a single energy range model is due to few factors. Firstly, at energies below~$\sim$100 GeV, considerably fewer pixels are fired in the calorimeter, therefore the amount of information is significantly lower than at~$\sim$TeV and higher energies. As we will show further in the paper, the accuracy of the calorimeter CNNs model at low energies is significantly worse than at high energies. Therefore it bottlenecks the training, biasing it towards  the optimisation for low-energy events. Secondly, at the highest energies, the effect of the BGO readout saturation starts taking place, changing qualitatively the typical picture of showers in the calorimeter~\citep{mikhail,Yue:2020hmj}. The latter cannot be inferred from the lower (or middle-range) energy data. Therefore it requires a dedicated model with a particular attention to the saturated events. Finally, we choose not to split models into more than 3 intervals, as we did not observe any significant gain for alternative configurations. In particular, we have tried splitting the low-energy interval into subintervals, but no further improvement in accuracy was seen, at least in the energy range where we perform cosmic-ray measurements with DAMPE\footnote{Optimisation for $\sim$GeV region relevant for gamma rays is beyond the scope of this paper.}.

\newcommand{\tensorflow}{$\mathrm{TensorFlow}~2$}
\newcommand{\texttensorflow}{TensorFlow}
\newcommand{\bgoinferencetime}{$0.24$}
\newcommand{\bgotrainingtime}{$36$}
\newcommand{\nvidia}{$\mathrm{Nvidia}$}
\newcommand{\amd}{$\mathrm{AMD}~\mathrm{Opteron}$}

All the neural networks described in this paper are implemented in~\texttensorflow~2\footnote{https://www.tensorflow.org/} and trained on~\nvidia~2080 series GPUs. The training is done in batches of size 32. For the calorimeter CNNs, the fitting time is about~\bgotrainingtime~minutes/epoch for $10^6$ batches. Training sample size ranges between O($10^4$) and O($10^6$) batches depending on the available MC statistics in different energy ranges (see Table~\ref{tab:mc_samples}). The Adam stochastic gradient descent algorithm~\citep{kingma2017adam} in its default configuration is employed. The learning rate is initially set to $10^{-4}$ and controlled during the training by the~\texttensorflow's Reduce-on-Plateau method. During the data processing the model inference is done on conventional CPUs.  
Inference time amounts to~\bgoinferencetime~s/event, evaluated on~\amd~6274 processor\footnote{The same CPUs are used for the other CNNs inference benchmarks elsewhere in the paper.}. 
 
\begin{figure}
\begin{center}
\includegraphics[width=0.69\textwidth]{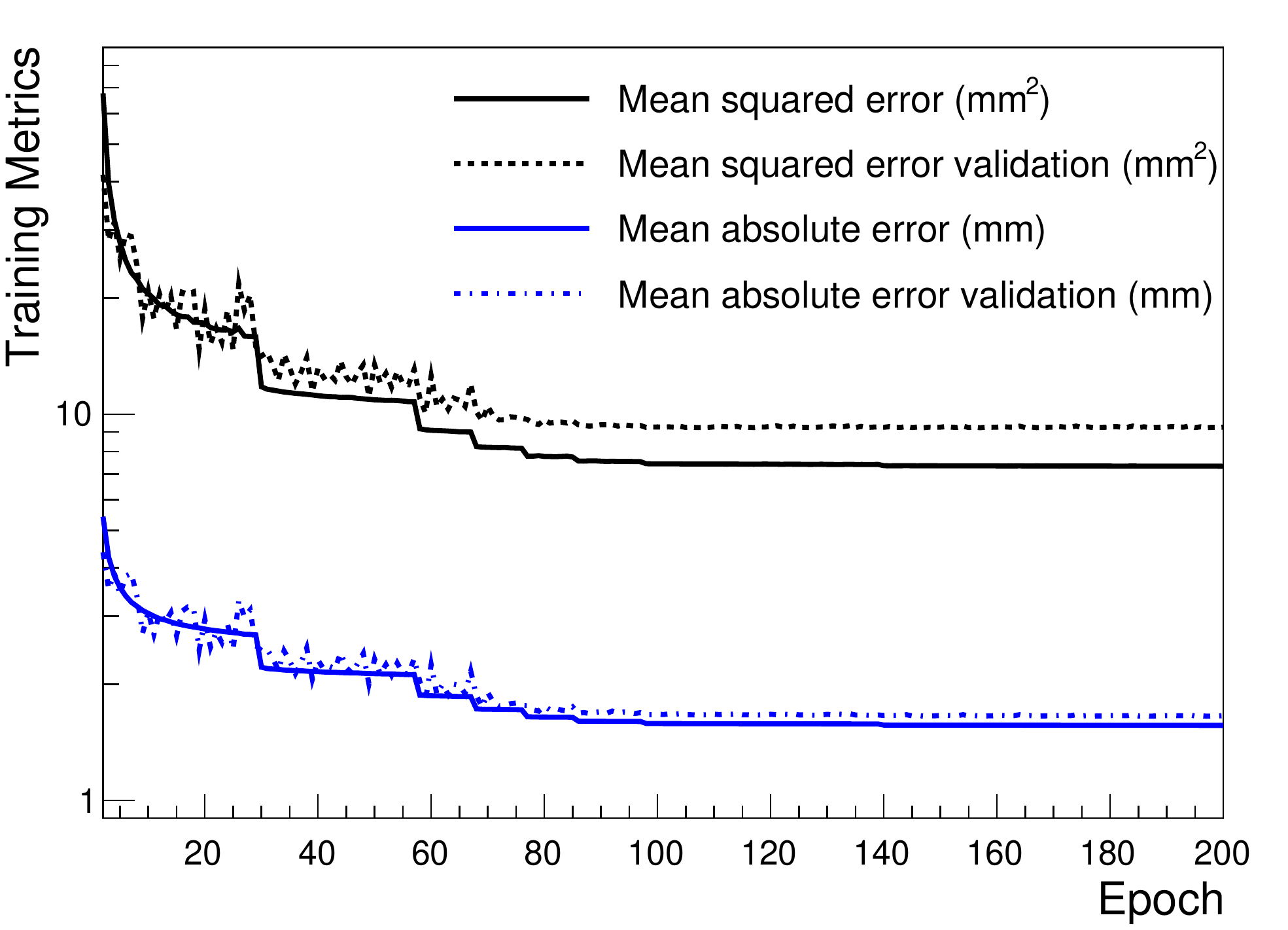}
\end{center}
\caption{Evolution of the mean squared error and mean absolute error in training for the calorimeter neural network model in the middle-energy
range (particle  kinetic energy larger than  1~TeV). The steps correspond to the reduction of the learning rate by a factor of two.}
\label{fig:loss}
\end{figure}

About 25$\times$10$^6$, 3$\times$10$^6$ and 1$\times$10$^6$ events were used for the training in the low-{, middle-} and high-energy range, respectively. The MC samples were divided into~\emph{training},~\emph{validation}, and~\emph{test} samples in an approximate proportions of 80\%--10\%--10\%. As shown in Figure~\ref{fig:loss}, the convergence of  the gradient descent algorithm in the network optimization is achieved in about 100 epochs. Note that a gap corresponding to some lack of generalization can be observed, which is mostly due to the limited MC statistics at the highest energies. We will quantify this effect in the further analysis as a systematic uncertainty. As we will show later,  its impact is nearly negligible. We have also tried reducing the complexity of the CNNs model as well as adding dropout layers to it, which resulted in a significantly worse performance of the network. The vertical dimension of the convolutional filters is set such that it spans two consequent layers of the calorimeter, both in $xz$ and $yz$ projections. We have also tested a network with convolutional filters of dimension 3$\times$5 instead of 4$\times$4, which resulted in a significantly lower accuracy. Finally, the majority of simulated data are generated with GEANT4. We have also added the FLUKA samples in the training, which marginally improved the performance at the highest energies thanks to the increased training statistics.

Since the shower-shape characteristics differ only marginally between different ions, the performance of the trained network on particles beyond helium (e.g. lithium, beryllium, boron, carbon, oxygen, iron) was found to be similar to that of proton and helium. We found no significant improvement if ions heavier than helium were added to the training. At the same time, it is well known that the shower shape characteristics for electromagnetic and hadronic interactions are fundamentally different. For this reason, we added simulated electrons corresponding to about 25\% of the training sample, to ensure high prediction accuracy of the network for all particle species. It is worth noting that adding electrons did not degrade the performance of the CNNs with respect to the hadronic showers.  We have also tried training a dedicated ``electromagnetic'' model by increasing the fraction of electrons to 80\% and found no significant performance improvement  on  electromagnetic showers compared to the baseline model. On the contrary, we have also tested the model with a relatively low electron content, about 3\% and found it to have significantly worse performance. Hence, we conclude that the training is not particularly dependent on the exact relative content of hadronic and electromagnetic showers, as long as they are of a comparable scale. 

\begin{figure}
\begin{center}
\includegraphics[width=0.49\textwidth]{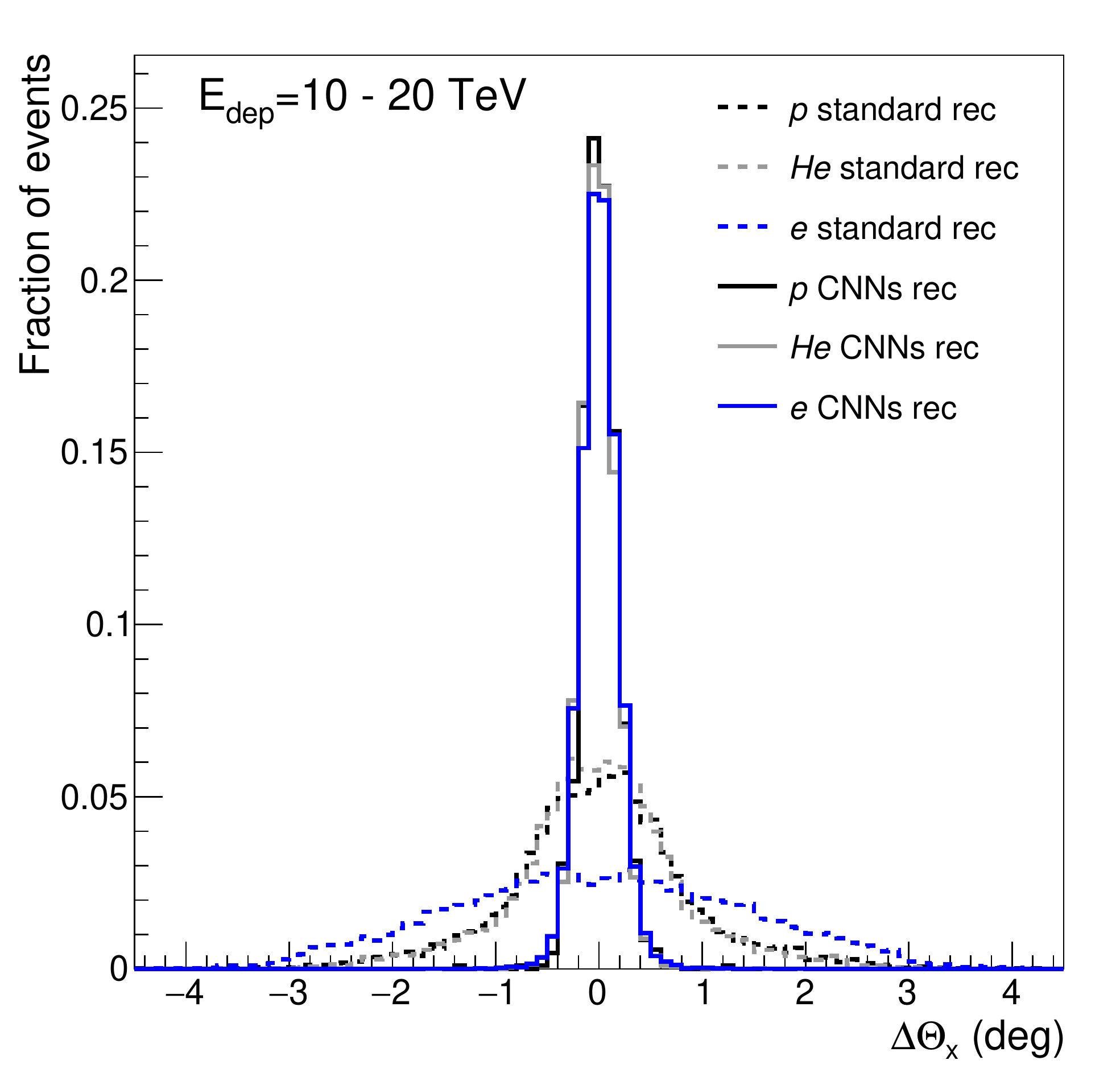}
\includegraphics[width=0.49\textwidth]{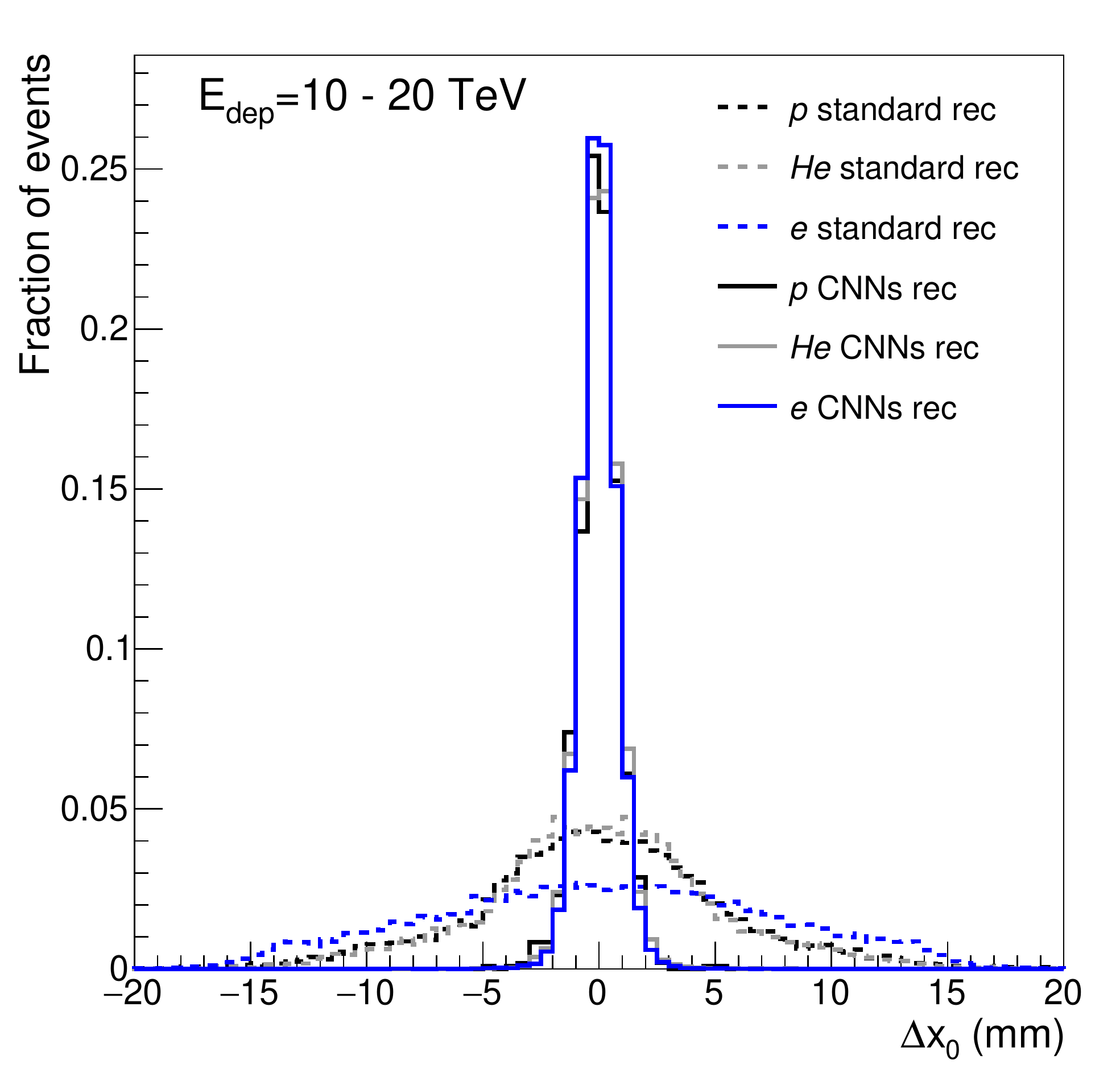}
\end{center}
\caption{Residual distributions of the azimuthal angle,  $\theta_x$ (left),  and intercept, $x_0$ (right) for MC events. The residual is obtained as the difference between the prediction of the CNNs model and the true particle direction. Similar distributions obtained with the standard BGO direction reconstruction~\citep{dampe_mission} are shown for comparison. While the results for the $xz$ plane of DAMPE are shown, the $yz$ distributions share the same behavior. Intercepts are calculated at the $z=0$ plane, which corresponds to the border between the BGO and STK subdetectors, see Figure~\ref{fig:event_displays}.} 
\label{fig:bgo_residuals}
\end{figure}

The performance of the developed algorithm is illustrated in Figure~\ref{fig:bgo_residuals}. For the sake of clarity, we convert the output variables of the network into conventional azimuth angles and intercept coordinates of a particle in two orthogonal projections of the DAMPE coordinate system. The distributions are derived from the~\emph{test} MC samples. 

\begin{figure}
\begin{center}
\includegraphics[width=0.49\textwidth]{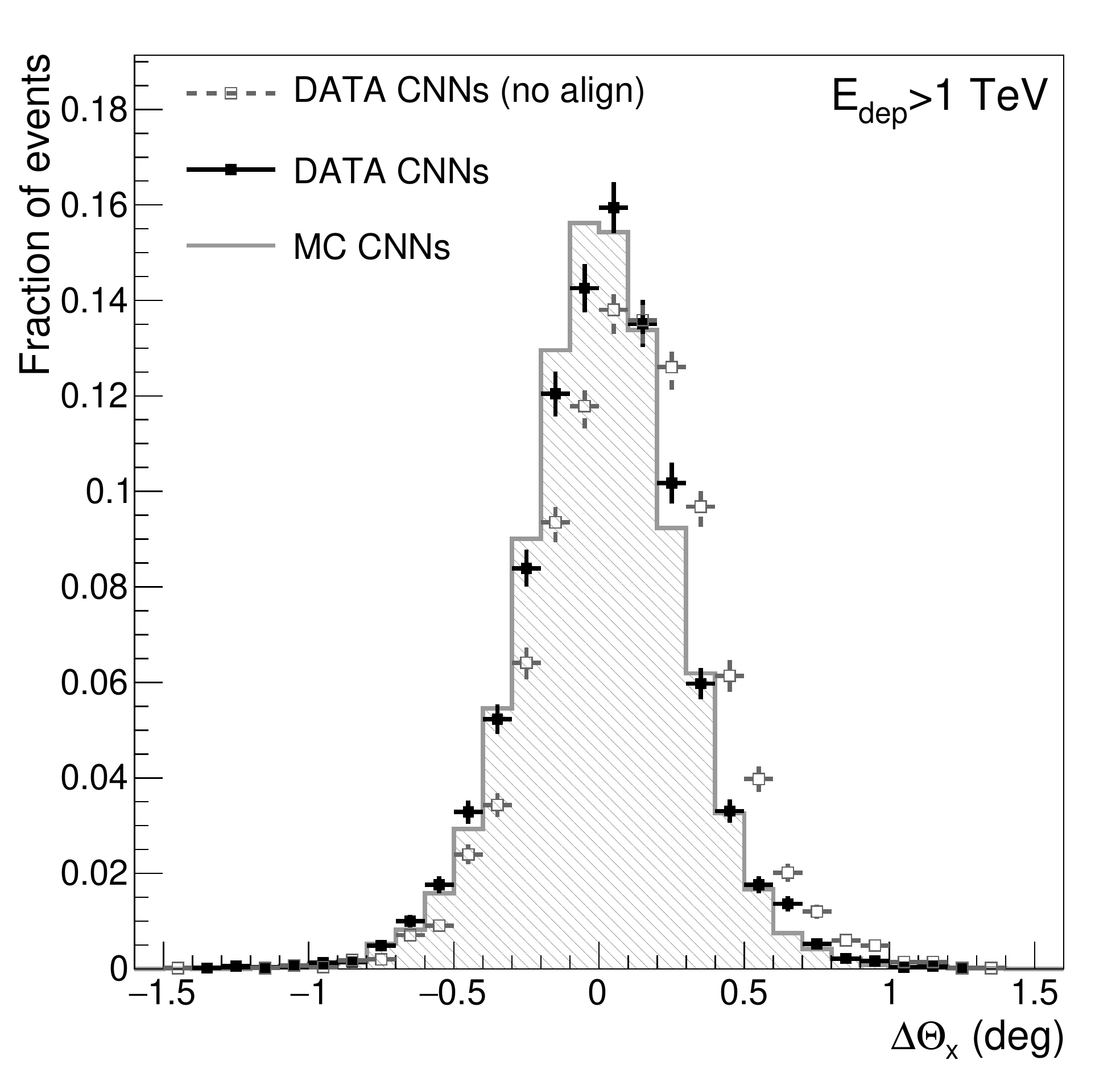}
\includegraphics[width=0.49\textwidth]{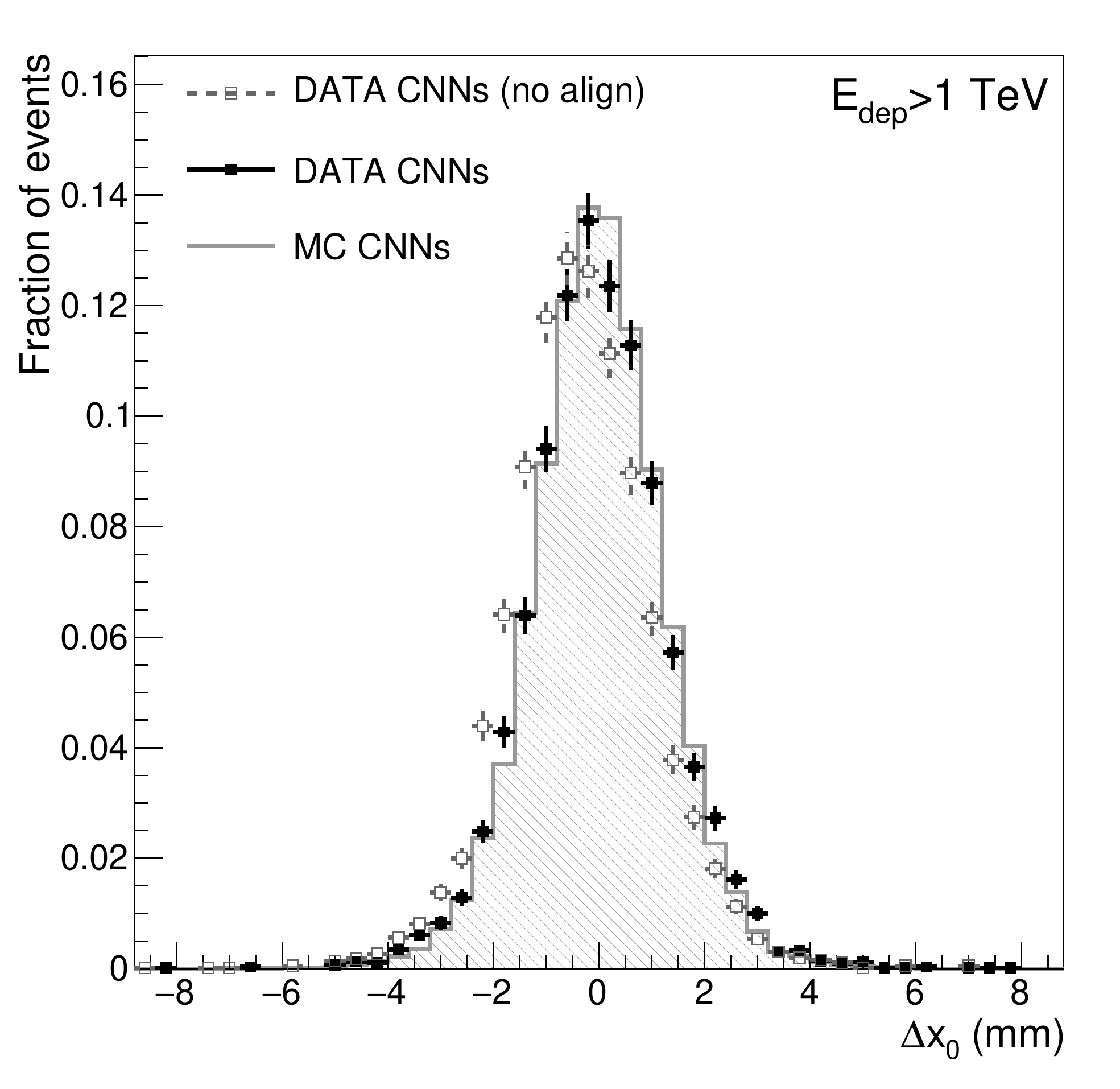}
\includegraphics[width=0.49\textwidth]{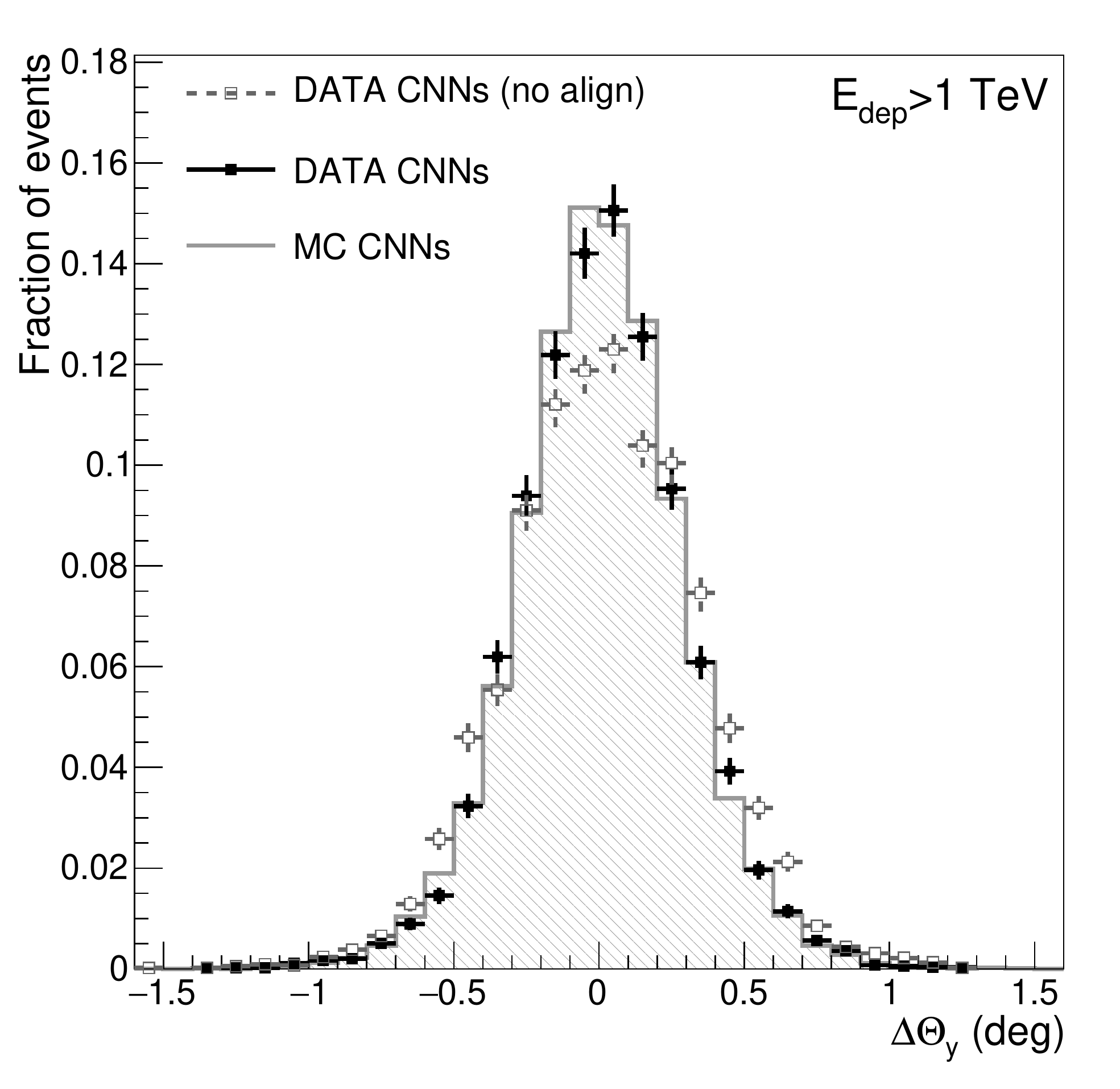}
\includegraphics[width=0.49\textwidth]{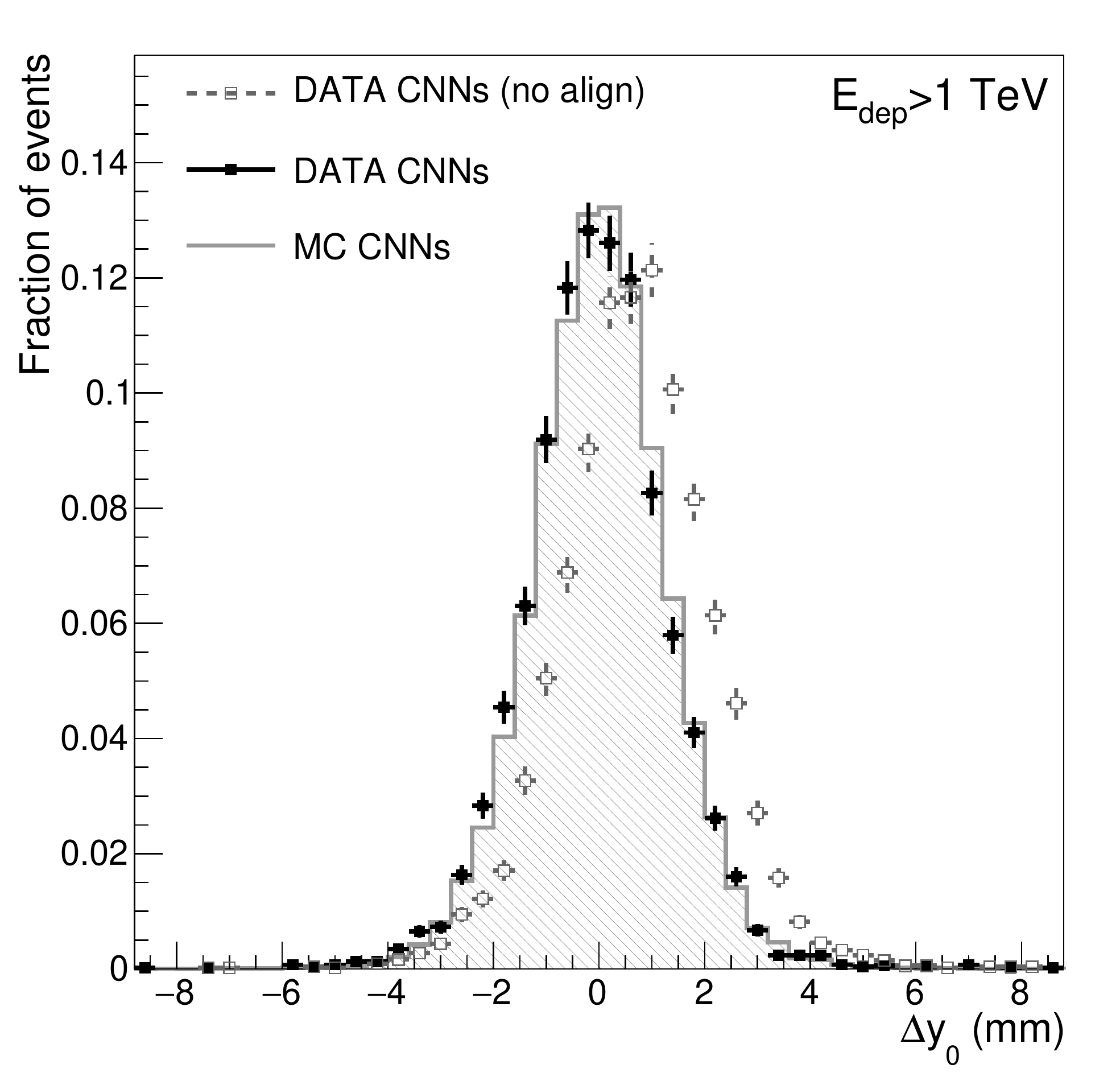}

\end{center}
\caption{Residual distributions of the azimuthal angle  (left) and intercept (right), in both DAMPE projections, for MC events and flight data with and without the additional layer in the CNNs model responsible for the BGO--STK alignment. The~\emph{clean} event selection is applied, requiring the presence of exactly one well-defined STK track, reconstructed with the standard procedure~\citep{dampe_mission}. This track is considered as a reference particle direction.}
\label{fig:bgo_residuals_data_mc}
\end{figure}

To combine the BGO direction prediction with the further trajectory finding in the tracker, precise alignment between the BGO and STK has to be performed.  While in simulation the alignment is perfect by definition, the trained CNNs model to be applied to the real data has to be corrected for possible misalignments between BGO and STK. To perform this task, we add another fully-connected layer of 4 neurons at the output of the network, while keeping the other adaptive parameters frozen, as illustrated in Figure~\ref{fig:bgo_cnn}, and separately train this layer directly on the data. The selection, in this case, is performed to ensure the presence of exactly one clearly defined track in the STK, obtained with the standard reconstruction algorithm, which we consider as a ``true'' particle direction in the corresponding training. Hereafter we refer to it as a~\emph{clean} selection. Furthermore, in Figure~\ref{fig:bgo_residuals_data_mc} we use the~\emph{clean} selection to evaluate the CNNs model direction reconstruction performance on the data. Mutually exclusive data samples are used for the BGO--STK alignment in the CNNs model training and  the evaluation of the residuals, each corresponding to about $10^4$ events. For illustration purposes, we also show the CNNs model prediction on the data if no BGO--STK alignment is applied. It is worth noting that the BGO--STK alignment does not depend on the particle energy. In other words, the additional layer (the right-most layer in Figure~\ref{fig:bgo_cnn}), being trained in one energy region of the data, works equally well at other energies, which is expected since the (mis-)alignment is a purely geometrical  effect.

Figure~\ref{fig:bgo_psf} shows the 68\% and 95\% containment radius using the developed CNNs particle direction prediction.  The direction of the primary particle to be compared with the CNNs prediction is either the MC truth direction or the   standard reconstructed STK track, obtained after applying the~\emph{clean} event selection. The latter is done in order to allow for the data/MC comparisons, as no true particle direction is known in the real data. The effect of the CNNs generalization gap is quantified as a systematic uncertainty. It is estimated as the difference between the results obtained on the \emph{training} MC sample and the~\emph{test} MC sample (which was excluded from the CNNs training). The impact of the generalization gap is observed only for hadronic showers with a deposited energy larger than $\sim$10~TeV.
     
\newcommand{\psffiguresize}{0.43} 
\begin{figure}
\begin{center}
\includegraphics[width=\psffiguresize\textwidth]{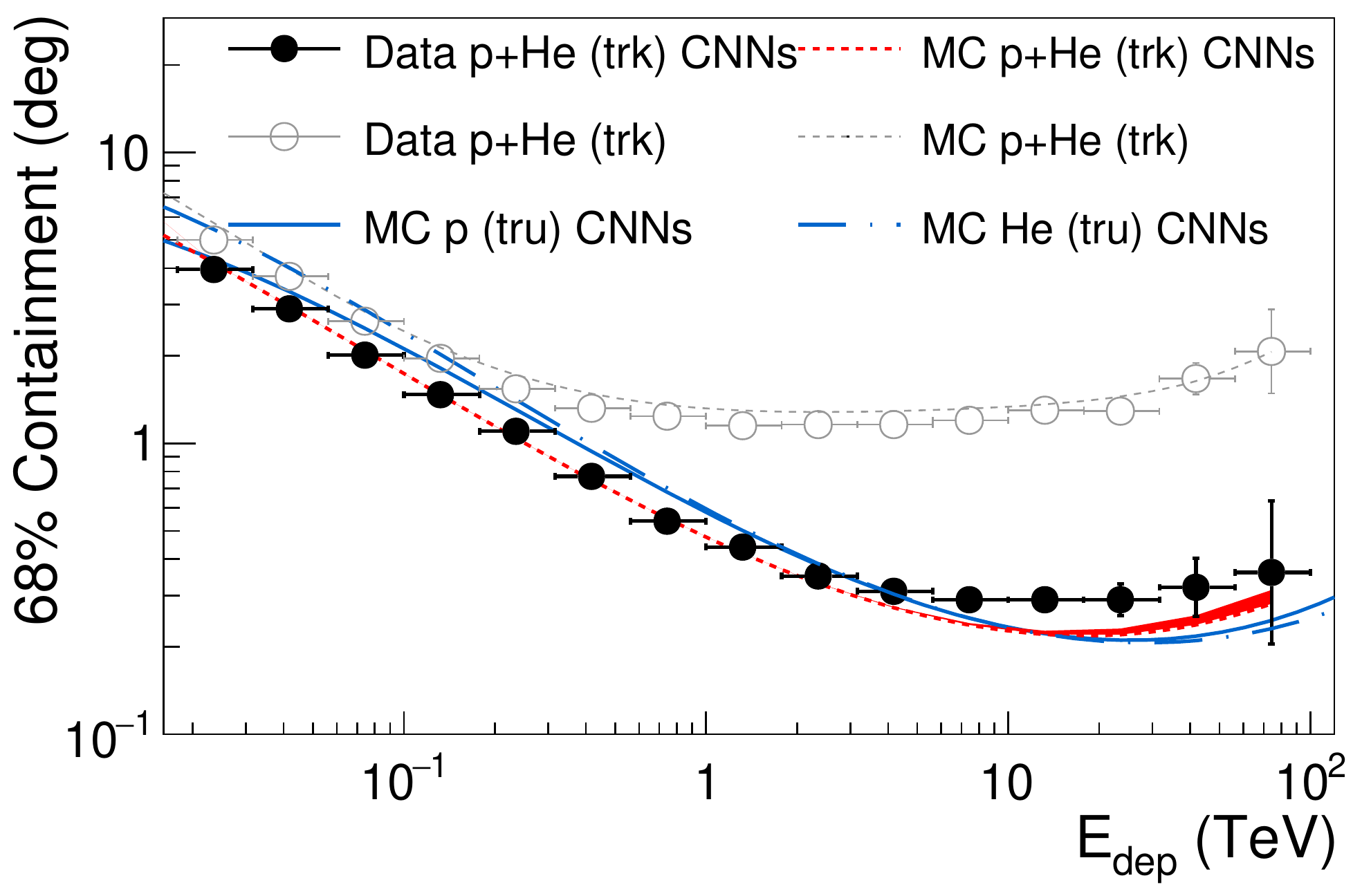}
\includegraphics[width=\psffiguresize\textwidth]{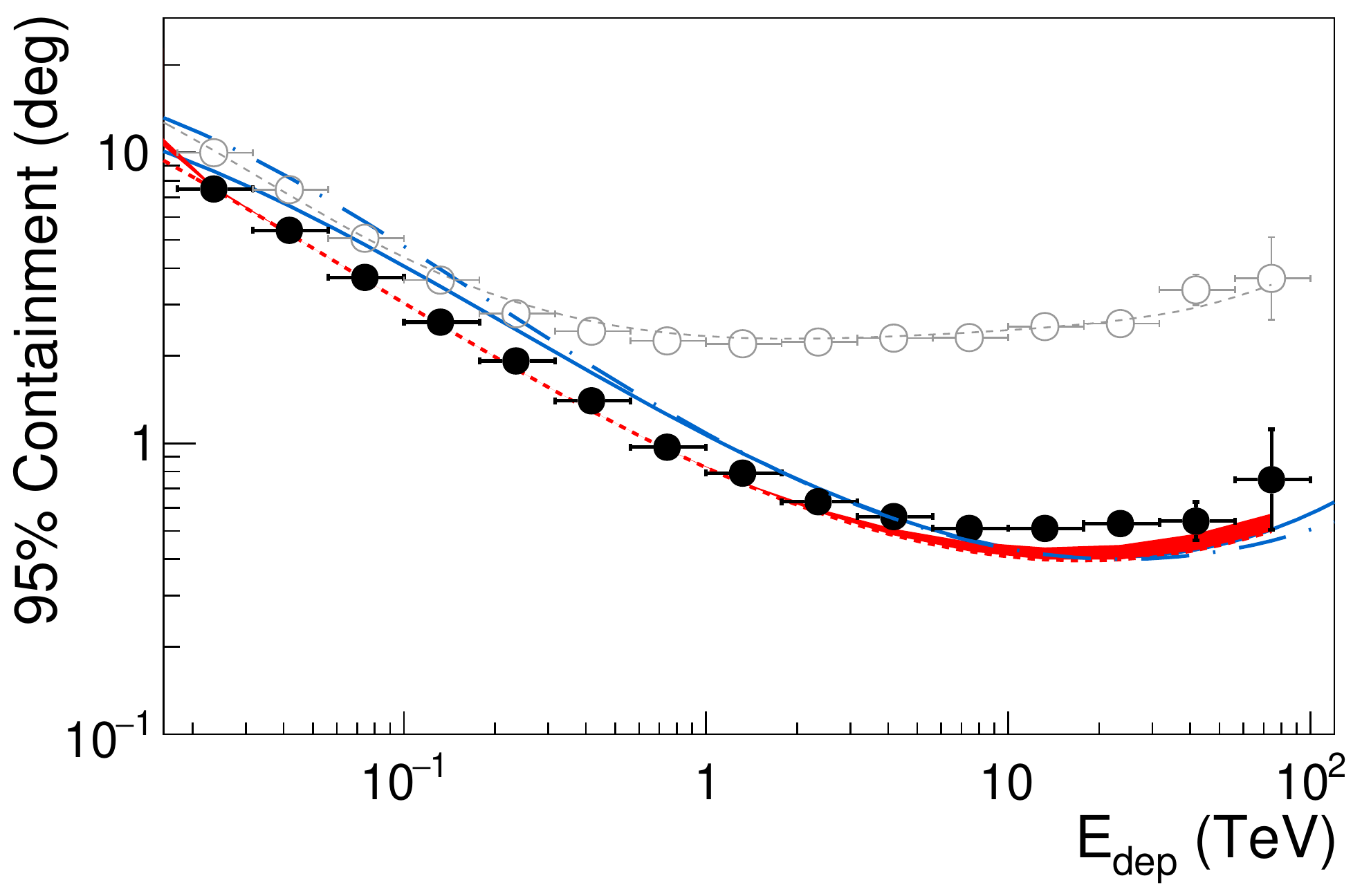}
\includegraphics[width=\psffiguresize\textwidth]{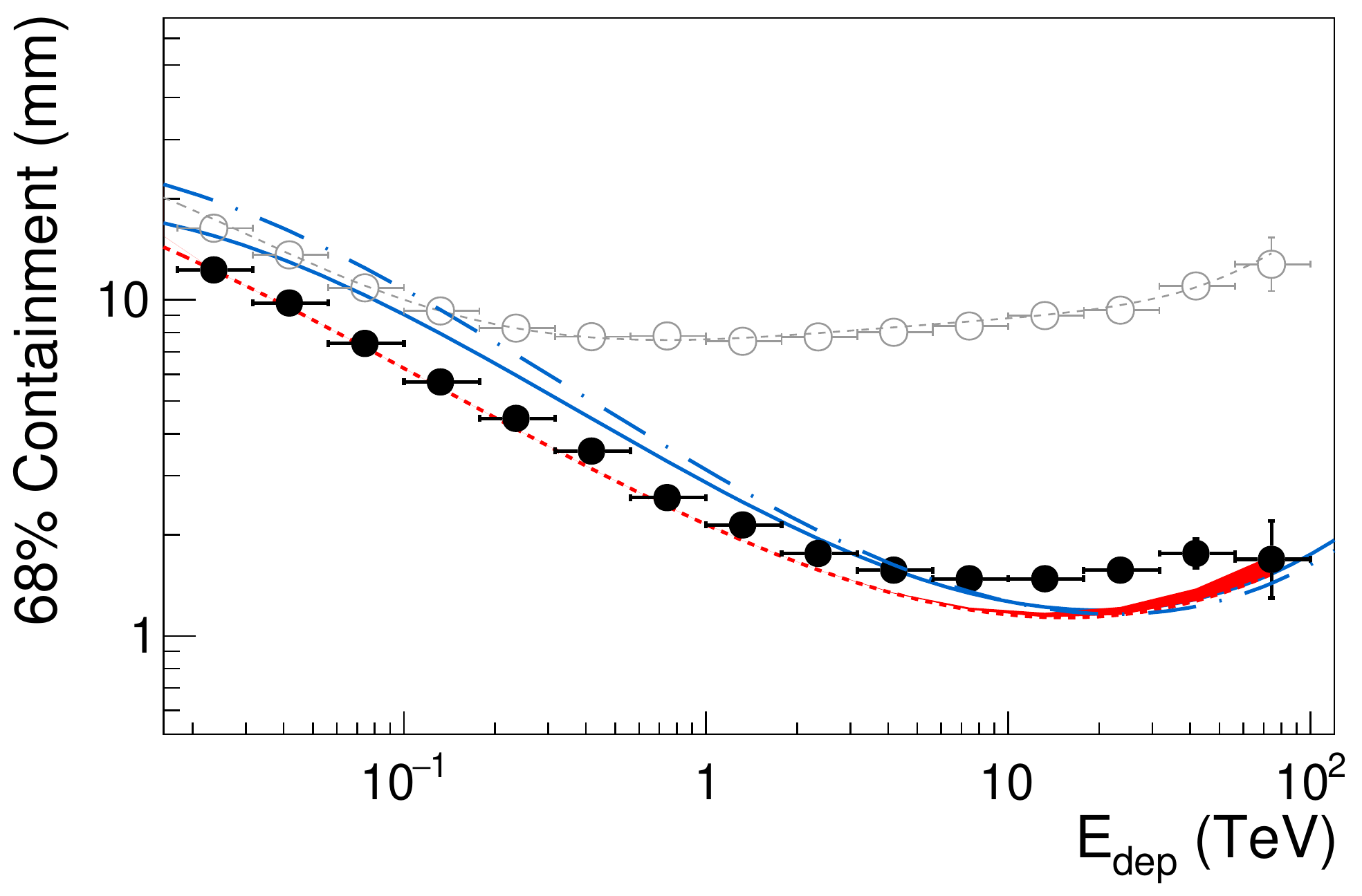}
\includegraphics[width=\psffiguresize\textwidth]{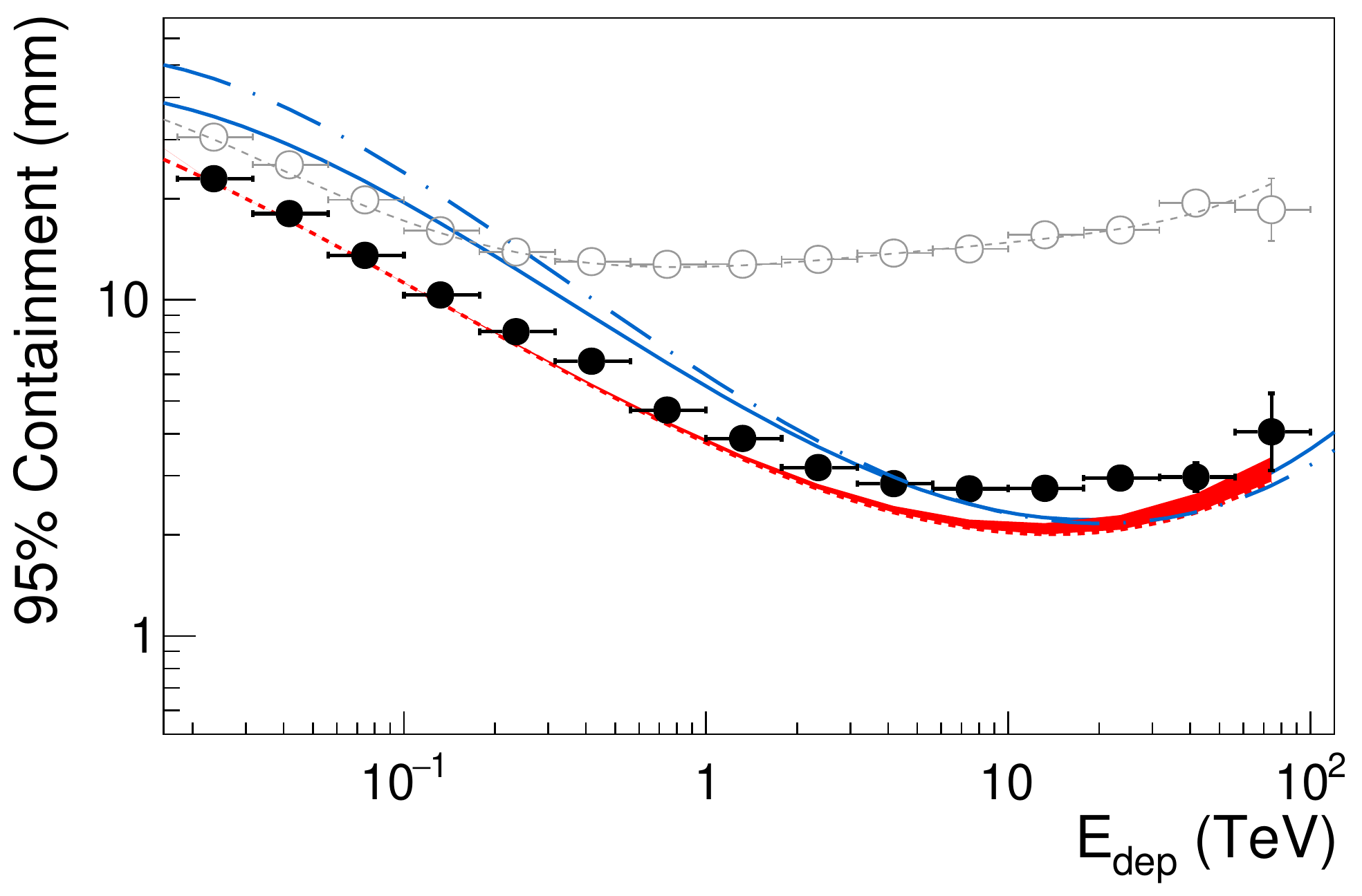}

(a)

\vspace{5mm}

\includegraphics[width=\psffiguresize\textwidth]{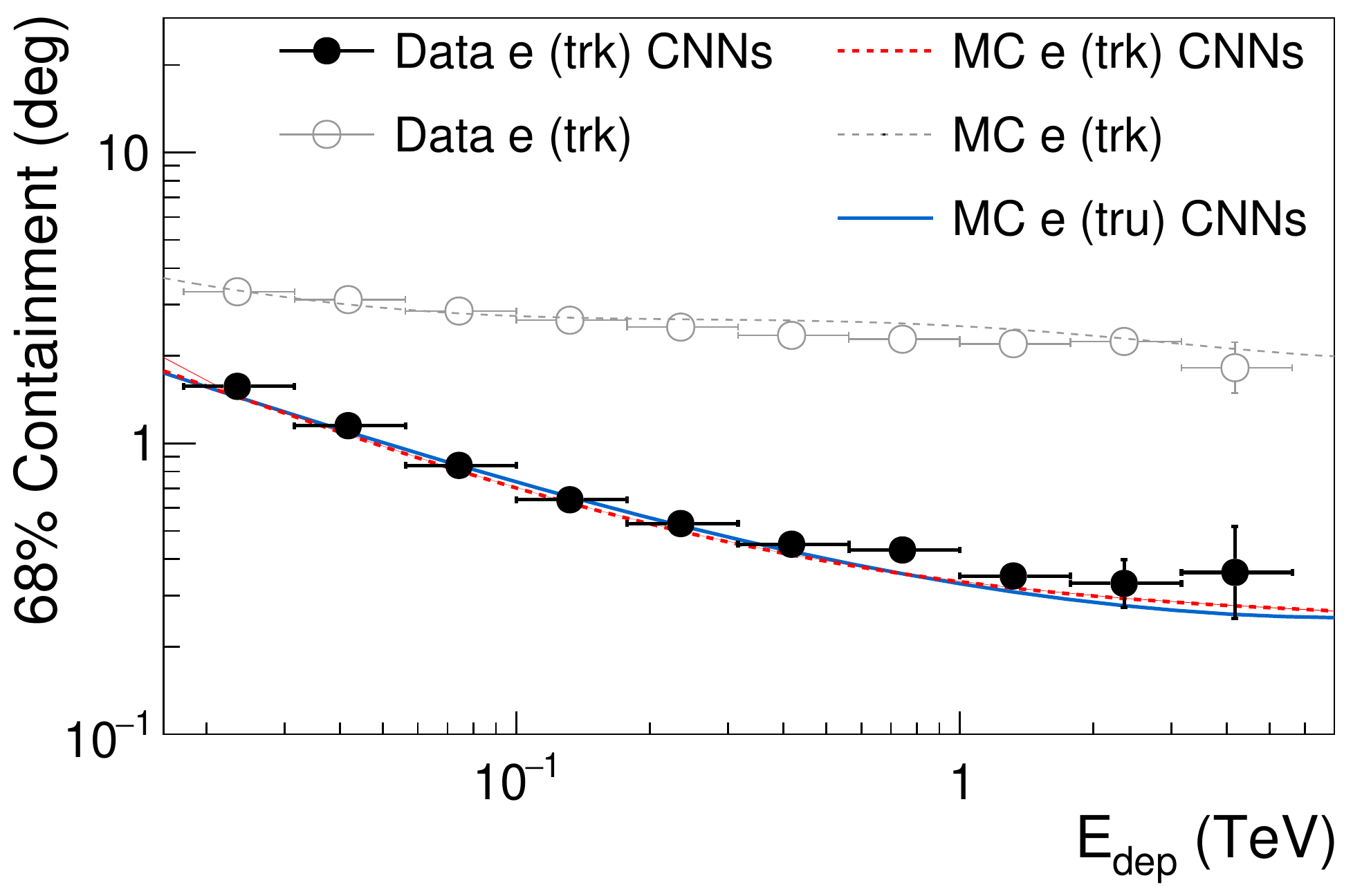}
\includegraphics[width=\psffiguresize\textwidth]{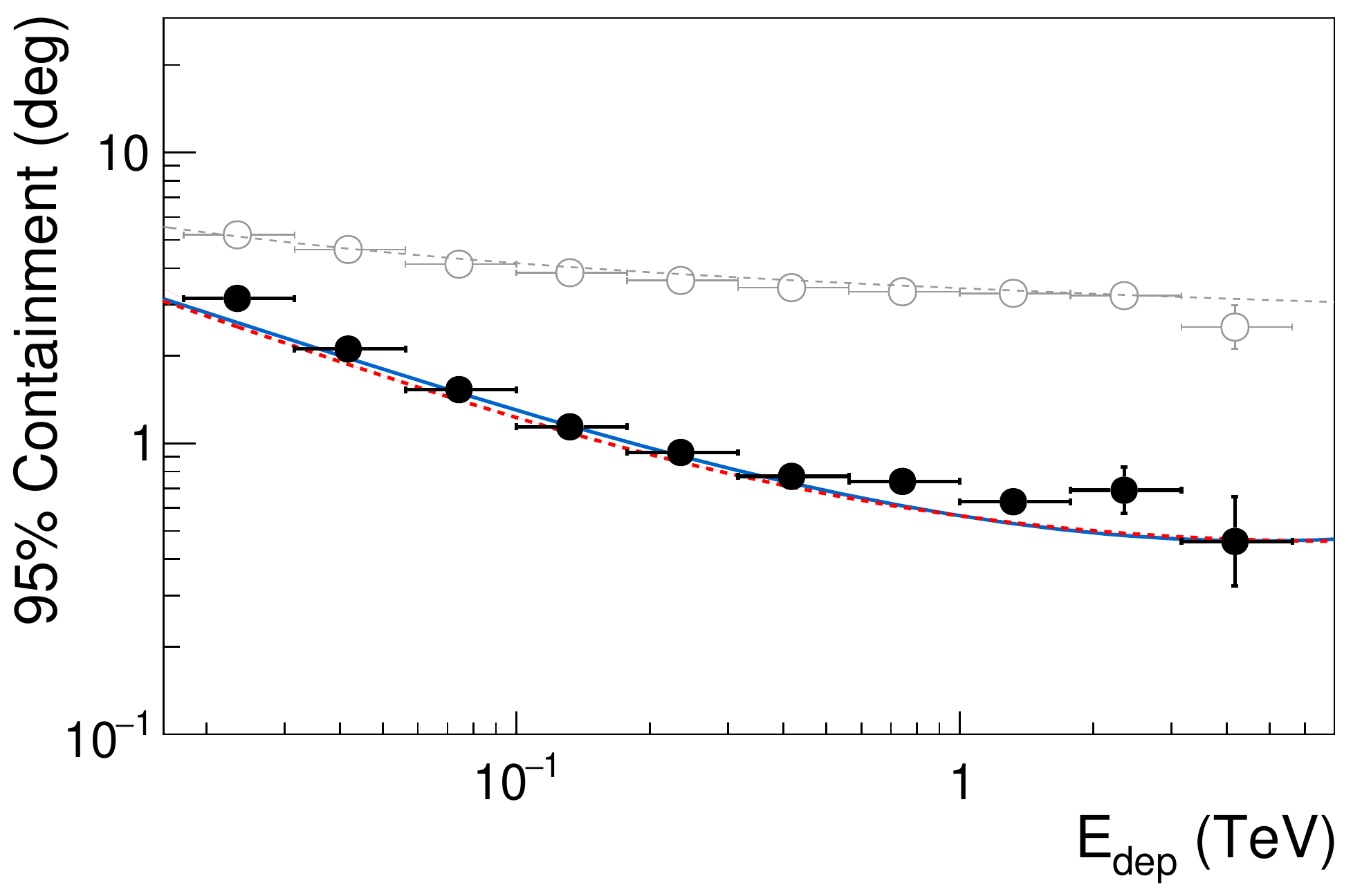}
\includegraphics[width=\psffiguresize\textwidth]{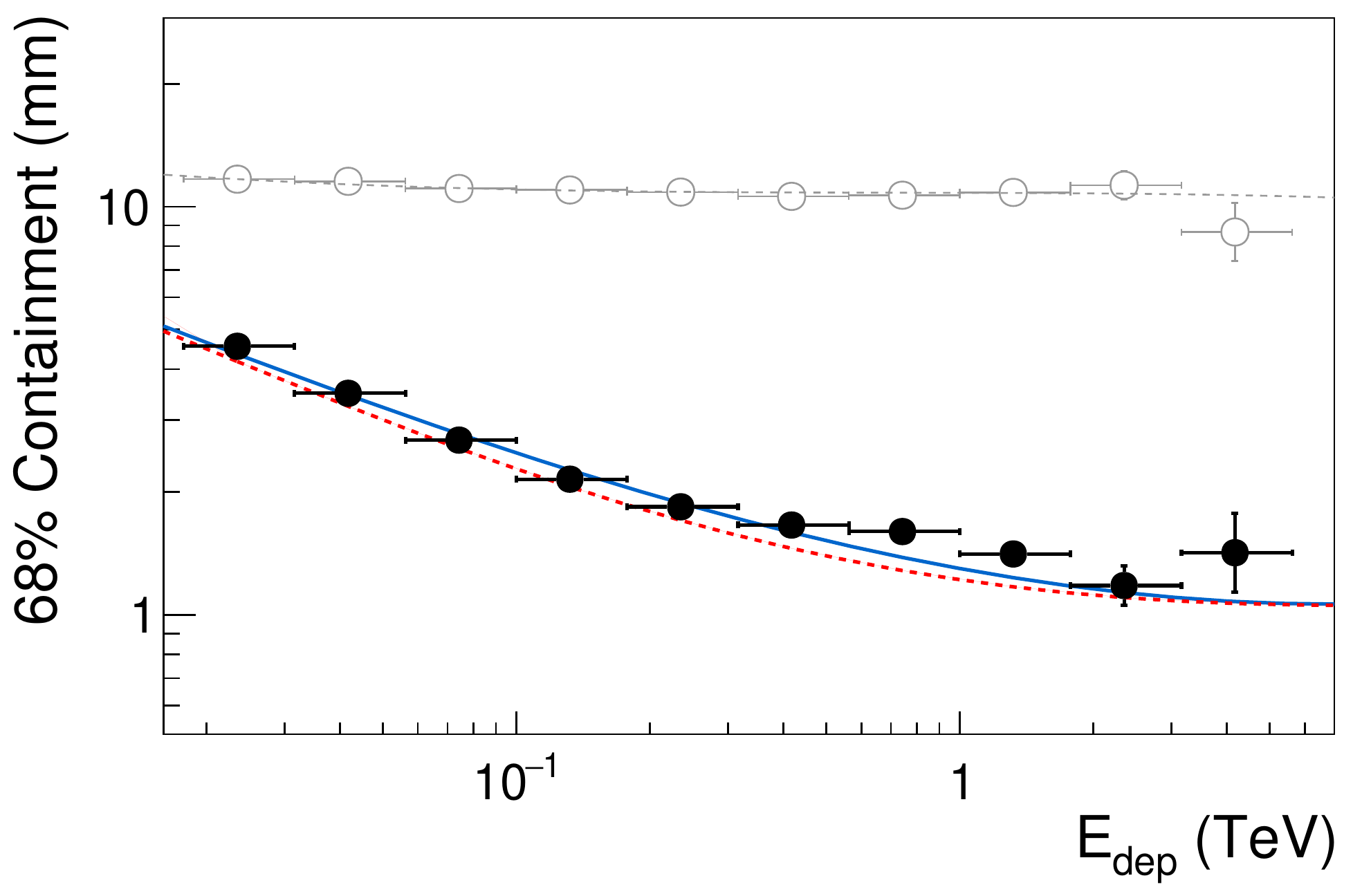}
\includegraphics[width=\psffiguresize\textwidth]{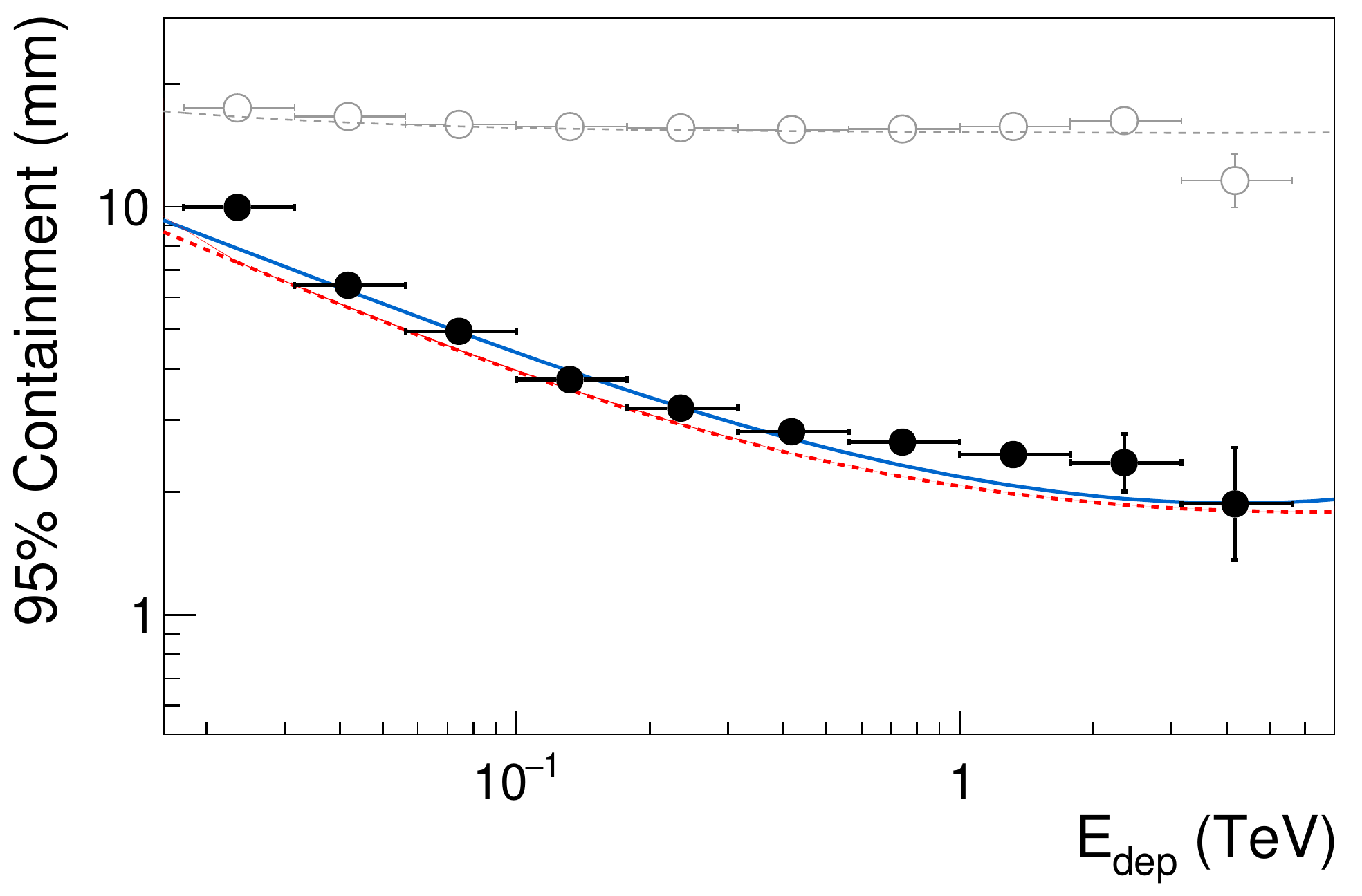}

(b)
\end{center}
\caption{68\% and 95\% containment of particle direction, in terms of azimuthal angle, $\sqrt{\Delta\Theta_x^2+\Delta\Theta_y^2}$, and intercept, $\sqrt{\Delta x_0^2+\Delta y_0^2}$, obtained with the calorimeter CNNs model, as a function of the total deposited energy in BGO: (a) proton and helium, (b) electrons. Either the  true particle direction ({\bf tru}) or the~\emph{clean} STK track  ({\bf trk}) is taken as a reference. A combined proton plus helium ({\bf trk} reference) MC is shown in (a) for comparison with the data. The systematic errors due to the CNNs generalization gap are shown with shaded bands. The results for the standard DAMPE calorimeter reconstruction are overlaid for comparison. }
\label{fig:bgo_psf}
\end{figure}  

As shown in Figure~\ref{fig:bgo_psf}, a significant improvement over the standard~\emph{shower-axis} algorithms can be observed. In particular, the 68\% angular (position) containment for hadronic showers  is lower than 0.4\degree (1.7~\mm) at 100 TeV of deposited energy, which is more than 5 (7) times better than with the standard algorithm (Figure~\ref{fig:bgo_psf}--a).  For the electromagnetic showers, the 68\% angular (position) containment reaches about  0.35\degree (1.4~\mm) at 5 TeV, with the corresponding improvement of about 6 (8) times with respect to the standard DAMPE algorithm. Some relatively small discrepancy between data and MC can be observed at the highest energies, which is likely attributed to the possible imperfections in the BGO simulation. It is interesting to note that the angular resolution at TeV and higher energies (0.35--0.4\degree),  while being obtained purely from the calorimeter, is not much worse than the typical angular resolution of dedicated tracking subdetectors (0.05--0.5\degree) of cosmic- and gamma-ray missions, including DAMPE itself. Moreover, it is better than the typical gamma-ray pointing accuracy of DAMPE, Fermi-LAT, and CALET at the GeV scale, where the bulk of gamma-ray sources is observed~{\citep{dampe_mission,Fermi-LAT:2013mql,CALET:2018bqg}. At the same time, we also note that the  advantage of  the CNNs algorithm is marginal at low energies. The latter is expected  since at low energies the shower is concentrated in very few pixels of the calorimeter image. In turn, the CNNs advantage manifests clearly at higher energies, as the amount of information per particle image increases. 
 
Finally, the developed CNNs approach represents a significant improvement over the standard methods. We hypothesise that the architecture of the model can be improved even further, in particular by exploiting more efficiently the correlations between signals in alternating BGO layers, with more complex CNNs or, alternatively, Graph Neural Networks. We leave this subject for future studies.

\section{Neural network tracker direction prediction}
\label{sec:tracker}
The above prediction of the calorimeter CNNs algorithm  serves as a seed for the particle direction reconstruction with the tracker. Even if simply combined with the conventional Kalman algorithm~\citep{Tykhonov:2017uno,Tykhonov:2017rdq} it is expected to improve the accuracy compared to the standard DAMPE track reconstruction. However, the problem of the correct track identification from the ensemble of Kalman track candidates would remain open. Instead, our goal is to develop an algorithm which provides a single particle trajectory as close as possible to the real one. The developed algorithm is also based on CNNs,  as illustrated in Figure~\ref{fig:stk_cnn}. First, the direction prediction from  the calorimeter CNNs model is projected onto the tracker, selecting the hits within a certain window, hereafter called the Region of Interest (RoI)\footnote{We use a window of $\pm$10~\mm, corresponding to $\geq$~99\% containment of a true particle direction.}. Next, a Hough transform~\citep{Hough:1959qva} is done converting the selected hits into the lines on a Hough image. Values on the image axes correspond to the offsets in pixel units from the predicted calorimeter trajectory in the top and bottom layers of the tracker, respectively:
\begin{equation*}
\delta_{x,\text{top}} \equiv (x_\text{top} - x_\text{top}^{\text{BGO CNNs}})/50 ~\text{\textmu m} + 200,
\end{equation*}
where $x_\mathrm{top}$ is the position in the top $x$ layer of the STK and $x_\mathrm{top}^\mathrm{BGO\:CNNs}$ is the corresponding prediction of the calorimeter CNNs model. Similar definitions holds for $\delta_{x,\text{bot}}$, $\delta_{y,\text{top}}$ and  $\delta_{y,\text{bot}}$. The image represents a 20$\times$20~\mm~window, with a pixel resolution of 50~\micron\footnote{Pixel resolution is chosen to match the position resolution of the STK~\citep{Tykhonov:2017uno}.}. In the event of the BGO CNNs model prediction being perfectly correct, the true position of a particle would be placed directly in the center of the image. 
 
\begin{figure}
\begin{center}
\includegraphics[width=0.97\textwidth]{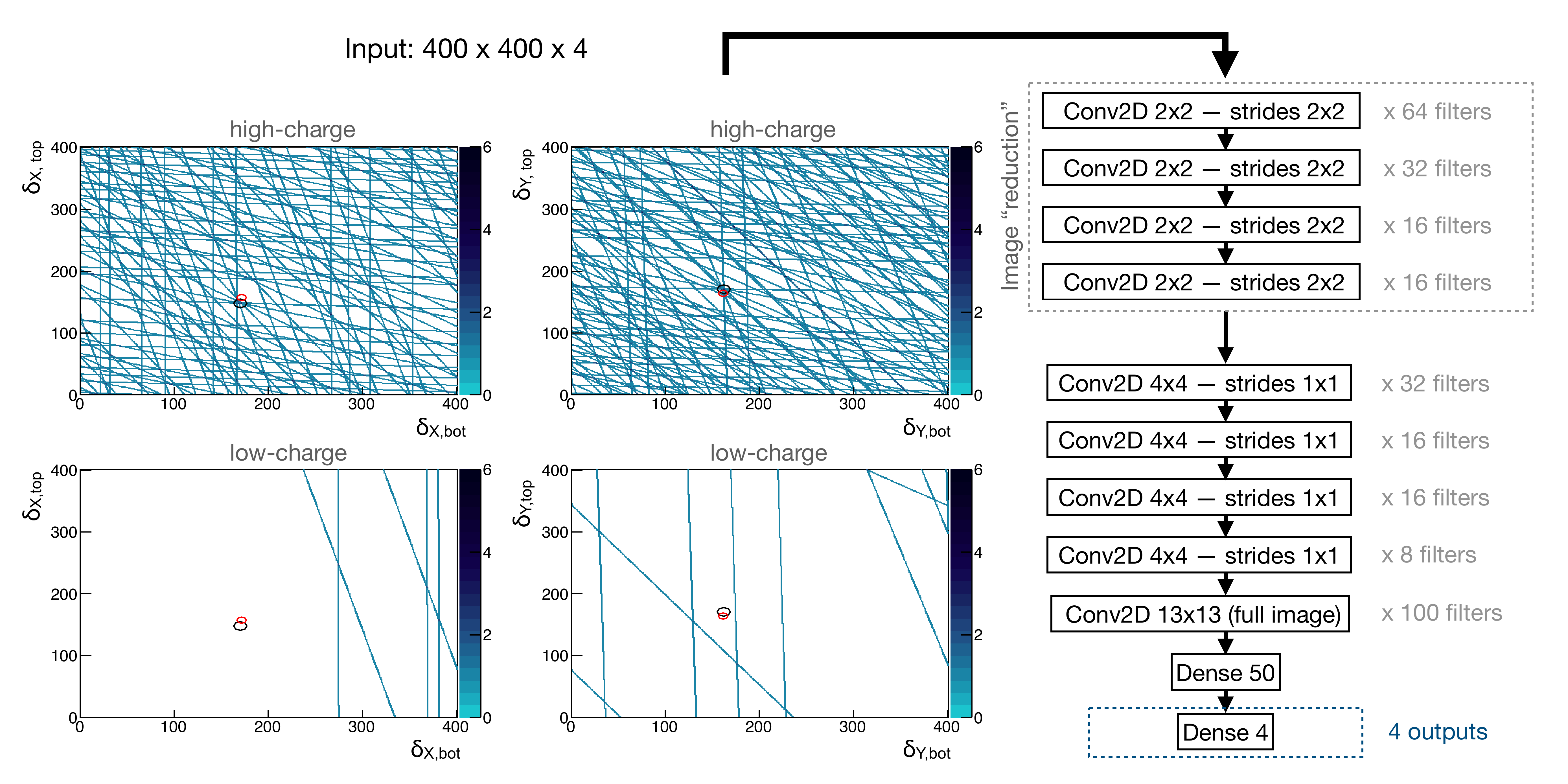}
\end{center}
\caption{Hough image of a typical Helium event and the architecture of the tracker convolutional neural network. Big (black) and small (red) circles represent the true and the reconstructed trajectory of a primary particle,  respectively. Similar to the calorimeter CNNs model, the ReLU activation function is used in all layers except for the last one, which has a linear activation. The 400$\times$400 dimension is determined by the chosen RoI size and pixel resolution.}
\label{fig:stk_cnn}
\end{figure}   
 
 It is worth noting that, contrary to the BGO, we do not use raw (non-transformed) images of the STK for the track reconstruction.  The reason is that the nature of a particle's passage through the tracker is fundamentally different from the one in the calorimeter. In particular, the raw tracker image does not necessarily show a pre-shower profile correlated with the primary particle direction. The true particle track may be hidden among secondary-particle hits with higher signals. In fact, we have also  tried to develop a CNNs model which uses raw STK images but no satisfying solution was found\footnote{The raw-image CNNs in turn prove to be useful for a classification type problem, as will be shown in Section~\ref{sec:classifier}.}. Another possible alternative could potentially lie in the Graph Neural Networks domain~\citep{Duarte:2020ngm}, which we do not cover in this paper. We opt for the Hough transform as a simple but powerful enough way of  structuring topologically the hits. As we will show later in the paper, the Hough transform combined with the CNNs allows to achieve an excellent particle trajectory reconstruction in DAMPE. 

 \newcommand{\houghinferencetimecpu}{0.35}
 \newcommand{\kalmantimecpu}{3.4} 
  \newcommand{\houghitrainingtimegpu}{8}
 \newcommand{\cnninferencetotalfivemodels}{1.6}
Similar to the calorimeter network, 8-bit precision is used to store information in each pixel. Two tracker projections are mapped on separate images. Moreover,  each projection is split further into two images, consisting of hits with STK signal (charge) either below or above a threshold $\mathrm{Z_{thr}}$, respectively\footnote{The performance of the algorithm is not sensitive to the exact choice of the $\mathrm{Z_{thr}}$ value. We have tested values in the range from 1 to 2, found no significant difference, and chosen $\mathrm{Z_{thr}}=\sqrt{2}$. At the same time the algorithm performance is better than in the case of no splitting.}. In other words, for each projection, there is one image with the STK hits more likely corresponding to protons and the other one with the hits which potentially correspond to helium or heavier ions. In this way we partially encode the STK signal information into the image. As a result, the input image has a dimension 400$\times$400$\times$4. The internal STK alignment is applied to correct for the hit positions  in the tracker~\citep{Tykhonov:2017uno,Tykhonov:2018stq}. Note that in order to  avoid potential problems due to a (mis-)modeling of the readout saturation for heavy ions, we do not add more detailed STK signal information into the Hough image. Moreover, even for low-charge particles, the STK signal in the MC simulations may be slightly different from the flight data~\citep{Tykhonov:2017uno}. The described image design allows us to mitigate the effect of the simulation inaccuracy, such that the trained model can adapt correctly to the real data.

The image is provided as an input to the dedicated regression network, whose goal is to predict the true ``position''  of a particle on the image. The building blocks of the CNNs model are depicted in Figure~\ref{fig:stk_cnn} right.  The fitting time of the network is about~\houghitrainingtimegpu~hours/epoch for $10^6$ batches. The training is also done in three energy ranges, as described in Section~\ref{sec:calo}. Similar to the BGO calorimeter, the  STK tracker starts saturating at the highest energies due to the increased density of secondary particles and the limitations of the data reduction algorithms in the STK readout electronics~\citep{Dong_2015}. Hence, the same argumentation holds for having  different models focused on specific energy ranges, for both the BGO calorimeter and the STK tracker. The CPUs inference time of the tracker CNNs model is about~\houghinferencetimecpu~s/event\footnote{We expect the inference time to decrease dramatically if the inference is done in batches of events, using GPUs instead of CPUs. In the current DAMPE software framework, the data processing is done on an event-by-event basis.}, comparable to that of the calorimeter model. For reference, the average time consumed by conventional track pattern recognition in the reconstruction of DAMPE flight data is~\kalmantimecpu~s/event. 
 
Once the particle direction is obtained from the STK CNNs model, the STK hits are assigned to it using a simple distance matching, as described in Section~\ref{sec:reconstruction}, forming the final track. Note that unlike the conventional Kalman approach, we do not perform  fitting of the hits. There is no need for a further track selection since only one track is provided per event.

In summary, we note that the particle direction prediction is done with two CNNs algorithms chained together. The first one infers the particle direction from the calorimeter, while the second one yields a more precise prediction from the tracker, using the prediction provided by the calorimeter CNNs model as RoI. At energies below $\sim$1~TeV, however, the precision of the calorimeter CNNs model is not sufficient to provide an accurate RoI for the tracker CNNs model\footnote{For events with deposited energies below $\sim$1~TeV, the probability that the true particle direction falls into the $\pm$10~\mm~RoI does not meet the requirement of 99\%.}. Hence we add an additional intermediate step to account for the lower accuracy of the calorimeter inference. In this step we use a  coarse Hough CNNs model, identical to the one described above but with a lower pixel resolution, 400~\micron~instead of 50~\micron, and a wider window, $\pm$80~\mm~instead of $\pm$10~\mm. As a result, at low energies the particle direction prediction happens in three steps: the calorimeter CNNs model prediction, the coarse tracker inference (400~\micron), and finally the precise tracker inference (50~\micron)\footnote{The coarse tracker CNNs model is added to the reconstruction chain if the deposited particle energy in BGO is lower that 1~TeV, which approximately corresponds to a true particle energy of about $\sim$3~TeV~\citep{dampe_science,dampe_prl}.}.

\newcommand{\efficiency}{$\epsilon$}
As a figure of merit of the tracking algorithm, we use the combined reconstruction and selection efficiency (\efficiency) of the track as follows. First, we perform the event selection as described in Section~\ref{sec:reconstruction}, using the true particle direction as the ``track''. Next, we make distributions of the STK signal (Figure~\ref{fig:signal_psd_stk_truth}) in different energy bins and define the regions of 85\% (95\%) signal containment for proton and helium peaks in each bin. Then we repeat the procedure using instead of the true direction of the particle, the direction given by the track reconstruction and identification algorithm under test. In this procedure, the number of signal events are again counted in the same 85\% (95\%) containment interval derived from the true track direction. We consider three algorithms for the comparison: (1) standard reconstruction with the~\emph{ideal} identification; (2) standard reconstruction with the~\emph{standard} identification; (3) the developed CNNs algorithm. Finally, the efficiency is defined as the ratio of the number of events in the 85\% (95\%) window obtained with one of the three algorithms, to the number of events (in the same window) obtained with the true track. The results are shown in Figure~\ref{fig:trk_eff} and Table~\ref{tab:trk_eff}. The standard track reconstruction with the~\emph{ideal} identification has an efficiency in the 96\%-98\% range up to 100 TeV and reduces to about 70--75\% and 50--55\% at PeV for proton and helium, respectively. The drop of the reconstruction efficiency at few hundred TeV is explained by  the decreased accuracy of the standard BGO direction reconstruction at highest energies, caused by the saturation of the BGO readout~\citep{mikhail,Yue:2020hmj}. The combined standard  track reconstruction and identification efficiency is on average about 80\% and 90\% for proton and helium respectively, below 100 TeV, and drops sharply towards PeV energies (below 35\% and 40\% respectively),  as it becomes increasingly hard to identify the correct track with the classical selection methods~\citep{dampe_science,dampe_prl}. For the developed CNNs algorithm, the tracking efficiency is higher than 98\% at energies up to a few hundred TeV, where it reduces slightly to about 96--97\% at PeV. The estimated uncertainty related to the CNNs generalization gap (both calorimeter and the tracker CNNs models) is negligible below 500~TeV. At higher energies, it does not  exceed 1\% and 2\% for helium and proton, respectively.

\newcommand{\efffiguresize}{0.85}
\begin{figure}
\begin{center}
\includegraphics[width=\efffiguresize\textwidth]{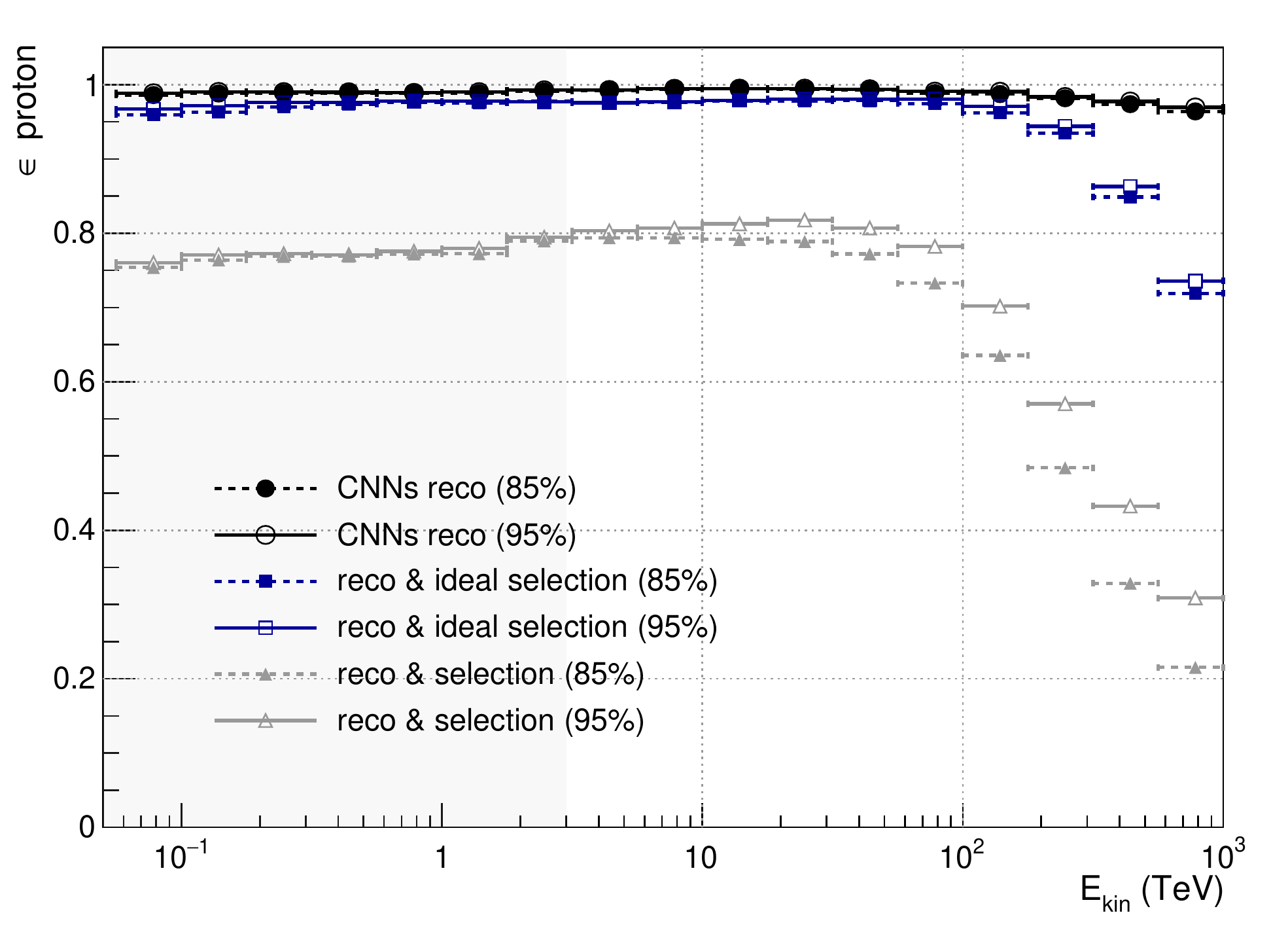}
\includegraphics[width=\efffiguresize\textwidth]{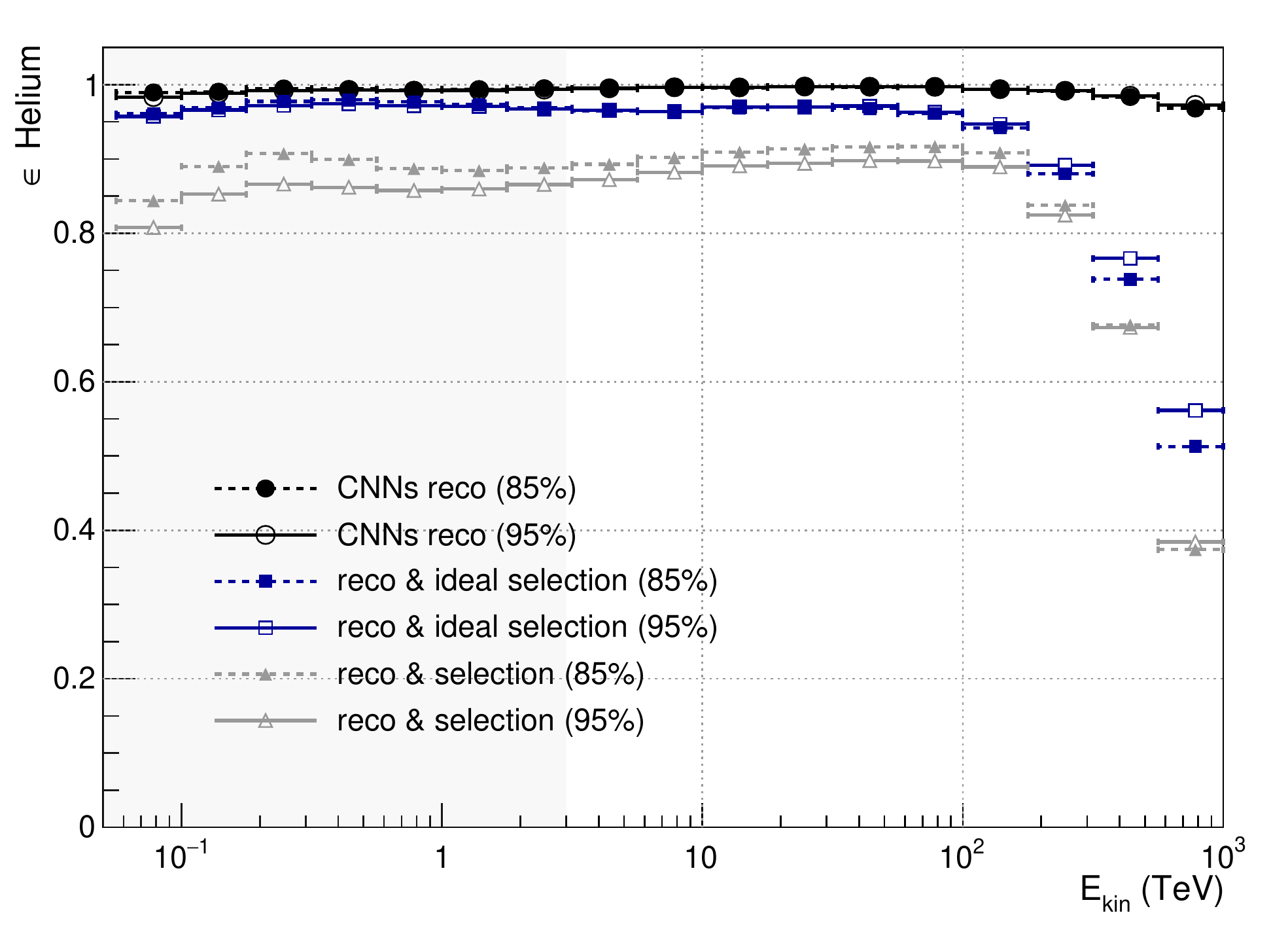}
\end{center}
\caption{Efficiency of the track reconstruction and identification derived from MC as a function of the particle kinetic energy: (circles) the developed CNNs algorithm; (squares) standard track reconstruction with the \emph{ideal} identification; (triangles) standard track reconstruction with the standard identification. Proton (top) and helium (bottom) cases are shown. The gray shaded area indicates the region where the two-step tracker CNNs prediction is used (1~TeV deposited energy roughly corresponds to $\sim$3~TeV particle kinetic energy~\citep{dampe_science,dampe_prl}).} 
\label{fig:trk_eff}
\end{figure}

\begin{table}
\begin{tabular}{r@{ -- }l|r@{}lr@{}l|r@{}lr@{}l|r@{}lr@{}l}
\hline
\multicolumn{2}{c|}{} &   \multicolumn{12}{c}{p Efficiency (\%)} \\
\multicolumn{2}{c|}{$\mathrm{E_{kin}}$ (TeV)}   &  \multicolumn{4}{c}{\bf reco \& std. sel.}  &   \multicolumn{4}{c}{\bf reco \& ideal sel.}   &  \multicolumn{4}{c}{\bf CNNs reco} \\
\multicolumn{2}{c|}{} & \multicolumn{2}{c}{(85\%)} &   \multicolumn{2}{c}{(95\%)}    & \multicolumn{2}{c}{(85\%)} &   \multicolumn{2}{c}{(95\%)}  & \multicolumn{2}{c}{(85\%)} &   \multicolumn{2}{c}{(95\%)}  \\
\hline
$   0.056$ & $   0.100$ & $ 75.4$ & $^{\pm0.1}$ & $ 76.0$ & $^{\pm0.1}$ & $ 96.0$ & $^{\pm0.1}$ & $ 96.7$ & $^{\pm0.1}$ & $ 98.6$ & $^{\pm0.0}$ & $ 98.9$ & $^{\pm0.0}$ \\
$   0.562$ & $   1.000$ & $ 77.2$ & $^{\pm0.2}$ & $ 77.6$ & $^{\pm0.2}$ & $ 97.6$ & $^{\pm0.1}$ & $ 97.8$ & $^{\pm0.1}$ & $ 98.8$ & $^{\pm0.0}$ & $ 98.9$ & $^{\pm0.0}$ \\
$   5.620$ & $  10.000$ & $ 79.4$ & $^{\pm0.2}$ & $ 80.7$ & $^{\pm0.2}$ & $ 97.6$ & $^{\pm0.1}$ & $ 97.7$ & $^{\pm0.1}$ & $ 99.4$ & $^{\pm0.0}$ & $ 99.5$ & $^{\pm0.0}$ \\
$  56.230$ & $ 100.000$ & $ 73.3$ & $^{\pm0.3}$ & $ 78.2$ & $^{\pm0.2}$ & $ 97.4$ & $^{\pm0.1}$ & $ 98.1$ & $^{\pm0.1}$ & $ 98.9$ & $^{\pm0.1}$ & $ 99.1$ & $^{\pm0.1}$ \\
$ 562.340$ & $1000.000$ & $ 21.5$ & $^{\pm0.4}$ & $ 30.9$ & $^{\pm0.4}$ & $ 71.9$ & $^{\pm0.4}$ & $ 73.6$ & $^{\pm0.4}$ & $ 96.4$ & $^{\pm1.7}$ & $ 96.9$ & $^{\pm1.4}$ \\
\hline
\multicolumn{2}{c|}{} &   \multicolumn{12}{c}{He Efficiency (\%)} \\
\multicolumn{2}{c|}{$\mathrm{E_{kin}}$ (TeV)}   &  \multicolumn{4}{c}{\bf reco \& std. sel.}  &   \multicolumn{4}{c}{\bf reco \& ideal sel.}   &  \multicolumn{4}{c}{\bf CNNs reco} \\
\multicolumn{2}{c|}{}  & \multicolumn{2}{c}{(85\%)} &   \multicolumn{2}{c}{(95\%)}    & \multicolumn{2}{c}{(85\%)} &   \multicolumn{2}{c}{(95\%)}  & \multicolumn{2}{c}{(85\%)} &   \multicolumn{2}{c}{(95\%)}  \\
 \hline
$   0.056$ & $   0.100$ & $ 84.4$ & $^{\pm0.1}$ & $ 80.8$ & $^{\pm0.1}$ & $ 96.1$ & $^{\pm0.1}$ & $ 95.7$ & $^{\pm0.1}$ & $ 99.0$ & $^{\pm0.0}$ & $ 98.3$ & $^{\pm0.0}$ \\
$   0.562$ & $   1.000$ & $ 88.7$ & $^{\pm0.2}$ & $ 85.7$ & $^{\pm0.2}$ & $ 97.7$ & $^{\pm0.1}$ & $ 97.1$ & $^{\pm0.1}$ & $ 99.3$ & $^{\pm0.0}$ & $ 99.2$ & $^{\pm0.0}$ \\
$   5.620$ & $  10.000$ & $ 90.2$ & $^{\pm0.2}$ & $ 88.2$ & $^{\pm0.2}$ & $ 96.3$ & $^{\pm0.1}$ & $ 96.4$ & $^{\pm0.1}$ & $ 99.7$ & $^{\pm0.0}$ & $ 99.7$ & $^{\pm0.0}$ \\
$  56.230$ & $ 100.000$ & $ 91.7$ & $^{\pm0.1}$ & $ 89.7$ & $^{\pm0.1}$ & $ 96.1$ & $^{\pm0.1}$ & $ 96.3$ & $^{\pm0.1}$ & $ 99.7$ & $^{\pm0.1}$ & $ 99.7$ & $^{\pm0.1}$ \\
$ 562.340$ & $1000.000$ & $ 37.4$ & $^{\pm0.3}$ & $ 38.4$ & $^{\pm0.3}$ & $ 51.3$ & $^{\pm0.3}$ & $ 56.1$ & $^{\pm0.3}$ & $ 96.8$ & $^{\pm0.9}$ & $ 97.3$ & $^{\pm0.9}$ \\
\hline
\end{tabular}
\caption{Efficiency of the track reconstruction and selection derived from MC using different techniques: standard track reconstruction with the standard identification (left column);  standard track reconstruction with the \emph{ideal} identification (middle column); the developed CNNs algorithm (right column). Proton (top) and helium (bottom) cases are shown. Errors for the first two algorithms are statistical only. Errors for the CNNs algorithm also include the uncertainties related to the CNNs generalization gap. For the sake of brevity, selected points from Figure~\ref{fig:trk_eff} corresponding to different energy decades are shown. For the  CNNs reconstruction algorithm, a two-step tracker CNNs prediction is used in the first two energy bins.}
\label{tab:trk_eff}
\end{table}  

In order to quantify the performance of the developed calorimeter and tracker CNNs separately, we evaluate the tracking efficiency using the following two options: 
\begin{itemize}
\item (a) calorimeter CNNs model combined with the the standard track pattern recognition based on the Kalman filter approach;
\item (b) conventional calorimeter shower axis direction reconstruction combined with the tracker CNNs model\footnote{In order to compensate for the lower accuracy of the standard calorimeter shower axis direction reconstruction, compared to that of the calorimeter CNNs, here we use the two-step tracker CNNs (coarse and precise) in the entire energy range.}.
\end{itemize}

\begin{figure}
\begin{center}
\includegraphics[width=\efffiguresize\textwidth]{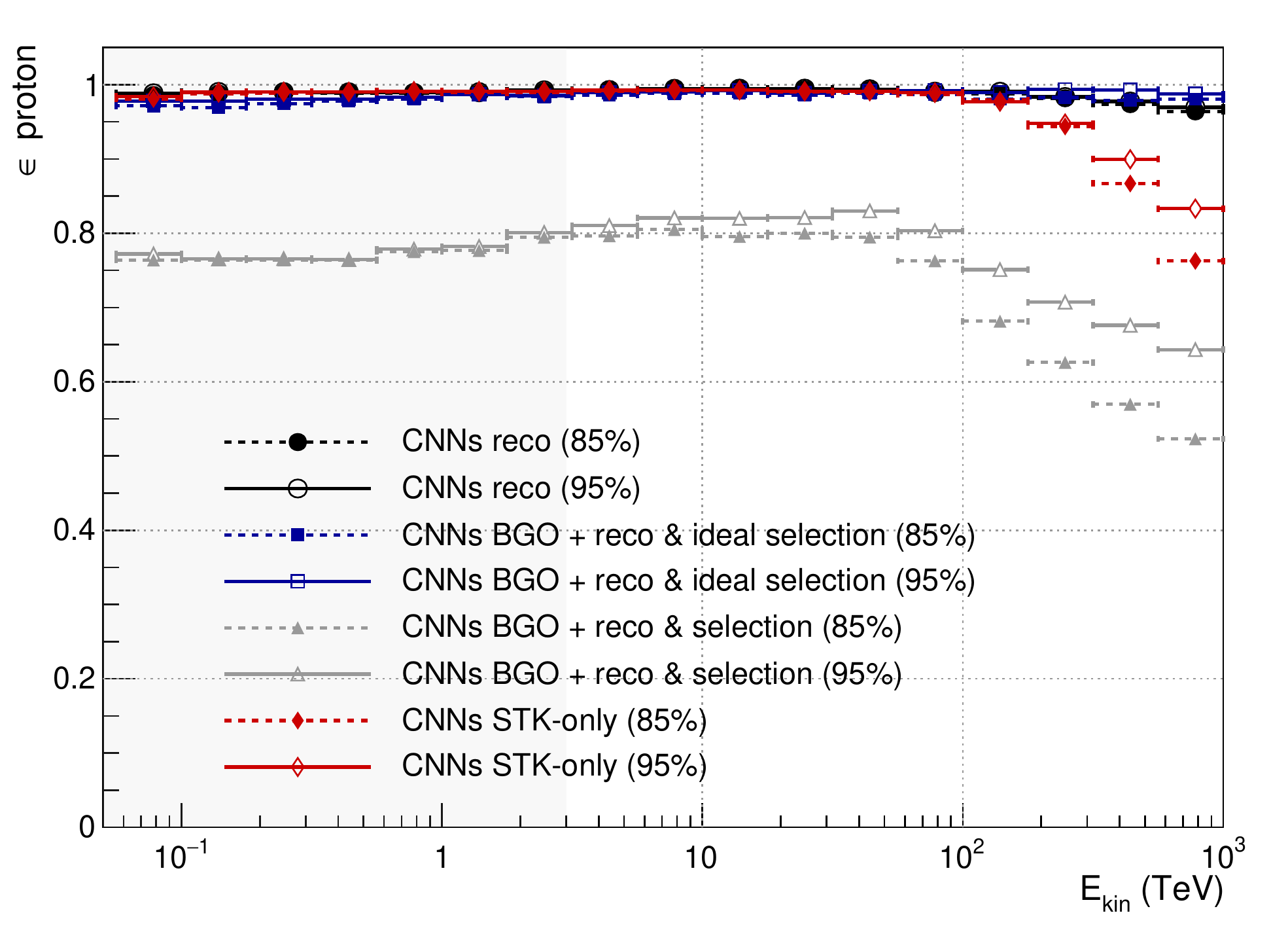}
\includegraphics[width=\efffiguresize\textwidth]{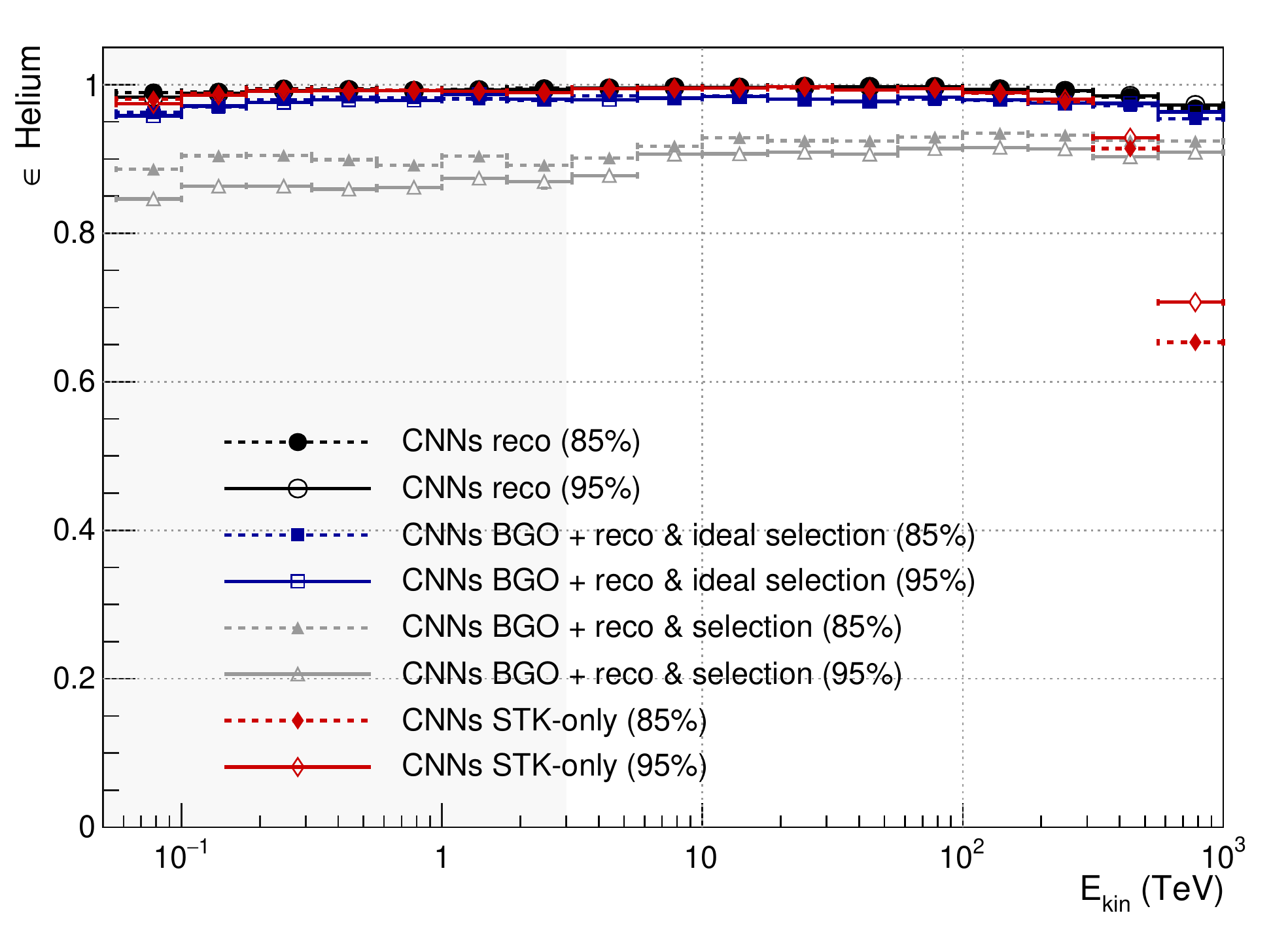}
\end{center}
\caption{Efficiency of the track reconstruction and identification derived from 
 MC as a function of the particle kinetic energy: 
 (circles) the combined chain of tracker and calorimeter CNNs (the same as in Figure~\ref{fig:trk_eff}); (squares and triangles) standard track reconstruction combined with the calorimeter CNNs model, using either the \emph{ideal} (circles) or standard (triangles) identification; (diamonds) tracker CNNs model combined with the standard calorimeter shower axis direction reconstruction. Proton (top) and helium (bottom) cases are shown. The gray shaded area indicates the region where the two-step tracker CNNs are used for the combined CNNs reconstruction.  
For the tracker-only CNNs model, the two-step prediction is used in the entire energy range.}
\label{fig:trk_eff_ablation}
\end{figure}

\begin{table}
\begin{tabular}{r@{ -- }l|r@{}lr@{}l|r@{}lr@{}l|r@{}lr@{}l}
\hline
\multicolumn{2}{c|}{} &   \multicolumn{12}{c}{p Efficiency (\%)} \\
\multicolumn{2}{c|}{$\mathrm{E_{kin}}$ (TeV)}   &  \multicolumn{4}{c}{\bf reco \& std. sel.}  &   \multicolumn{4}{c}{\bf reco \& ideal sel.}   &  \multicolumn{4}{c}{\bf CNNs STK} \\
\multicolumn{2}{c|}{}                                              &  \multicolumn{8}{c}{\bf + CNNs BGO}  &  \multicolumn{4}{c}{} \\ 
\multicolumn{2}{c|}{} & \multicolumn{2}{c}{(85\%)} &   \multicolumn{2}{c}{(95\%)}    & \multicolumn{2}{c}{(85\%)} &   \multicolumn{2}{c}{(95\%)}  & \multicolumn{2}{c}{(85\%)} &   \multicolumn{2}{c}{(95\%)}  \\
\hline
$   0.056$ & $   0.100$ & $ 76.4$ & $^{\pm0.5}$ & $ 77.2$ & $^{\pm0.4}$ & $ 97.1$ & $^{\pm0.2}$ & $ 97.8$ & $^{\pm0.2}$ & $ 98.1$ & $^{\pm0.0}$ & $ 98.4$ & $^{\pm0.0}$ \\
$   0.562$ & $   1.000$ & $ 77.5$ & $^{\pm0.5}$ & $ 77.9$ & $^{\pm0.4}$ & $ 98.1$ & $^{\pm0.2}$ & $ 98.3$ & $^{\pm0.1}$ & $ 99.1$ & $^{\pm0.0}$ & $ 99.1$ & $^{\pm0.0}$ \\
$   5.620$ & $  10.000$ & $ 80.5$ & $^{\pm0.5}$ & $ 82.1$ & $^{\pm0.4}$ & $ 98.8$ & $^{\pm0.1}$ & $ 99.0$ & $^{\pm0.1}$ & $ 99.2$ & $^{\pm0.0}$ & $ 99.3$ & $^{\pm0.0}$ \\
$  56.230$ & $ 100.000$ & $ 76.3$ & $^{\pm0.6}$ & $ 80.3$ & $^{\pm0.5}$ & $ 98.7$ & $^{\pm0.1}$ & $ 99.2$ & $^{\pm0.1}$ & $ 98.7$ & $^{\pm0.1}$ & $ 99.0$ & $^{\pm0.1}$ \\
$ 562.340$ & $1000.000$ & $ 52.3$ & $^{\pm0.5}$ & $ 64.3$ & $^{\pm0.5}$ & $ 98.0$ & $^{\pm1.7}$ & $ 98.7$ & $^{\pm1.4}$ & $ 76.3$ & $^{\pm0.4}$ & $ 83.3$ & $^{\pm0.3}$ \\
\hline
\multicolumn{2}{c|}{} &   \multicolumn{12}{c}{He Efficiency (\%)} \\
\multicolumn{2}{c|}{$\mathrm{E_{kin}}$ (TeV)}   &  \multicolumn{4}{c}{\bf reco \& std. sel.}  &   \multicolumn{4}{c}{\bf reco \& ideal sel.}   &  \multicolumn{4}{c}{\bf CNNs STK} \\
\multicolumn{2}{c|}{}                                              &  \multicolumn{8}{c}{\bf + CNNs BGO}  &  \multicolumn{4}{c}{}\\ 
\multicolumn{2}{c|}{}  & \multicolumn{2}{c}{(85\%)} &   \multicolumn{2}{c}{(95\%)}    & \multicolumn{2}{c}{(85\%)} &   \multicolumn{2}{c}{(95\%)}  & \multicolumn{2}{c}{(85\%)} &   \multicolumn{2}{c}{(95\%)}  \\
 \hline
$   0.056$ & $   0.100$ & $ 88.6$ & $^{\pm0.3}$ & $ 84.6$ & $^{\pm0.3}$ & $ 96.3$ & $^{\pm0.2}$ & $ 95.8$ & $^{\pm0.2}$ & $ 98.0$ & $^{\pm0.1}$ & $ 97.4$ & $^{\pm0.1}$ \\
$   0.562$ & $   1.000$ & $ 89.2$ & $^{\pm0.3}$ & $ 86.2$ & $^{\pm0.3}$ & $ 98.2$ & $^{\pm0.1}$ & $ 97.9$ & $^{\pm0.1}$ & $ 99.3$ & $^{\pm0.0}$ & $ 99.2$ & $^{\pm0.0}$ \\
$   5.620$ & $  10.000$ & $ 91.7$ & $^{\pm0.7}$ & $ 90.6$ & $^{\pm0.7}$ & $ 98.2$ & $^{\pm0.3}$ & $ 98.2$ & $^{\pm0.3}$ & $ 99.5$ & $^{\pm0.0}$ & $ 99.5$ & $^{\pm0.0}$ \\
$  56.230$ & $ 100.000$ & $ 92.9$ & $^{\pm0.5}$ & $ 91.4$ & $^{\pm0.5}$ & $ 98.0$ & $^{\pm0.3}$ & $ 98.3$ & $^{\pm0.2}$ & $ 99.5$ & $^{\pm0.1}$ & $ 99.5$ & $^{\pm0.1}$ \\
$ 562.340$ & $1000.000$ & $ 92.4$ & $^{\pm0.6}$ & $ 90.9$ & $^{\pm0.6}$ & $ 95.4$ & $^{\pm1.0}$ & $ 96.3$ & $^{\pm0.9}$ & $ 65.3$ & $^{\pm0.3}$ & $ 70.7$ & $^{\pm0.3}$ \\
\hline
\end{tabular}
\caption{Efficiency of the track reconstruction and selection derived from MC 
using different combinations of the developed CNNs algorithms and standard techniques: (left and middle columns) standard track reconstruction combined with the calorimeter CNNs model, using either the standard (left column) or \emph{ideal} (middle column) identification; tracker CNNs model combined with the standard calorimeter shower axis direction reconstruction (right column). Proton (top) and helium (bottom) cases are shown. Errors include statistical uncertainties and the uncertainties related to the CNNs generalization gap, summed in quadratures. 
 For the sake of brevity, selected points from Figure~\ref{fig:trk_eff_ablation} corresponding to different energy decades are shown. For the tracker-only CNNs model, the two-step
prediction is used in all energy bins.}
\label{tab:trk_eff_ablation}
\end{table}  
The corresponding results are shown in Figure~\ref{fig:trk_eff_ablation} and Table~\ref{tab:trk_eff_ablation}. 
For case (a), the track reconstruction efficiency assuming the ideal selection algorithm is equivalent to that of the combined calorimeter and tracker CNNs approach. This result confirms that the calorimeter CNNs algorithm by itself enables a near $\sim$99\% track reconstruction efficiency even if combined with the standard Kalman pattern recognition. In this case, however, the problem of the identification of the correct track remains open, especially for protons. In particular, a significant drop of the proton track selection efficiency is observed at the highest energies, which  is related to the increased multiplicity of hits in the tracker when approaching the PeV regime. On the other hand, the combined calorimeter and tracker CNNs approach provides similar track reconstruction performance and at the same time solves the track identification problem. Moreover, as noted above, even with the conventional (CPUs) hardware, the time required for a single inference of the tracker CNNs model is one order of magnitude smaller than that required by the standard track pattern recognition in DAMPE.

For case (b), the efficiency is identical to the case when both calorimeter and tracker CNNs are used, but only until~100~TeV. At higher energies, a sharp decrease of the efficiency is observed. The latter is related to the resolution degradation of the conventional shower axis direction reconstruction, caused by the saturation of the BGO calorimeter~\citep{mikhail,Yue:2020hmj}. 

With the above argumentation, the two CNNs algorithms can be considered complementary: the tracker CNNs model provides a relatively fast and reliable way of reconstructing the primary particle track, without the need of further track identification, while the calorimeter CNNs model recovers the drop of the tracking efficiency at the highest energies. An open question remains whether it is possible to perform a purely tracker-based CNNs reconstruction of particle trajectory without relying on the calorimeter or, at least, not requiring a dedicated calorimeter CNNs model. This could be potentially achieved by introducing at the highest energies an additional (third) step to the tracker CNNs reconstruction, with a large pixel size. In other words, a ``very coarse'' Hough model could be added to the CNNs chain prior to the coarse (400~\micron) and the precise (50~\micron) ones. Our estimates show that, if the conventional shower axis direction reconstruction is used at 100~TeV--PeV, a 99\%-containment RoI would correspond to a window of $\pm$160~\mm, twice bigger than that of the current coarse tracker model. While not strictly needed for achieving the goals of this paper, the purely tracker-based particle trajectory reconstruction at 100~TeV and higher energies could be an interesting subject for future research, in particular for the HERD detector. At this point, we can conclude that the combination of the calorimeter and tracker CNNs efficiently solves the problem of track reconstruction and identification with DAMPE. 
  
Finally, we have also tested the developed approach with particles heavier than helium. For intermediate mass ions like boron, carbon, nitrogen and oxygen, the tracking efficiency of the CNNs algorithm is in the 96\%-98\% range. We found that including these particles in the tracker CNNs model training only marginally (less than 1\%) improves their tracking efficiency. For iron, the tracking efficiency at the highest energies is about 92\% if no iron is included in the training, and more than 96\%  if it is included. At low energies, the iron tracking efficiency drops dramatically to 50\%--60\% if no iron MC is used in the training, and restores to about 95\% if it is included. Hence, we conclude that the tracker CNNs models are weakly sensitive to the differences between cosmic-ray particle species. The only strong dependence on particle type is observed for iron at low energies, which is likely attributed to the coarse CNNs model. Further optimisation of CNNs models for cosmic-ray analyses with elements heavier than helium is beyond the scope of this paper and will be considered in the future research.  
  
Validations of the complete CNNs reconstruction chain with the proton and helium data are shown in Section~\ref{sec:analysis}.

\section{Neural network inelastic classifier}
\label{sec:classifier}

Note that for the analysis in Section~\ref{sec:reconstruction}, in particular in Figures~\ref{fig:signal_psd_stk_truth} and~\ref{fig:signal_stk_standard}, we have applied the selection criterion requiring no inelastic interaction in PSD,  imposed on the MC truth level. Obviously, such a cut cannot be applied to real data. Therefore, to perform the rejection of events that interact  inelastically before the tracker, we have developed a dedicated classifier based on the CNNs approach. Hereafter it is referred to as the~\emph{inelastic} classifier.  

The classifier consists of two independent CNNs models, as shown in Figure~\ref{fig:stk_vertex_cnn}. For the first model, we use a network which takes as input a raw image of the STK. Similar to the BGO image in Figure~\ref{fig:bgo_cnn}, the raw STK image is constructed as a mixture of two projections. Since there are 12 layers (6 layers per projection) and each layer is read out by two groups of electronic boards (3072 readout channels per group), the STK image has a total dimension of 24$\times$3072 pixels. For the second model, we recycle the network architecture from Figure~\ref{fig:stk_cnn}, replacing the last ``Dense~4'' with the  ``Dense~1'' layer. The pixel resolution of the input Hough image is 400~\micron. Finally, we multiply the outputs of the two networks. Each of the two models demonstrate comparable performance if used standalone. At the same time, a simple multiplication of the outputs of two models yields better performance than any of them used separately. We deliberately choose not to mix the Hough and raw STK images in a single network, in order to maintain the architecture modularity and its interoperability for future applications. The performance of the total~\emph{inelastic} classifier is shown in Figure~\ref{fig:vertex_roc},  benchmarked against a typical PSD charge consistency cut.
 \begin{figure}
\begin{center}
\includegraphics[width=0.97\textwidth]{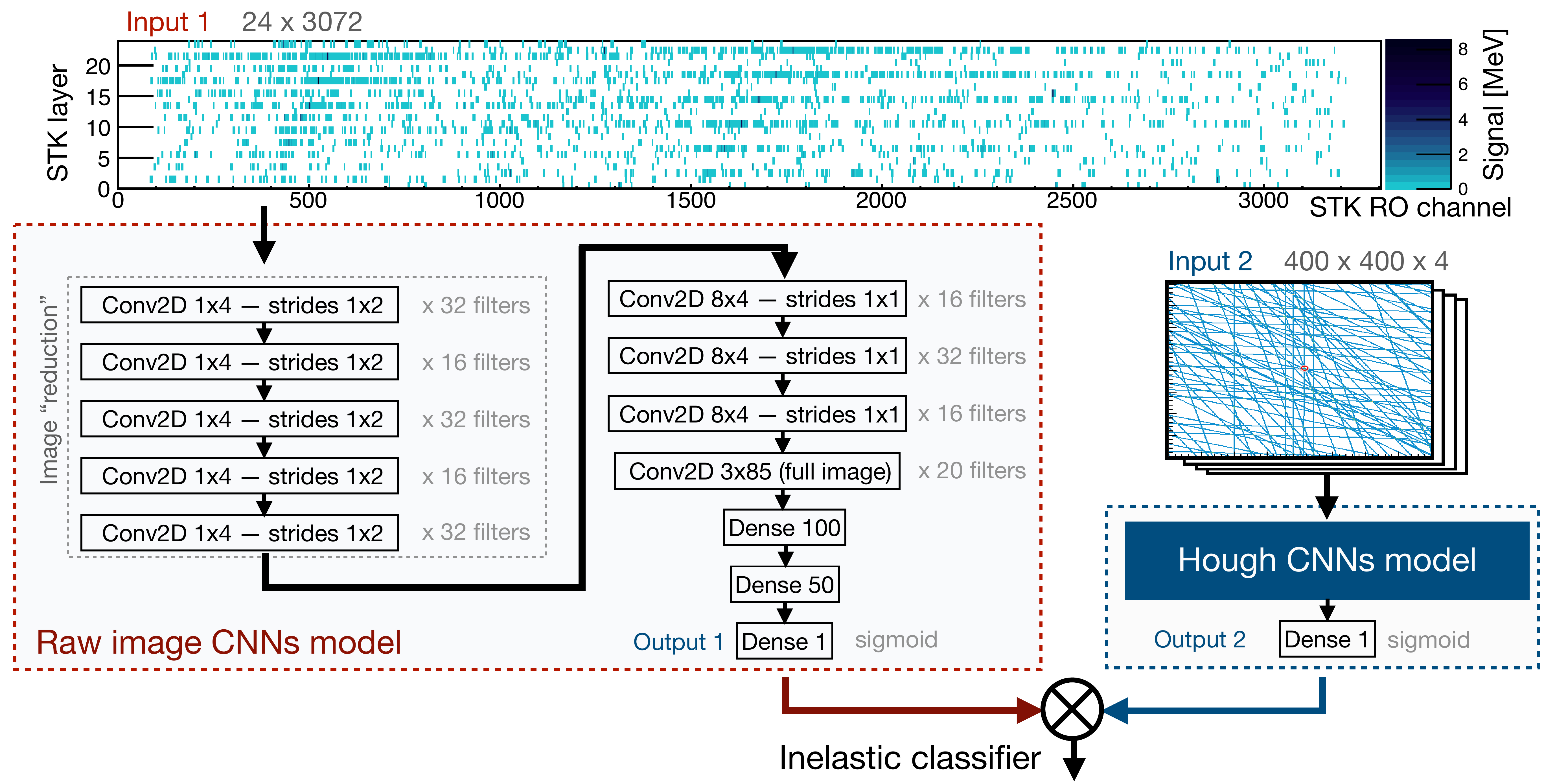}
\end{center}
\caption{A raw STK image of a typical proton interacting inelastically inside PSD and the architecture of the~\emph{inelastic} classifier CNNs. The blue box on the right corresponds  to the network in Figure~\ref{fig:stk_cnn} with the last layer replaced with ``Dense~1''.  The ReLU activation function is used in all layers except for the last one, which is activated by a sigmoid function. The total classifier is the product of the outputs of the two networks.}
\label{fig:stk_vertex_cnn}
\end{figure}   

\begin{figure}
\begin{center}
\includegraphics[width=0.49\textwidth]{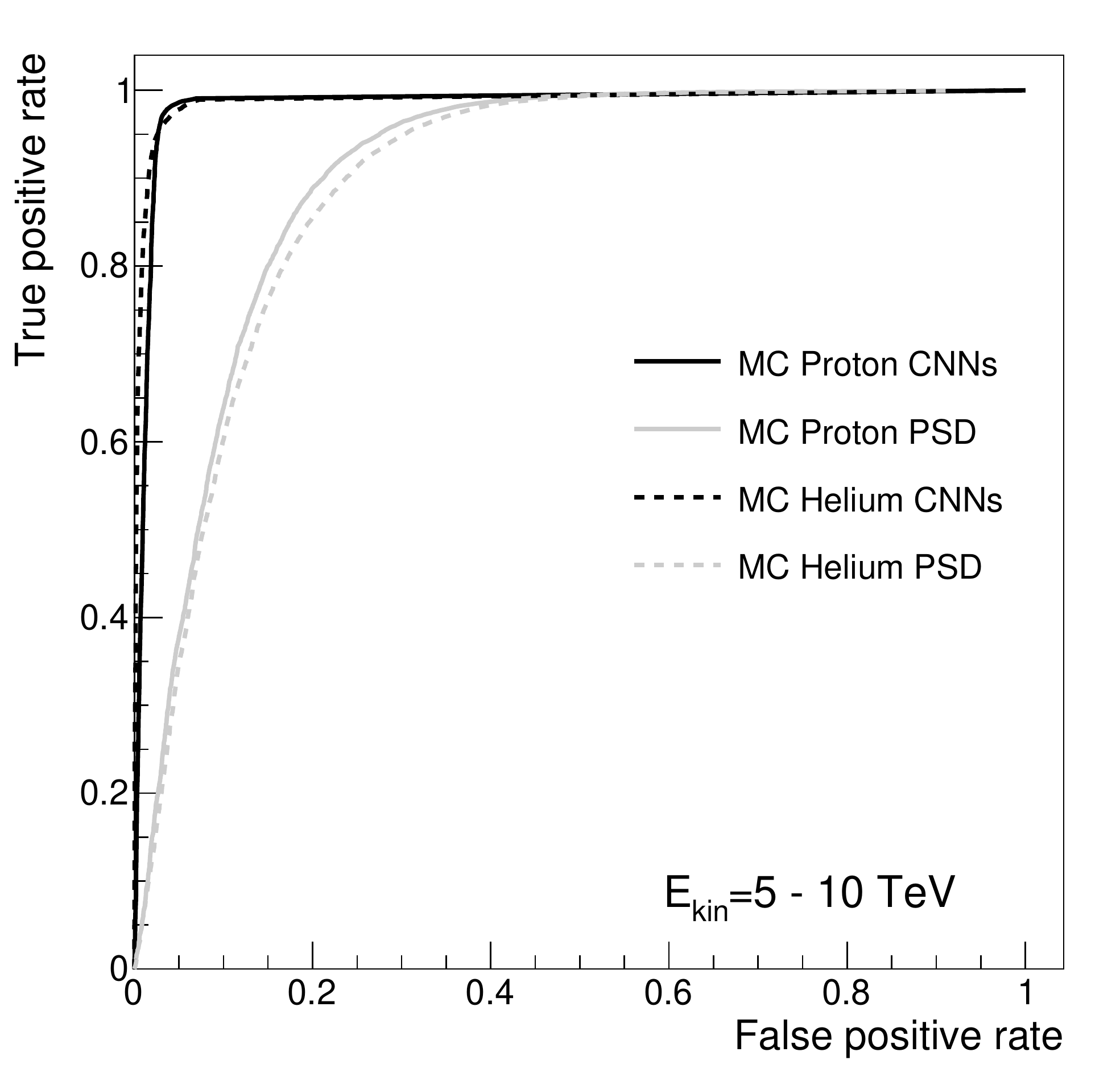}
\includegraphics[width=0.49\textwidth]{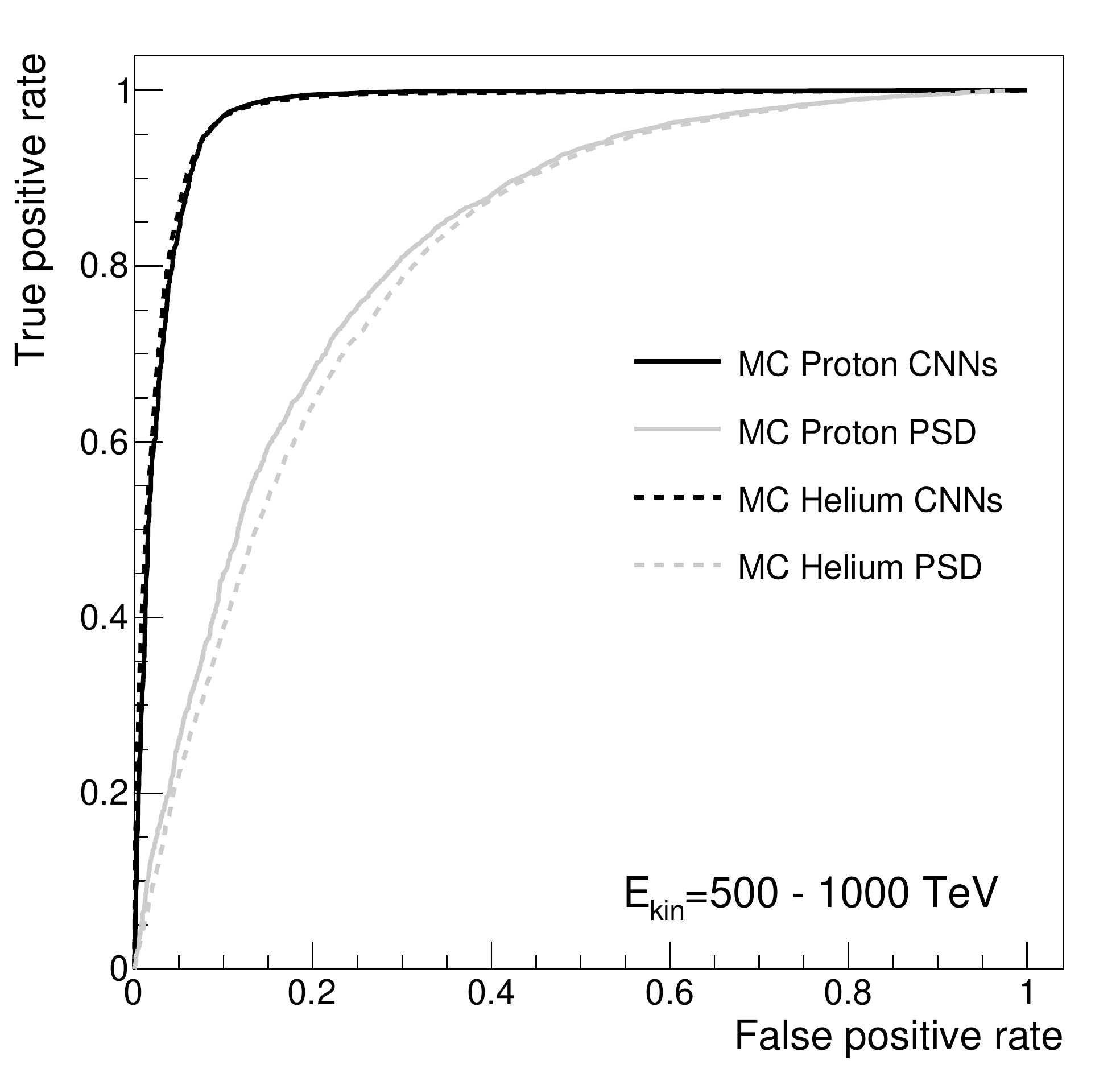}
\end{center}
\caption{Sample Receiver Operating Characteristic (ROC) curve of the~\emph{inelastic} CNNs classifier for selecting events that do not interact inelastically before the tracker (i.e. inside PSD or the passive support structures before the STK). For comparison, another ROC curve is shown for the typical PSD charge consistency cut, imposed on the difference between the highest and the lowest signal in the different PSD bars crossed by the particle. Two particle kinetic energy ranges are considered.}
\label{fig:vertex_roc}
\end{figure}   

\newcommand{\hoghvertextrainingtime}{12}
\newcommand{\rawvertextrainingtime}{6}
\newcommand{\hoghvertexinferencetime}{0.35}
\newcommand{\rawvertexinferencetime}{0.28}

The training of the two models is performed using the same training and validation MC sets as for the CNNs in the previous sections.  We do not split the training in different energy ranges, since no significant gain in accuracy is observed in this case.  The binary cross-entropy is used as a loss function. The models are fitted separately on the same input data. Notably, the two classifier models tend to a better generalization if some dropout is added. We tested 5\%, 10\% and 20\% for both networks. The former two options demonstrated comparable accuracy, while the latter one degraded the performance. Hence we have chosen a 10\% dropout for the training. The performance of the total classifier evaluated on the test and training samples did not show any significant deviation from one another, indicating the absence of overfitting. The Receiver Operating Characteristic (ROC) curves in Figure~\ref{fig:vertex_roc} are evaluated on the entire sample. 

It is worth noting that unlike the regression problem of the particle trajectory reconstruction, the classification of the inelastically interacting events can be successfully done with the raw tracker image. Notably, a pattern of an interacting particle can be spotted even by eye in Figure~\ref{fig:stk_vertex_cnn}. Therefore the CNNs used with a raw image is the right tool for classification in this case. 
 
 A peculiar alternating structure in the number of convolutional filters (Figure~\ref{fig:stk_vertex_cnn}) is introduced in order to gain maximum performance, while maintaining an optimal model complexity. In particular, if  32 or more filters are used in all convolutional layers, including the last one, the model  exhibits less generalization,  which degrades the accuracy by about~1--2\%.  Alternatively, a lower but fixed number of filters also yields inferior performance. We have noticed that the model is particularly sensitive to the number of filters in the very first layer. We have also tried increasing the horizontal span of the convolutional filters. The latter, however, did not result in any further improvement, hinting that inelastic interactions manifest more clearly through the  localised fine patterns in the raw STK image, rather than large-scale correlations between the alternating STK projections. 

%
%

In addition to the presented model, we have also tested if a neural network which takes PSD data as an input can serve as an~\emph{inelastic} classifier. In this case, we used a network architecture inspired by the calorimeter CNNs model in Section~\ref{sec:calo}. However, the performance of such network was never close to the tracker CNNs model. 
 This result is expected since the majority of the information about the particle passage through PSD is concentrated in at most 4 PSD bars crossed by the particle (see Figure~\ref{fig:event_displays}). Moreover, as described in Section~\ref{sec:analysis}, at high energies the PSD signal gets severely contaminated, obscuring the footprint of the particle absolute charge and its inelastic interaction, if present.

The training time of the Hough-image and raw-image models for $10^6$ batches is~\hoghvertextrainingtime~hours/epoch and~\rawvertextrainingtime~hours/epoch respectively. An increased time for the  Hough model compared to the regression case in Section~\ref{sec:tracker} is explained by the addition of the dropout between all the layers. The inference (CPUs) time of two models is~\hoghvertexinferencetime~s/event and~\rawvertexinferencetime~s/event respectively. The use of the~\emph{inelastic} classifier with data is demonstrated in the next section.

\section{Application case: proton and helium analyses}
\label{sec:analysis}

In this section, we use the developed CNNs algorithms to perform the absolute particle charge identification in the STK. That is, we repeat the event selection described in Section~\ref{sec:reconstruction}, replacing the standard track reconstruction with the one developed in this work. Also, we add a selection criterion on the~\emph{inelastic} classifier to require that no inelastic interaction occurs in the PSD. 
We choose a cut value to maintain the true positive rate of selecting non-interacting events higher than 95\% in the entire energy range. The results are shown in Figure~\ref{fig:stk_signal_cnn_data_mc}.
The distributions at a deposited energy of 1~TeV and above are shown -- the distributions at lower energies have similar behavior -- a good agreement between flight data and simulation is observed. To account for minor differences between flight data and simulation\footnote{As an example of minor data/MC difference of the STK signal, see Figure 7 in~\citep{Tykhonov:2017uno}.} smearing corrections of peak positions/widths in proton and helium MC samples have been performed. The systematic uncertainty of the proton and helium charge identification efficiencies, related to the smearing correction, is conservatively estimated to be within 1\%. Thanks to the clean proton and helium peak identification with the CNNs approach, the relative background cross-contamination for either proton or helium analyses at all energies does not exceed 1\% (2\%), at 90\% (95\%) STK signal selection efficiency. Note that in the case of the standard selection, the corresponding backgrounds at the highest energies reach 3\% (5\%) for proton and 10\% (13\%) for helium, respectively. The effect of applying the~\emph{inelastic} classifier can be clearly observed in Figure~\ref{fig:stk_signal_cnn_data_mc}, as it helps eliminate the middle and far ``tails'' around the proton and helium peaks\footnote{About 1--2\% of events yield  inelastic interactions in PSD with highly collimated secondary particles which cannot be resolved in the STK.}.

\begin{figure}
\begin{center}
\includegraphics[width=\phefiguresize\textwidth]{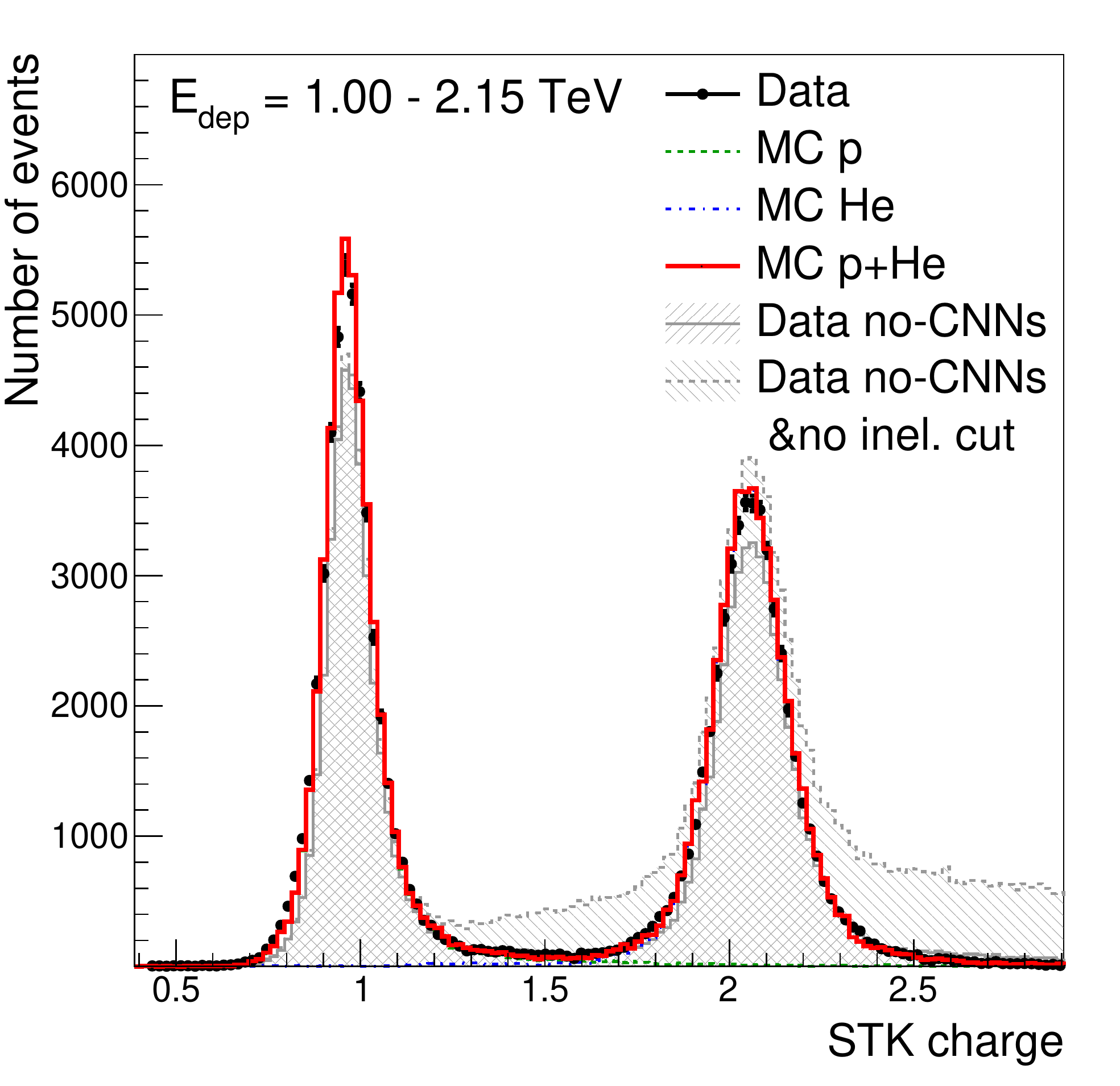}
\includegraphics[width=\phefiguresize\textwidth]{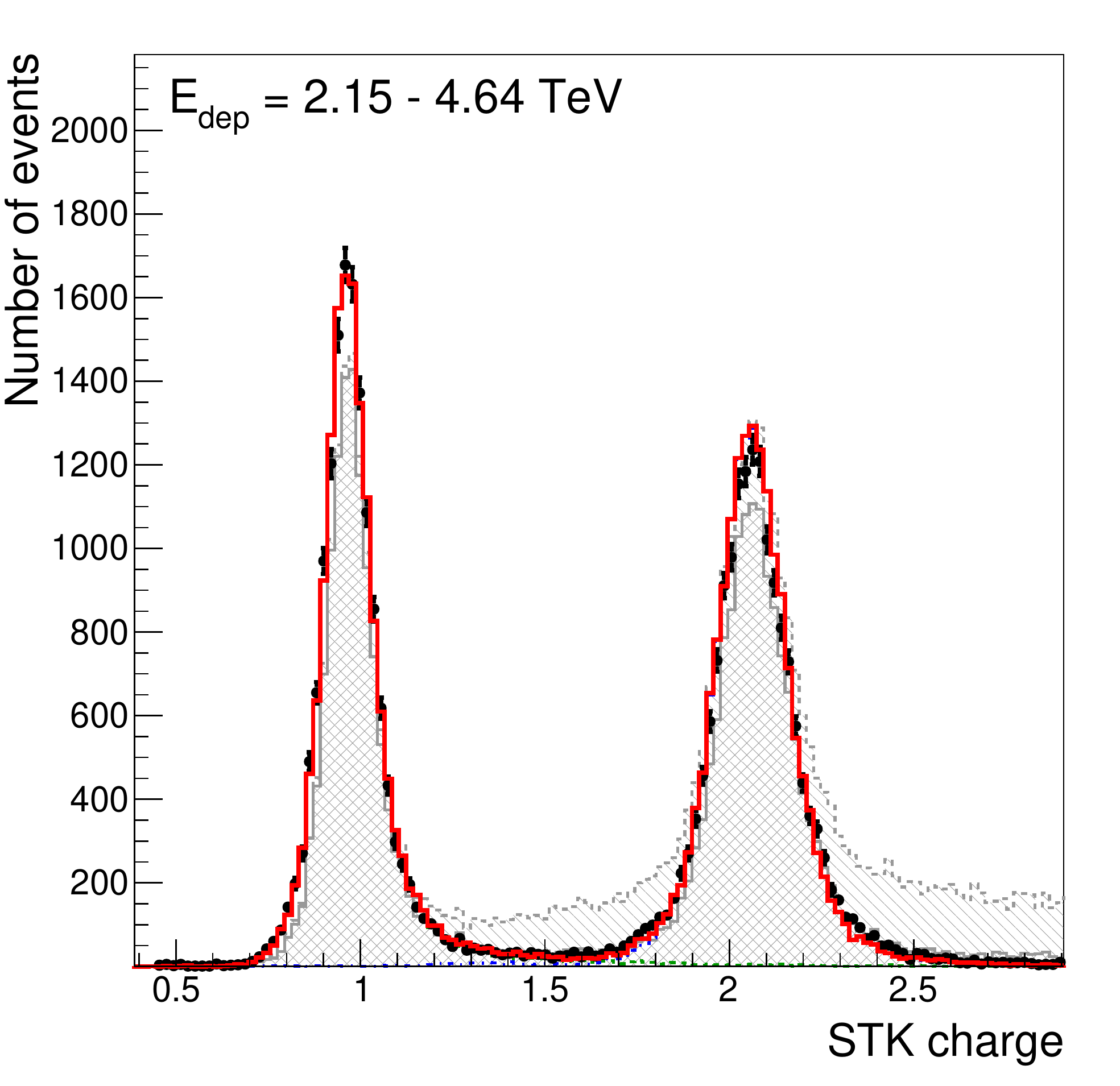}
\includegraphics[width=\phefiguresize\textwidth]{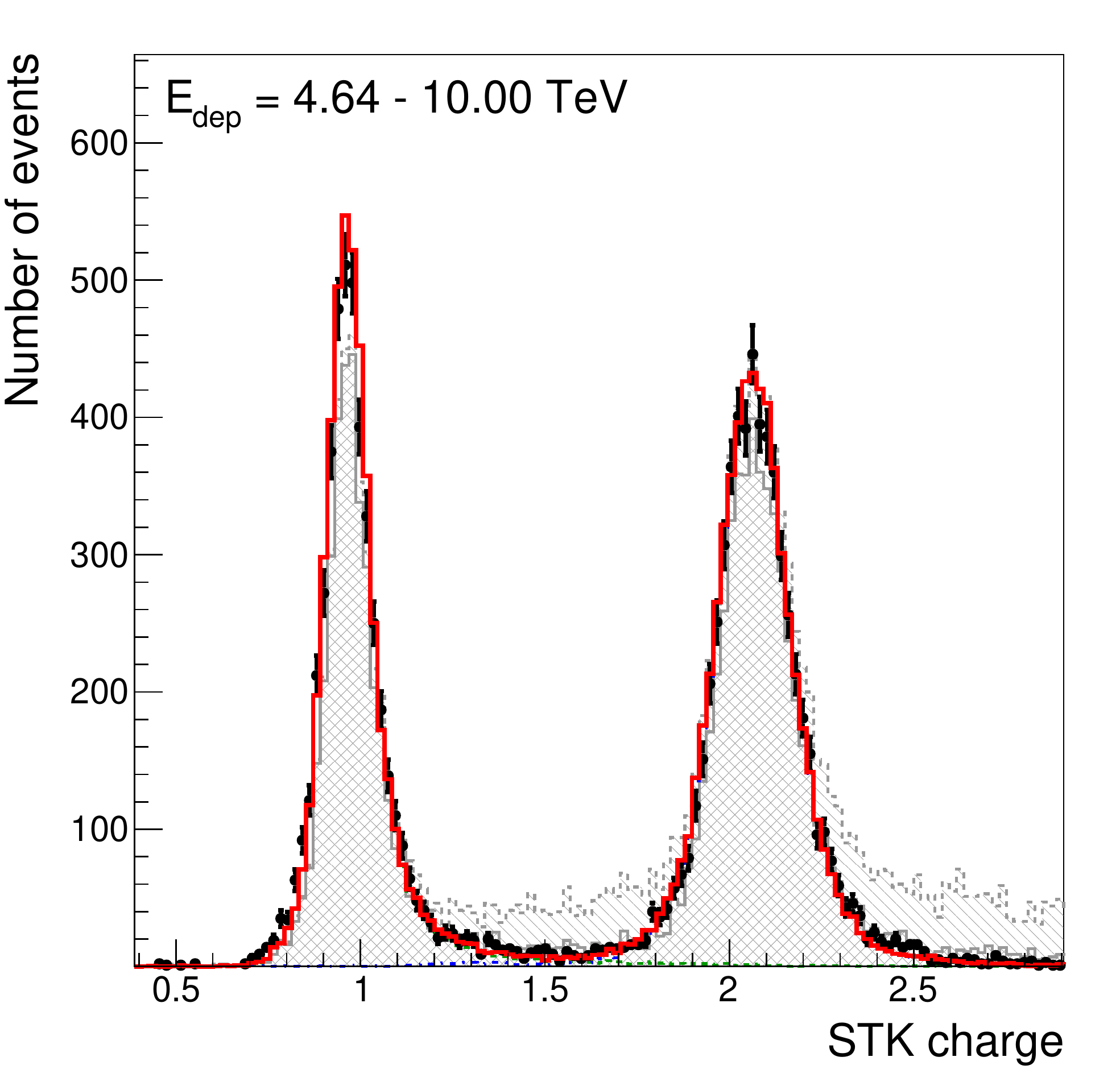}
\includegraphics[width=\phefiguresize\textwidth]{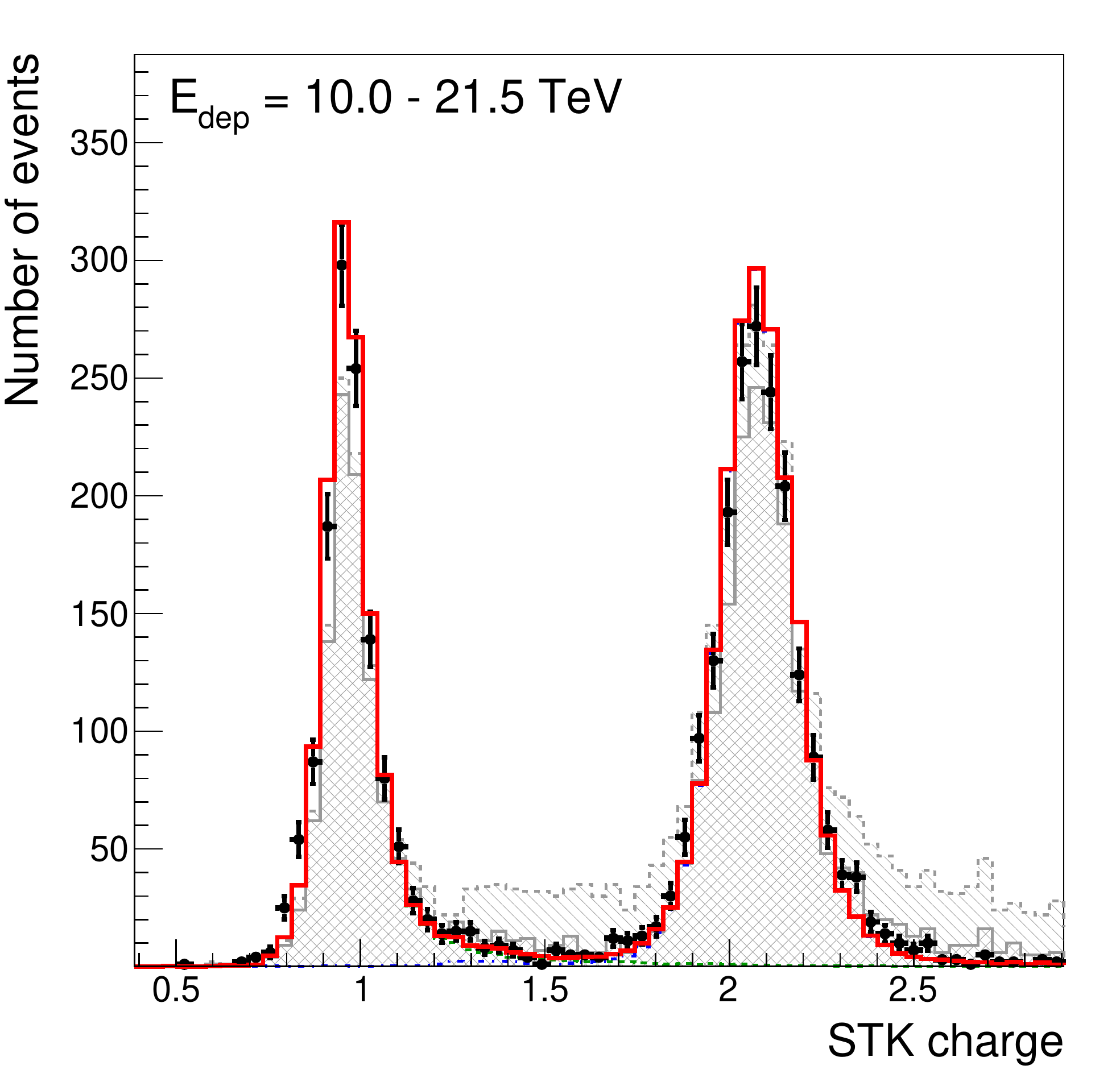}
\includegraphics[width=\phefiguresize\textwidth]{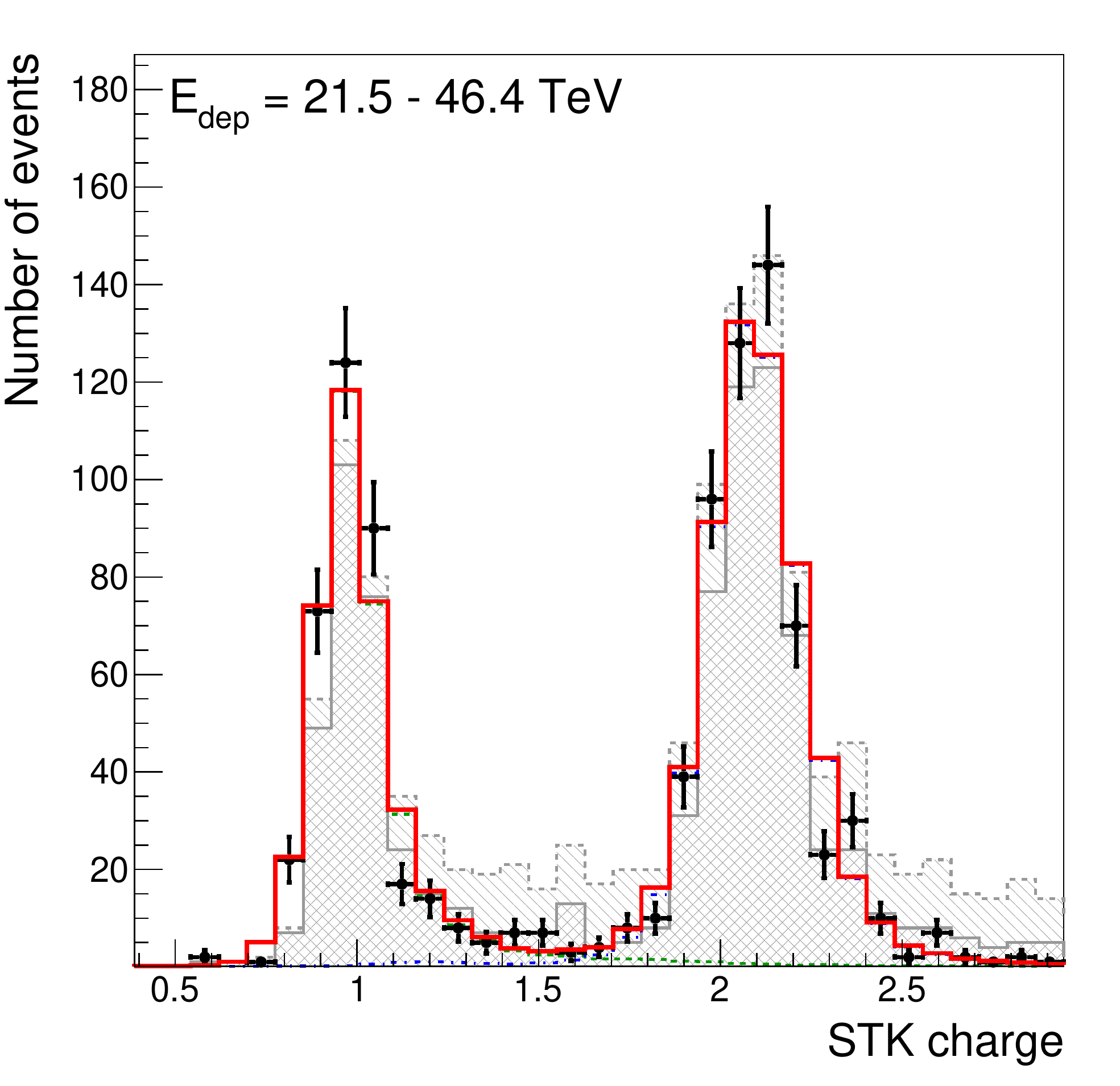}
\includegraphics[width=\phefiguresize\textwidth]{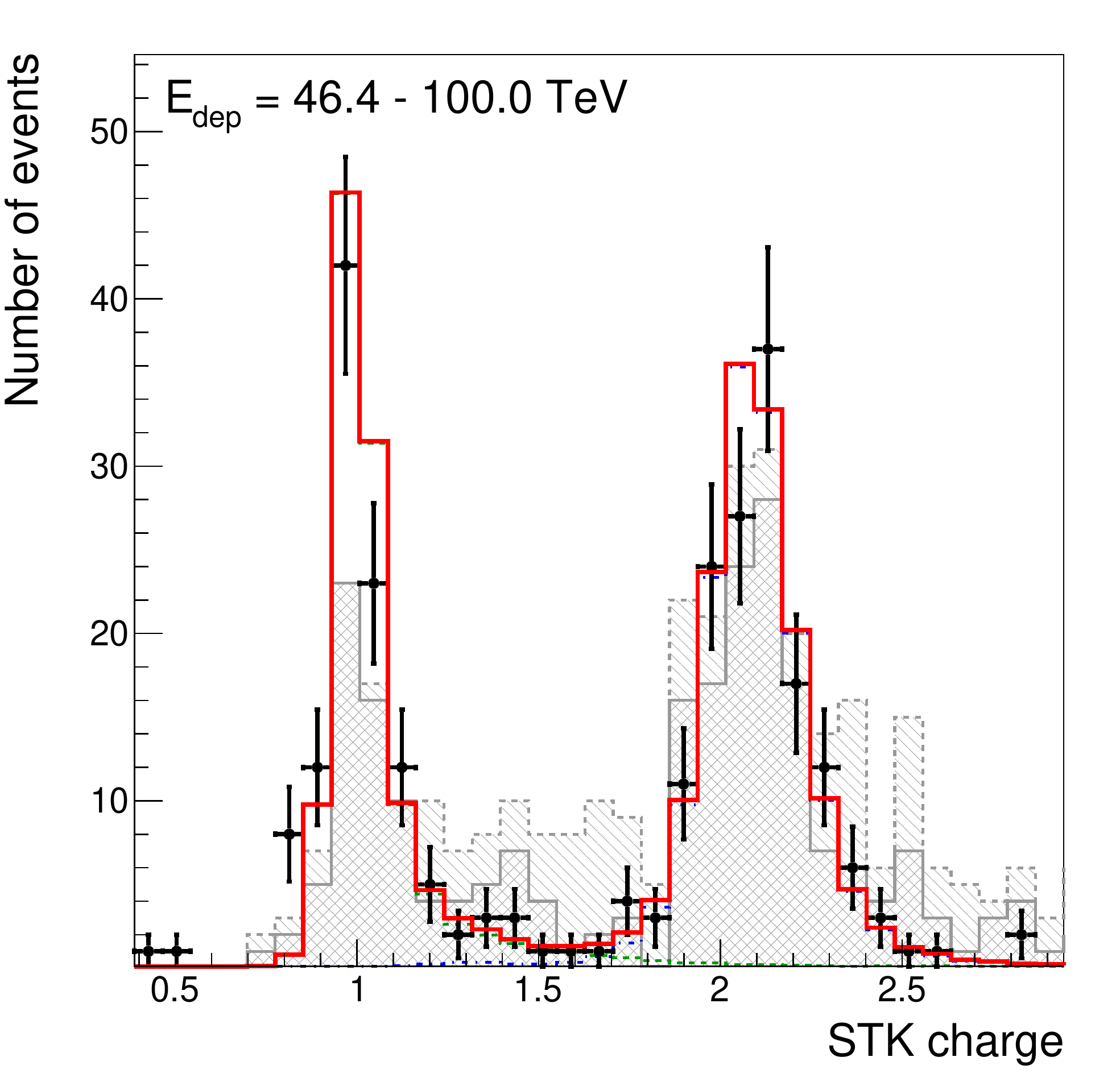}
\end{center}
\caption{Particle absolute charge measured with the STK using the track reconstructed by the CNNs model chain. The selection cut requiring no inelastic interaction inside the PSD is imposed. For comparison, similar data distributions using the standard track reconstruction and identification (same as in Figure~\ref{fig:signal_stk_standard}--right) are shown, with and without the~\emph{inelastic} cut.}
\label{fig:stk_signal_cnn_data_mc}
\end{figure} 

Next, we perform  validations of the CNNs track reconstruction on flight data. While there is no way of knowing the true particle direction outside the simulation, we can still validate the CNNs algorithm using the standard track reconstruction algorithm as a reference. Namely, we evaluate the ratio $r$ of the number of reconstructed events for the developed CNNs algorithm and the standard approach. The resulting comparison is shown in Figure~\ref{fig:trk_eff_data_mc}. This ratio is equivalent to the ratio of ``CNNs reco'' and ``reco \& selection'' efficiencies from Figure~\ref{fig:trk_eff}:
\begin{equation*}
r\equiv\frac{ \mathrm{N}_{\mathrm{CNNs}}}{\mathrm{N}_{\mathrm{std}}} =\frac{\mathrm{N}_{\mathrm{CNNs}}/\mathrm{N}_{\mathrm{tru}}}{\mathrm{N}_{\mathrm{std}}/\mathrm{N}_{\mathrm{tru}}}  = \frac{\epsilon_{\mathrm{CNNs}}}{\epsilon_{\mathrm{std}}}.
\end{equation*}
Note that the two approaches are completely independent, hence specific problematic behaviors (if any) of the CNNs approach that can potentially arise, for example, due to some imperfections in the simulation accuracy,  would likely manifest in such ratio.  It can be seen that the improvement of the developed CNNs track reconstruction over the standard approach is fully consistent with the MC prediction. The error bands are mostly attributed to  the standard track reconstruction, selection, and particle identification, namely to the track selection uncertainty at the lowest energies and background estimation uncertainty at the highest energies.  From this comparison, we conservatively estimate the systematic uncertainty of the CNNs tracking approach to not exceed 1--2\%.

\begin{figure}
\begin{center}
\includegraphics[width=0.55\textwidth]{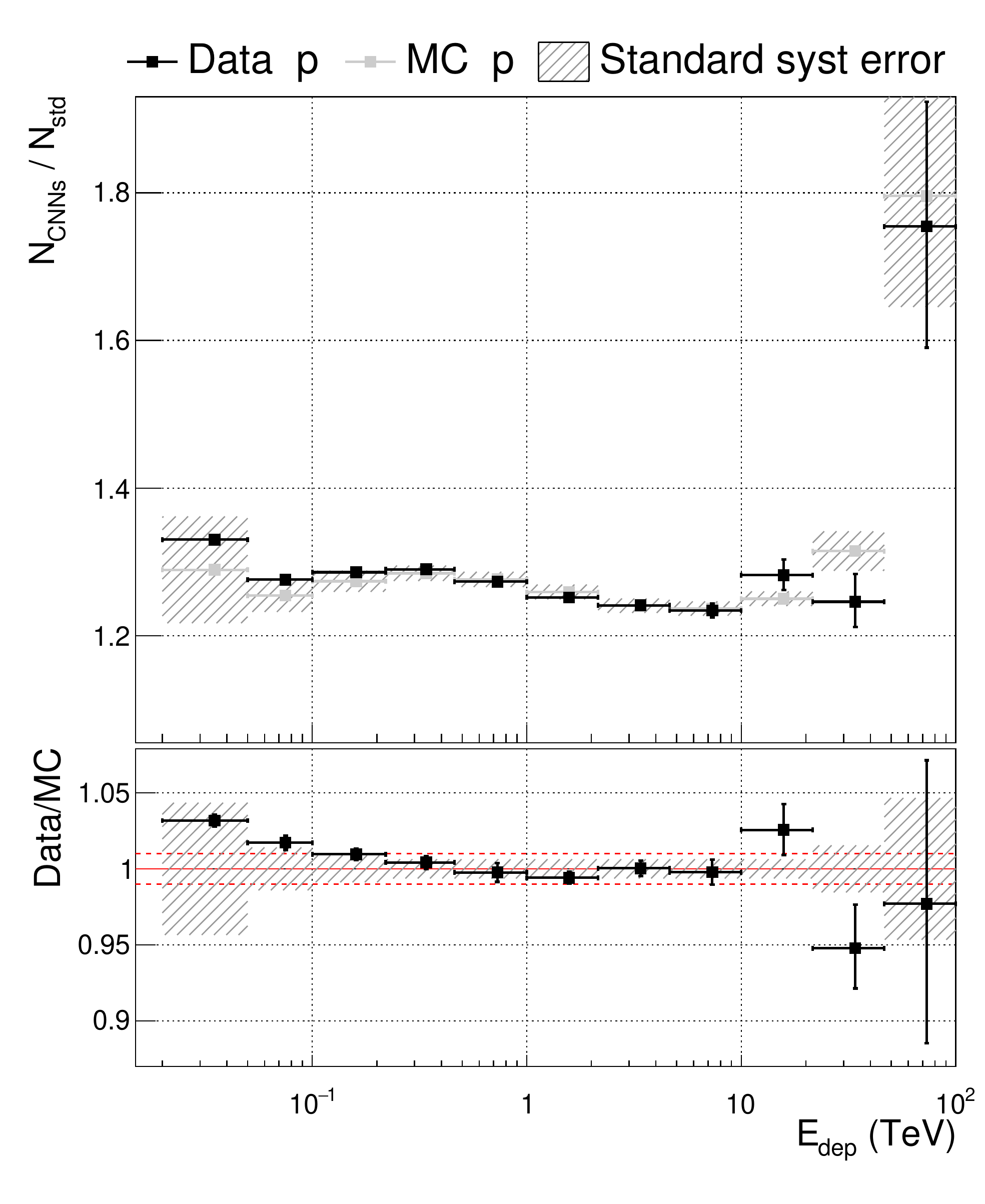}
\includegraphics[width=0.55\textwidth]{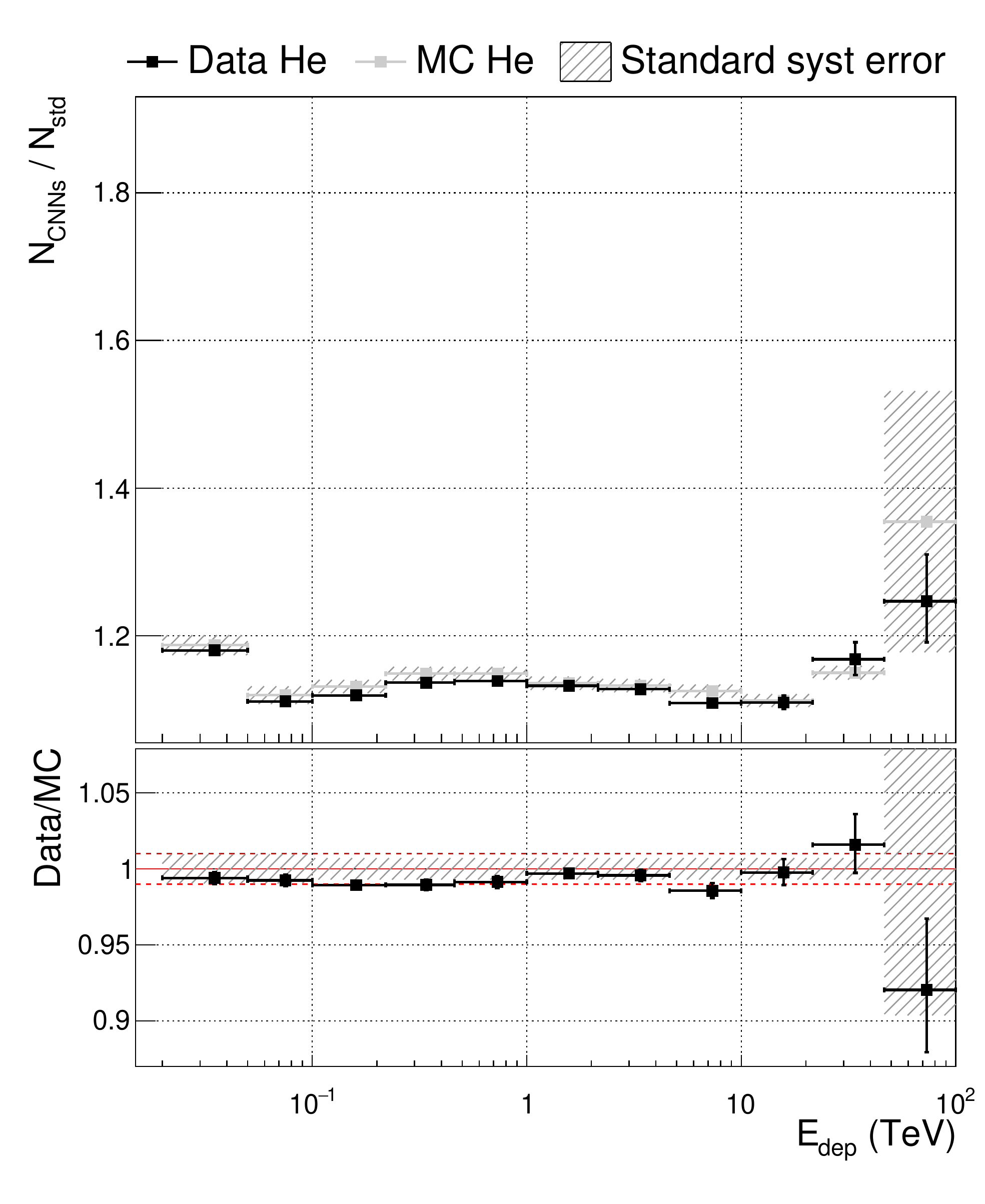}
\end{center}
\caption{The ratio of the number of reconstructed events with the developed CNNs algorithm and the standard track reconstruction and selection algorithm, as a function of deposited energy in the detector. It is obtained for the 90\% STK signal containment intervals for both proton and helium distributions (Figure~\ref{fig:stk_signal_cnn_data_mc}). The ratios for  85\% and 95\% containment have similar behavior. Red dashed lines indicate the region of $\pm$1\% data/MC deviation, to guide the eye.}
\label{fig:trk_eff_data_mc}
\end{figure} 

Finally, we summarise the characteristic execution time of the developed algorithms implemented in the DAMPE data processing pipeline. Up to five CNNs inferences are performed per event: one for the calorimeter, two for the tracker trajectory reconstruction, and two for the~\emph{inelastic} classifier. The total CPUs time consumption for the five model inferences is about~\cnninferencetotalfivemodels~s/event. Notably, the standard track reconstruction algorithm of DAMPE takes on average~\kalmantimecpu~s/event.

\section{Summary and discussion}
\label{sec:conclusion}

Particle trajectory reconstruction represents one of the key challenges on the way of extending the direct cosmic-ray measurements towards the PeV landmark.  To tackle this problem, we have developed and implemented a new tracking paradigm for the calorimetric cosmic-ray detectors in space. Our approach is based on the deep learning methods, in particular Convolutional Neural Networks (CNNs). The reconstruction of the cosmic ray trajectory is performed in multiple steps: first, a ``rough'' prediction of the impinging particle direction is inferred by the CNNs from the image of the calorimeter subdetector; the calorimeter prediction defines the RoI in the tracker, where the further reconstruction is done; then, the Hough-transformed image of the tracker RoI is used for inferring the precise trajectory, using another dedicated CNNs model. Finally, specific signal hits in the tracker are assigned to the predicted trajectory based on a simple distance matching, forming the reconstructed track of the impinging cosmic ray. In addition, to enhance the analysis we have also developed a CNNs-based classifier to reject events with early pre-showers, for which the cosmic ray identity (absolute charge) cannot be determined with the tracker. Unlike the conventional approach, where multiple candidate tracks are constructed and then the best track is selected on the analysis level, our algorithm yields one track from the beginning -- a selection in the further analysis is not needed. 
 
We have benchmarked our algorithm for the case of proton and helium selection. The developed algorithm demonstrates excellent tracking efficiency, about 98\% in the entire energy region of the DAMPE data. At the same time, the absolute charge mis-identification is very low, in the 1--2\% range. While implemented for DAMPE, the designed approach is equally relevant for next-generation experiments, like HERD~\citep{herd_2021_icrc_gargano,herd_2021_icrc_Pacini}. Moreover, HERD is targeting cosmic ray measurements at energies higher than DAMPE. Hence the problem of track reconstruction will manifest itself on a bigger scale, and the outlined CNNs approach appears as a credible potential solution.

The two key elements of the approach, the tracker and the calorimeter CNNs are complementing each other. The  former allows to perform a precise particle trajectory reconstruction in the tracker, given that the RoI is provided by the calorimeter, while the latter one enables to recover the loss of track reconstruction efficiency at the highest energies, caused by the calorimeter saturation. It is worth noting that the set of the tracker CNNs models is sufficient by itself for reaching the target performance up until $\sim$100~TeV, even if combined with the classical calorimeter shower axis direction reconstruction. We choose not to mix the tracker and the calorimeter data in a single CNNs model in order to facilitate: the cross-calibration of the algorithms based on different subdetectors; the interoperability of models for other applications, beyond the proton and helium analyses; the validation of the models with the data;  the optimal integration of the  developed algorithms in the data processing pipeline.
    
Aside from the tracker CNNs, the obtained results for the calorimeter CNNs direction prediction look interesting~\emph{per se}. The developed algorithm at the highest energies outranks the conventional approach of analytical fitting of the shower axis, by at least a factor of 5. In particular, the 68\% angular containment of electromagnetic showers at TeV and higher energies is better than 0.4\degree. Even more intriguing, the CNNs approach applied to a finer and thicker calorimeter having a 3-D granularity, like that in HERD, is naturally  expected to yield even better results. A possible question at this point could be whether  a purely calorimeter-based direction reconstruction of electromagnetic showers could replace a dedicated tracker, at least at the highest energies. An answer to this question would be important in the design phase of new instruments, in particular with respect to their gamma-ray detection capability and the need for dedicated photon converters. This remains a subject for future research. 
 
\section{Acknowledgment}

The DAMPE mission was funded by the strategic priority science and technology projects in space science of the Chinese Academy of Sciences.  In Europe, the experiment and data analysis is supported by the Swiss National Science Foundation  and the National Institute for Nuclear Physics (INFN), Italy. Development of the deep learning techniques and related data analysis is funded by the European Union's Horizon 2020 research and innovation programme (grant agreement No 851103).

Ukrainian authors, M.~Deliyergiyev, A.~Kotenko, and A.~Tykhonov are indebted to the resilience and courage of the Armed Forces of Ukraine, for keeping their loved ones safe during the course of this work, without which  the presented results would have been impossible   to achieve.



\bibliographystyle{elsarticle-num}
\bibliography{paper_arxiv}








\end{document}